# Population change, age structure, and socio-economic performance

Running title: Population change, age structure, and socio-economic performance

Corey J. A. Bradshaw[1,2,*], Shana M. McDermott[3], Matthew E. Oliver[4]

[1]Global Ecology | *Partuyarta Ngadluku Wardli Kuu*, College of Science and Engineering
Flinders University, Adelaide, South Australia, Australia
[2]Australian Research Council Centre of Excellence for Indigenous and Environmental Histories and Futures, Cairns, Australia
[3]Department of Economics, University of Missouri, Columbia, Missouri, USA
[4]School of Economics, Georgia Institute of Technology, Atlanta, Georgia, USA

*Corresponding author: Corey J. A. Bradshaw; e-mail: corey.bradshaw@flinders.edu.au; tel: +61 400 697 665

**ORCID**

**CJAB**: 0000-0002-5328-7741; **SMM**: 0000-0002-1005-5481; **MOE**: 0009-0001-5183-2582

## ABSTRACT

Concerns about declining or ageing populations often centre on the possibility that fewer people or older age structures could weaken economies, strain fiscal systems, and reduce living standards. However, population change reflects multiple demographic processes, including fertility, mortality, and migration, and its relationship with socio-economic performance depends on capital accumulation, human capital, productivity, institutions, and policy responses. We examine national data at the global scale to test whether, and to what extent, overall population growth or older population age structures are associated with economic and social outcomes using hybrid machine-learning approaches. Across nine indices of socio-economic performance, we find no evidence that slower-growing or older populations perform worse on average. Instead, countries with low or negative overall population growth and older age structures often have higher values for many indicators, and within-country time series show similar broad patterns for most responses. Our results suggest that demographic change is not inevitably associated with socio-economic decline, and that further research is needed to identify the institutional, economic, policy, and demographic mechanisms underlying these global patterns.

## Introduction

Public and policy debates about demographic change often focus on whether slower population growth, population decline, or population ageing could weaken economic performance, fiscal sustainability, or living standards. As of 2019, 55 countries had instituted explicit measures to raise birth rates (United Nations Department of Economic and Social Affairs Population Division, 2021), reflecting widespread policy concern about low fertility and population ageing. A commonly cited concern is that slower overall population growth, including decline, could reduce the share of working-age adults and thereby lower aggregate output, typically measured as gross domestic product (GDP), with potential implications for per-capita living standards. Another concern is that older populations might have different spending patterns from younger populations, including lower purchases of some durable goods, which could dampen domestic consumption (Hurd and Rohwedder, 2022). At the same time, rapid population growth has also been associated with pressure on natural resources, infrastructure, and labour markets (Alexandratos, 2005; Bloom and Freeman, 1986; O'Sullivan, 2020). These different concerns illustrate that demographic change can raise challenges under both faster- and slower-growth scenarios, depending on the mechanisms involved and the institutional context.

This aggregate measure reflects the combined effects of fertility, mortality, and migration, and our analyses do not distinguish among these components. We therefore do not interpret associations with overall population change as evidence for or against any specific fertility, mortality, or migration policy. Likewise, we use ‘ageing’ to refer to population age structure, measured here through the ratio of people aged ≥ 65 years to those aged 16–64 years, rather than to a causal demographic process.

Despite the importance of demographic change for economic, social, health, and environmental systems (Bradshaw et al., 2021; Bradshaw et al., 2026a; Bradshaw et al., 2026b; Saraswati et al., 2024), few studies have tested the relationships among overall population change, age structure, and measures of national wealth and well-being at global and regional scales. Even fewer have examined these relationships across a broad suite of socio-economic indicators that extend beyond GDP, or considered how productivity, institutional quality, and well-being co-vary with demographic structure (Bloom et al., 2015; Faruqee and Mühleisen, 2003; Oliver, 2015; Reher, 2007). This gap matters because GDP is an incomplete measure of economic performance and does not fully capture national wealth, distributional conditions, environmental pressures, or well-being (Costanza et al., 2009; Kubiszewski et al., 2013; Oxfam International, 2012) (Appendix I).

In this paper we address this limitation by examining a broader set of socio-economic indicators that capture multiple dimensions of economic and social performance relative to variation in country-level population growth and age structure. Throughout, we use 'population change' or 'population growth' to refer to net change in total national population size. Although our analyses do not establish econometric causality, they provide a global assessment suggesting broad patterns. Specifically, we test two descriptive hypotheses. First, if overall population change is systematically associated with socio-economic performance, then national indices of wealth, income inequality, productivity, political stability, well-being, and health should vary with the rate of total population change. Second, if population age structure is systematically associated with socio-economic performance, then these indicators should vary with the percentage of older cohorts in the population. These hypotheses are associational rather than causal: they test whether broad global and within-country patterns are consistent with common claims about demographic change, not whether fertility, mortality, migration, or ageing independently causes socio-economic outcomes.

We compiled global data on wealth, income inequality, productivity, corruption, freedom, human development, well-being, and healthy life expectancy from international sources, including the World Bank, Penn World Table (Feenstra et al., 2015), Transparency International, Freedom House, the United Nations, the Gallup World Poll, and the World Health Organization. We then examined relationships between these response variables and two demographic predictors: the mean rate of total population change and national age structure. We used resampled boosted regression trees for cross-country analyses, complemented by within-country time-series, regional, and lag-aligned analyses. We intend for these methods to identify broad, nonlinear, and potentially heterogeneous patterns.

Overall, we find no evidence that countries with slower overall population growth or older age structures perform worse according to the socio-economic indicators we examine. In many analyses, these countries perform better on average, although these associations should not be interpreted as evidence that slower growth or ageing causes better outcomes. The results instead suggest that demographic change alone is an insufficient explanation for national socio-economic performance, and that future research should examine the mechanisms that might underlie these patterns, including age-specific skills, human capital, migration, productivity, institutions, public policy, and historical development trajectories. We therefore interpret our findings as evidence against deterministic claims that slower population growth or ageing necessarily entails socio-economic decline, rather than as support for any particular population policy.

## Background and related literature

The relationships between demographic trends and socio-economic outcomes are often interpreted through the lens of traditional economic models that link population size to economic expansion. However, the relationship between population and economic performance is complex compared to the assumptions of neoclassical economic growth models (Bloom et al., 2024). For example, early models such as Ramsey (1928) and Solow (1956) allow population size to influence economic outcomes, whereas subsequent work emphasised the importance of physical and human capital accumulation and technological progress in determining long-run living standards (Cass, 1965, 1966; Koopmans, 1965; Koopmans, 1969).

Even in modern macroeconomic frameworks such as standard dynamic stochastic general equilibrium models used in policy analyses, long-run growth is typically driven by an assumed rate of technological progress, and population change mainly affects output rather than the long-run growth rate itself (Christiano et al., 2005; Smets and Wouters, 2003). In models incorporating endogenous innovation and research and development (Aghion and Howitt, 1992; Romer, 1990), demographics can play a larger role. For example, in semi-endogenous growth models (Jones, 1995), a larger population might increase the pool of researchers and thereby affect the pace of technological progress. More broadly, macroeconomic research suggests that long-term economic performance depends heavily on capital per person (both physical and human) and technological progress, rather than population size alone (Aghion and Howitt, 1992; Grossman and Helpman, 1991; Ramsey, 1928; Romer, 1990; Solow, 1956). In other words, population change does not automatically lead to changes in economic outcomes — what matters more is how societies invest in and share their capital and skills (Bloom et al., 2024). That said, recent studies such as Jones (2022) and Maestas et al. (2023) have argued that demographic change might still present growth challenges under certain conditions, underscoring the need for empirical tests at a global scale.

The academic literature therefore already provides a nuanced account of demographic-economic relationships, but public and political debates often simplify this complexity into claims that population growth is either necessary for prosperity or inherently harmful. Policy debates can therefore outpace the empirical evidence, particularly when aggregate population change is treated as a proxy for fertility, labour supply, ageing, migration, or productivity (Bloom et al., 2024; Klancher Merchant and Brown, 2024; Master, 2025; Population Connection, 2025).

Governments have adopted both pronatalist and antinatalist policies at different times, often invoking economic, fiscal, social, or environmental rationales. For example, Hungary has used a range of fertility-supporting policies over several decades, Russia has adopted policies intended to raise fertility, and China shifted in 2021 toward measures supporting couples to have more children after decades of fertility-restricting policy (United Nations Department of Economic and Social Affairs Population Division, 2021; Velikaya and Rodionova, 2025). Conversely, many countries have previously pursued policies designed to slow population growth, often in the name of development, poverty reduction, environmental management, or maternal and child health. We do not evaluate the desirability, effectiveness, or motivations of these policies. Rather, these examples illustrate that demographic policy is politically contingent and that claims about the economic consequences of demographic change require careful empirical evaluation.

Although the effects of demographic change on long-term economic growth are mixed and depend critically on capital accumulation and productivity, demographic change can directly influence fiscal systems and labour markets. Ageing populations are often discussed in relation to the sustainability of health care systems, pensions, and social security (Auerbach and Kotlikoff, 1987; Bloom et al., 2015; Faruqee and Mühleisen, 2003). These concerns are not unfounded, especially given that the share of people aged 65 and older is projected to rise from 10% globally in 2022 to 16% by 2050 (United Nations Department of Economic and Social Affairs Population Division, 2022).

A substantial literature has also examined possible fiscal and institutional dimensions of demographic change, such as financial sustainability of local governments (Buendía-Carrillo et al., 2025; Lara-Rubio et al., 2026; Rodríguez Bolívar et al., 2017; Rodríguez Bolívar et al., 2016), the efficiency of public service delivery (Alcaide Muñoz et al., 2024), and long-term fiscal obligations (Buendía-Carrillo et al., 2020). However, these studies focus primarily on government finances and institutional performance rather than broader measures of socio-economic outcomes.

What is clear is that demographic change can affect economies through more than one channel. Although fewer workers are expected to create (especially low-skill) labour shortages and dampen aggregate consumption growth (Brucker Juricic et al., 2021; Henkens et al., 2008; Nielsen and Fang, 2007), lower fertility combined with stable or rising capital per person might be associated with higher per-capita income and increased investment in education and skills and even ease resource and environmental pressures (Casey and Galor, 2017; Lee and Mason, 2010; Maja and Ayano, 2021; Saraswati et al., 2024). For example, Garloff and Wapler (2016) found no evidence that an increasing number of retirees in an occupation in Germany created higher demand for younger workers, mainly because middle-aged cohorts changed occupations in response.

Empirical evidence on the broader economic implications of demographic change is also mixed. Studies of high-income nations such as Japan (Ohashi et al., 2019), Italy (Zambon et al., 2020), and Australia (Bradshaw and Brook, 2016b) suggest that declining fertility and slower population growth do not necessarily imply weaker economies. In some cases, lower fertility has been associated with increased investment in human capital and higher per-capita consumption (Lee and Mason, 2010; Prettner et al., 2013), as well as changes in environmental pressures (Casey and Galor, 2017). Conversely, rapid population growth can pose its own challenges when economic and institutional capacity fails to keep pace. For example, Angola's 3% annual population growth since 1970 has expanded its population from 6 million to 33 million today, but its poverty rate increased by 15% between 2008 and 2018, leaving many young people in precarious conditions despite overall economic expansion (Oxford Analytica, 2020; World Bank, 2023a, b, c). Together, the literature suggests that the relationships between population change and economic performance depends on broader economic and institutional conditions.

## Methods

### Population trends

We obtained country-specific population data from the World Population Prospects 2022 database (United Nations Department of Economic and Social Affairs Population Division, 2022) (Fig. 1). These population data capture net changes in total national population size and therefore combine births, deaths, and migration. Our measure of population change should not be interpreted as a fertility-specific measure, nor can it distinguish immigration-driven, refugee-driven, mortality-driven, or fertility-driven population change. The dataset provides age- (by year) and gender-specific population sizes by country from 1950–2021. For each country, we calculated the instantaneous exponential rate of annual change (*r*):

$$r_{\Delta t} = \log_e \frac{N_{t+1}}{N_t}$$

where $N_t$ = total population size in year *t*. From these annual transitions, we calculated two means and their standard deviations: (*i*) mean and standard deviation of *r* from 1950–2021, and (*ii*) mean and standard deviation of *r* from 2012–2021 (Fig. 1). The latter parameters indicate more recent trends than the full time series for each country.

### Age structure

Using the same dataset described in the previous section, we took the sum of *N* in age categories ≥ 65 years as a proportion of the sum of *N* in age categories 16–64 years. This metric is one variant of the dependency ratio, which in this case ignores the youth cohort (0–15 years) and compares the 'unemployed' ('dependent') to the 'employed' portion of the population. This ratio is a descriptive measure of population age structure. It does not imply that all people aged ≥ 65 years are economically dependent, nor that all people aged 16–64 years are employed or economically independent. So-called ‘ageing’ populations have a higher dependency ratio than those dominated by younger cohorts (Fig. 1).

### Wealth

As we outlined in the Introduction, the standard measure of economic output — GDP per capita — is suboptimal for characterising the true wealth of a nation’s citizens (although we also applied our analyses to GDP per capita — see Appendix I). We therefore obtained the unchained domestic comprehensive wealth index from the World Bank as an index of national wealth (Fig. 2a). The aggregate national comprehensive wealth index is calculated using the Törnqvist volume index (Kohli, 2004) applied to wealth categories such as produced capital, non-renewable and renewable natural capital, human capital, and net foreign financial assets. The unchained index computes a weighted geometric mean of the quantity relatives for each asset (except financial assets) between periods, using nominal value shares as weights. These unchained indices are then chained to form a time series with 2019 as the base year, and the final chained index is expressed in monetary terms by applying the chained Törnqvist index to the base year’s aggregate nominal value. The domestic comprehensive wealth index excludes net foreign financial assets, but it is still strongly correlated ($R^2$ = 0.846) with per-capita GDP (purchasing power parity-adjusted; Appendix II, Fig. S3).

### Income inequality

National-level indices of wealth and productivity do not equate to an equal distribution of that wealth among citizens, so we obtained income inequality data from the World Bank (data.worldbank.org), measured by the Gini coefficient (Appendix III). The coefficient measures the extent to which the distribution of income among households within an economy deviates from a perfectly equal distribution, expressed as a percentage (0 = perfect equality; 100 = perfect inequality) of the maximum area under the Lorenz curve (cumulative percentages of total income received *versus* cumulative number of recipients). To maximise the sample size of countries to examine for a relationship, we took the mean Gini coefficient from 2010 to provide a sample of 162 countries (Fig. 2b).

### Productivity

We considered eleven different measures of productivity (Table 1; Appendix IV). Of these, labour productivity, per-capita capital stock, per-capita output-side real GDP, and per-capita services levels were all highly correlated ($R^2 > 0.90$) with per-capita GDP (Appendix IV, Fig. S9, S14, S15, S18), so we disregarded these metrics. Value-added manufacturing, gross capital formation, and gross savings showed no relationship with dependency ratio (Appendix IV, Fig. S10, S12, S20) or mean rate of population change (Appendix IV, Fig. S11, S13, S21), so we also disregarded these variables. Total factor productivity was weakly related to dependency ratio and rate of population change in the direction observed for other productivity measures (see below), so we report these results in the Supplementary Information only (Appendix IV, Fig. S16, S17). This left gross domestic expenditure on research and development, per-capita patent applications, and the human capital index as relevant measures of productivity we used in the analyses.

Gross domestic expenditures on research and development (World Bank) include both capital and current expenditures in four main sectors: business enterprise, government, higher education, and private non-profit (Fig. 2c). Per-capita patent applications (World Bank) are filed through the Patent Cooperation Treaty procedure or with a national patent office for exclusive rights for an invention (product or process that provides a new way of doing something or offers a new technical solution to a problem) (Fig. 2d). The human capital index (Feenstra et al., 2015) (Fig. 2e) is based on mean years of schooling and returns to education. For gross domestic expenditure on research and development, we took the logit of the most recent value per country. For per-capita patent applications, we took the $\log_{10}$ of the quotient of the most recent year available per country and total population size in that year to obtain a per-capita value. For the human capital index, we took the most recent value per country.

### Corruption

Even in wealthier and more productive countries, the existence of political corruption — the "... misuse of public office for private gain," where private gain can accrue either to the individual office or to groups or parties to which the individual belongs (Bardhan, 1997) — can erode economic performance (Serritzlew et al., 2014) and environmental outcomes (Oliver and Shogren, 2015; Tacconi and Williams, 2020). In general, perceived corruption tends to be lower in economically developed, long-established liberal democracies with a free and widely read press, a high share of women in government, and a history of openness to trade (Treisman, 2007). We therefore accessed the 2024 corruption perception index (logit-transformed; Fig. 3a) from Transparency International (transparency.org) to examine its relationship to changes in national-level human demographics (higher values = less corruption). The logit of the corruption perception index was moderately correlated with per-capita GDP (logit-log $R^2$ = 0.527; Appendix V, Fig. S28) and the logit of the freedom index (logit-logit $R^2$ = 0.556; Appendix V, Fig. S30; see below).

**Freedom**

As an additional measure of political stability and human rights, we obtained the aggregate (total) score of all categories assessed in the Freedom House *Freedom in the World* report (freedomhouse.org), measuring dimensions of a country's electoral process, political pluralism and participation, functioning of government, freedom of expression and of belief, associational and organisational rights, rule of law, and personal autonomy and individual rights. The index covers the period 2013–2025, so we used the most recent (2025) data as a response variable (Fig. 3b). The 2025 freedom score is correlated ($R^2$ = 0.273) with political spectrum (global parliament index), indicating that more left-wing governments today tend to have a higher freedom score (slope 95% confidence interval: -0.891 to -0.522; Appendix V, Fig. S31).

**Well-being and development**

We obtained four measures of national well-being and development: (*i*) the Human Development Index (United Nations Development Programme, 2024a) (latest year available = 2023; Appendix VI), (*ii*) the 'planetary pressures-adjusted' Human Development Index (United Nations Development Programme, 2024b) (Fig. 3c), (*iii*) total energy supply per capita in 2024 (Appendix IV), and (*iv*) a composite well-being index derived from the Gallup World Poll (Blanchflower and Bryson, 2024) (Appendix VII). Published since 1990 by the United Nations Development Programme (United Nationsl Development Programme, 1990), the Human Development Index is a composite measure of three dimensions of human development: health (measured as life expectancy), education (measured as mean years of schooling), and standard of living (measured as per-capita gross national income) (United Nations Development Programme, 2024a). The composite index itself is a geometric mean of the normalised indices across the three dimensions. Importantly, the index captures only part of the multidimensional definition of ‘human development’, because it does not reflect inequalities, poverty, security, empowerment, or environmental integrity (United Nations Development Programme, 2024a).

Because the Human Development Index includes a national index of gross income, it is strongly correlated ($R^2$ = 0.883) with purchasing power parity-adjusted per-capita GDP (Appendix VI, Fig. S32). To capture a more independent measure of well-being, we also retrieved the ‘planetary pressure-adjusted’ Human Development Index (United Nations Development Programme, 2024b) (Fig. 3c). This variant adjusts the Human Development Index by discounting the index for pressures on the planet to reflect a concern for intergenerational inequality. The adjusted index is the human development measures adjusted by per-capita carbon-dioxide emissions (production) and material footprint to account for global-scale environmental degradation (Bradshaw et al., 2021). The planetary pressure-adjusted Human Development Index still correlates positively with purchasing power parity-adjusted per-capita GDP, but the relationship is much weaker ($R^2$ = 0.432; Appendix VI, Fig. S34).

Energy supply is central to economic development (Oliver and Upton Jr, 2022), despite the legacy of fossil-fuel exploitation as a primary driver (Wackernagel et al., 2021) of the well-documented ecological and extinction crises gripping the planet (Bradshaw et al., 2021; Ceballos and Ehrlich, 2023; Ellis, 2023). At the same time, developed nations’ attempts to address such issues through international climate agreements and their associated energy market interventions (e.g., emissions taxes, economic support for clean energy technologies) could slow GDP growth (Cifci and Oliver, 2018), which opponents often cite as justification for political opposition, even though per-capita measures of well-being might be less affected. In less-developed economies, access to clean and affordable energy is a well-known driver of human and economic development (Acheampong et al., 2021; Rao and Pachauri, 2017). We therefore obtained the 2024 national per-capita total energy supply (GJ person$^{-1}$) data for 78 countries from the Energy Institute’s *Statistical Review of World Energy* (energyinst.org).

We also sought a fourth well-being index that did not explicitly include national measures of economic performance. Based on data from approximately four million respondents in 164 countries, the composite well-being ranking (Blanchflower and Bryson, 2024) (Appendix VII) incorporates four positive well-being measures: life satisfaction, enjoyment, smiling, and being well-rested, and four negative measures: pain, sadness, anger, and worry (Blanchflower and Bryson, 2024). We used the overall rank per country (Table 9 in Blanchflower and Bryson, 2024), where countries are ranked from highest (lowest rank) to lowest (highest rank) well-being. Because the composite rank included individual US states, we re-ranked the index for countries after excluding US states. The composite well-being rank is only weakly correlated ($R^2 = 0.228$) with purchasing power-adjusted per-capita GDP (Appendix VII, Fig. S39).

**Life expectancy**

We considered a final metric of 'well-being' not captured by the previous measures of prosperity — healthy life expectancy at birth from the World Health Organization (data.who.int; Fig. 3d). This measures the mean number of years that a person can expect to live in 'full health,' accounting for years lived in less than full health due to disease and/or injury.

**Analysis**

*Resampled boosted regression trees*

To test for evidence of relationships between the various socio-economic response variables and indices of demographic change (population age structure and growth rates), we employed boosted regression trees (a hybrid machine-learning approach). Boosted regression trees are well-suited to our study due to their ability to model complex, nonlinear relationships and interactions among predictors automatically, without requiring prior specification of these relationships as is necessary in multivariate linear or logistic regression (Zhang et al., 2019). Boosted regression trees build an ensemble of decision trees iteratively that together improve predictive accuracy (Elith et al., 2008). They also offer advantages in managing missing data and generating variable importance metrics to enhance interpretability. Boosted regression trees can also combine predictor variables of any type, so results are unaffected by different scales of measurement and outliers (Elith et al., 2008). Other advantages are that the results are insensitive to outliers, and the fitting process is stochastic such that it generally improves predictive performance of the final model (Breiman et al., 1984; Elith et al., 2008). In this context, the approach serves a similar purpose to multivariate regression in that it tests the joint influence of multiple predictors, but it does so non-parametrically, allowing for nonlinear relationships and complex interactions. Our use of this approach is primarily phenomenological in that we aimed to uncover patterns in the data without committing to a specific functional form, while recognising that, unlike parametric regression, boosted regression trees do not infer causality of underlying economic mechanisms.

The two main predictors in the boosted regression trees included the mean rate of population change $r$ (either 1950–2021 or 2012–2021, in separate analyses) and the dependency ratio following the hypotheses outlined in the Introduction — specifically, whether population growth and age structure are systematically associated with economic performance and human well-being. We also included a third predictor to account for the erosion of socio-economic stability and environmental integrity: total population size (in 2020). This captures potential effects of absolute scale that are not reflected in per-capita response (‘dependent variable’) (per-capita domestic comprehensive wealth index, labour productivity, per-capita capital stock, per-capita output-side real GDP, per-capita capital services levels, per-capita patent applications).

Given the potential for spatial autocorrelation to bias to the estimated relationships (because neighbouring countries are more likely to resemble each other just by virtue of their spatial proximity), we developed a resampling approach whereby we first categorised countries by major region (Africa, Asia + Oceania, Eurasia, Middle East, North America, South America + Caribbean). The combination of Asia + Oceania and South America + Caribbean was necessary to avoid regions with too few countries for resampling. We then took 10,000 random samples (without replacement) with each iteration's sample equal to 0.8 times the minimum number of countries per region (0.8 of 10 = 8) per region to build a random sample of 48 (6 regions × 8 countries) or 54 (6 regions × 9 countries for labour productivity, Gini coefficient, and healthy life expectancy to provide sufficient samples for analysis) countries to analyse per iteration. While this approach is unlikely to remove all spatial autocorrelation, it is sufficient to return the main relationships without excessive bias (Bradshaw et al., 2023) (i.e., it is a trade-off between a sufficient sample size to fit the models and too many closely neighbouring countries increasing spatial autocorrelation). In addition to the country-sampling procedure, we also took a random (Gaussian) deviate per iteration using the mean $r$ and its standard deviation either over the interval 1950–2021 (long-term trend) or 2012–2021 (more recent trend). We also took a uniform random deviate per iteration between the upper and lower bounds of healthy life expectancy.

For each iteration, we implemented the boosted regression tree models using the *gbm.step* function from the `dismo` R package (Blonder et al., 2025), setting a Gaussian distribution for the response variable. To optimise the models, we tuned several parameters, including the learning rate (0.001), tolerance (0.002), bag fraction (0.7), and tree complexity (2). The latter setting considers only first-order interactions and combines these effects if present into the relative influence scores. These settings balanced robustness of fit with processing speed. Relative influence $I$ is the relative influences of the individual inputs $x_j$ on the variation of $\hat{F}(x)$, the function that maps the explanatory variables $x$ to the response $y$. For a collection of $M$ decision (regression) trees $\{T_m\}_1^M$, the squared influence $\hat{I}^2$ is:

$$\hat{I}^2 = \frac{1}{M}\sum_{m=1}^{M}\hat{I}_j^2(T_m)$$

where $\hat{I}_j^2(T_m) = \sum_{t=1}^{J-1}\hat{\imath}_t(v_t = j)$, which is the summation over the nonterminal nodes $t$ of the $J$-terminal node tree $T$, $v_t$ = the splitting variable associated with node $t$, and $\hat{\imath}_t^2$ = the corresponding empirical improvement in squared error resulting from the split (Friedman, 2001). These squared influences sum to 100 over all $x$ explanatory parameters, analogous to the percentage of variation in the response variable $y$ explained by each parameter $x$.

We applied cross-validation to assess model performance and to reduce overfitting. To evaluate the goodness-of-fit of the models, we used the cross-validation correlation coefficient ($\beta_{CV}$), including its standard error calculated across all tree iterations. We also implemented a kappa ($\kappa$) limitation to the resampled selections (Bradshaw and Brook, 2016a). In this process, we retained only the resampled mean ranks within $\kappa$ standard deviations ($\sigma$) of the overall average mean, with $\kappa = 2$. After recalculating the mean and standard deviation of the ranks, we repeated this process several times. This iterative $\kappa\sigma$ 'clipping' approach, commonly used in image processing to remove artefacts, reduced the influence of outliers on the estimation of the mean rank across all 10,000 iterations. All R code and data required to repeat the analyses are available at doi:10.5281/zenodo.15826278.

*Time series analysis*

Geographic comparisons among countries using the most recent data do not capture the temporal dynamics in the response variables with changing demographic composition within a country. We therefore investigated the relationship between each response and the dependency ratio per country across years, ensuring that we only included countries with at least 15 years of data (13 in the case of the freedom score and the corruption perception index because of fewer years available overall for those responses) to ensure reliable sample sizes.

However, within-country time series of each response are temporally autocorrelated, so we calculated the Newey-West heteroscedasticity- and autocorrelation-consistent confidence interval of the slope ($\beta$) between each response and the dependency ratio using the *coeftest* function in the `lmtest` R package (Zeileis and Hothorn, 2002), setting the 'vcov' option to 'NeweyWest' with the `sandwich` R package (Zeileis, 2004, 2006; Zeileis et al., 2020). If the autocorrelation-consistent confidence interval did not include zero, we deemed there was evidence for a temporal relationship between the response and dependency ratio (with either a positive or negative $\beta$).

*Regional analysis*

We also tested whether the relationships between population age structure and socio-economic performance persisted when we compared countries within major world regions. We first allocated countries to six major regions: Africa, Asia–Oceania, Europe, the Middle East, North America, and South America–Caribbean. We combined Asia and Oceania, and South America and the Caribbean, to retain adequate sample sizes, consistent with the regional stratification we used in the resampled boosted regression tree analyses. For each region separately, we fitted ordinary least-squares regressions between the logit-transformed dependency ratio in 2020 and each transformed response. The ten responses were the domestic comprehensive wealth index ($\log_{10}$), research and development expenditure as a proportion of GDP (logit), per-capita patent applications ($\log_{10}$), Gini coefficient (logit), planetary pressure-adjusted Human Development Index (logit), healthy life expectancy at birth (untransformed), Freedom House freedom score (logit), corruption perception index (logit), composite wellbeing rank ($\log_{10}$), and purchasing-power-parity-adjusted GDP per capita ($\log_{10}$). For every regional regression, we recorded the sample size, slope and its standard error, Pearson's R, Spearman's $\rho$, $R^2$, adjusted $R^2$, root mean square error, evidence ratio, and the two-sided type I error for the slope. For the analysis, we ensured that at least three countries and variation in both the dependency ratio and response were available.

*Lag analysis*

We next assembled country-level annual panels for domestic comprehensive wealth (1995–2020), research and development expenditure, per-capita patent applications, Gini coefficient, planetary pressure-adjusted Human Development Index, healthy life expectancy at birth, freedom score, corruption perception index, and GDP per capita. Annual country-level data were unavailable for composite wellbeing rank, so no lag analysis was possible for that response. All response transformations matched those used in the regional analysis. As above, we assigned each country to one of six regions.

For each country-response combination, we tested lags from 0 to 10 years. A lag of $L$ years aligned the response in year $t$ with the dependency ratio in year $t - L$; hence, positive lags represent earlier population age structure preceding the measured outcome. At each lag, we fitted a linear regression between the transformed response and the lagged logit dependency ratio. We required at least 10 overlapping annual observations for domestic comprehensive wealth and at least 8 for all other responses, as well as variation in both variables. For every valid lag, we calculated the slope, its type I error, adjusted $R^2$, root mean square error, Spearman's $\rho$, Pearson's R, and the Akaike information criterion (AIC).

For each country-response combination, the best-scoring lag was the valid lag with the greatest adjusted $R^2$; we broke ties by the largest absolute Spearman's $\rho$, then the lowest AIC, then the shortest lag. We classified a non-zero lag as supported when it improved AIC by $\geq 2$ relative to the zero-lag model or improved adjusted $R^2$ by $\geq$ 0.02. This is a descriptive model-selection criterion, not a test that a demographic change causes a later socio-economic change. We summarised best lags globally and by region using the median and interquartile range and calculated the proportion of country-response combinations for which a non-zero lag was supported. To illustrate the aggregate relationship after temporal alignment, we realigned each country's annual observations by its selected lag. We then fitted pooled linear regressions to the aligned annual observations and, separately, to the country means calculated from the aligned observations. Finally, we used Kruskal–Wallis tests to compare selected lag lengths and improvements in adjusted $R^2$ among response variables and regions.

## Results

We show the results for the seven main responses describing wealth (domestic comprehensive wealth index) and income inequality (Gini coefficient), productivity (research and development expenditure, per-capita patent applications, human capital index), corruption (corruption perception index) and political freedom (freedom score), and well-being (planetary pressure-adjusted Human Development Index, healthy life expectancy at birth) in the main text, and the results for other potential responses (per-capita GDP, remaining productivity metrics, Human Development Index, composite well-being rank) in the Supplementary information.

### Wealth and income inequality

The $\log_{10}$ per-capita domestic comprehensive wealth index is related positively to the logit of the dependency ratio (Fig. 4a), but with countries in the Middle East departing from the expected relationship (Appendix II, Fig. S4). The boosted regression tree analyses showed that the dependency ratio had the highest relative influence (i.e., explained the most variance in the response) on wealth (43.5–82.5%) compared to $r_{\text{mean}}$ and population size (Appendix II, Fig. S5), and a clear threshold effect where wealth increased rapidly from a dependency ratio of ~ 0.09 (-2.3 on the logit scale) to ~ 0.16 (-1.65 on the logit scale) (Fig. 4a). In descriptive terms, countries with relatively older age structures tend to have higher national per-capita wealth on average (results are similar using per-capita GDP as the response; Appendix Ib).

The logit Gini coefficient as a measure of income inequality shows a weak negative relationship to the logit of the dependency ratio, but again with a few countries in the Middle East departing from the main trend (Appendix III, Fig. S7). The resampled boosted regression tree analysis revealed that of the three predictors we considered, $r_{\text{mean}}$ was the predictor most strongly associated with variation in the response (7.9–86.1% relative influence; Fig. 4b), while the dependency ratio contributed only a little less (7.0–79.5%; Appendix III, Fig. S8). Despite overall less variance explained due to higher uncertainty, the predicted relationship revealed that as $r_{\text{mean}}$ increased, so too did income inequality (Fig. 4d). In descriptive terms, countries with negative or slow total population growth tend to have higher income equality on average.

**Productivity**

All three productivity indices (research and development expenditure, per-capita patent applications, human capital index) revealed a strong positive relationship to the dependency ratio (Fig. 5; Appendix IV) as the dominant predictor (relative influence: 41.2–91.8%, 52.2–88.7%, and 65.2–92.0%, respectively; Appendix IV). However, there was little association with $r_{\text{mean}}$ (Fig. 5) or population size (Appendix IV). In descriptive terms, older national age structures are associated with higher productivity indicators on average.

**Corruption and freedom**

Corruption declined and freedom increased with the dependency ratio (Fig. 6), which had the highest relative association of all predictors we considered (35.9–89.5% and 39.7–87.5%, respectively; Appendix V). However, there was only a weak (corruption increases) or no (freedom) relationship with $r_{\text{mean}}$ (Fig. 6c,d).

**Well-being, development, and life expectancy**

Of the four well-being indices (Human Development Index, planetary pressure-adjusted Human Development Index — Appendix VI, total energy supply per capita — Appendix VI, and the composite well-being rank — Appendix VII), the planetary pressure-adjusted Human Development Index had the highest goodness of fit. We therefore only show those results in the main text, and those for the other well-being and development indices in the Supplementary information (Appendices V and VI).

The planetary pressure-adjusted Human Development Index increased with an increasing dependency ratio (Fig. 7a). The resampled boosted regression tree analysis showed that dependency ratio explained the most variance (Appendix VI, Fig. S37a), with a strong threshold effect where a country's well-being declined quickly for dependency ratios from ~ 0.06–0.10 (-2.2 to -2.7 on the logit scale; Fig. 7a). There was also a slight decrease in the planetary pressure-adjusted Human Development Index with an increasing population rate of change (Fig. 7c). In these descriptive models, nations with higher proportions of older cohorts therefore tend to have higher values of this well-being index.

Older countries as measured by higher dependency ratios had higher average healthy life expectancy than younger countries (Fig. 7b), and again countries in the Middle East departed from the main trend (Appendix VIII, Fig. S42). The resampled boosted regression tree analysis revealed that the dependency ratio explained the highest proportion of the total variance in the response (Appendix VIII, Fig. S43). The rate of population change had only a weak additional effect on life expectancy (Fig. 7d).

**Time series**

Within-country time series indicated that for eight of the nine main responses, most (> 50%) countries with temporal autocorrelation-corrected evidence for a non-zero slope had better outcomes as their populations aged (blue component of bars in Fig. 8). Only the corruption perception index had a slight majority (54%) of countries where there was a worse outcome with an increasingly older demographic (Fig. 8).

Of the countries with worse-outcome responses as their populations aged, 19 had ≥ 3 responses in this category: Belize (3 responses), Benin (3), Bolivia (3), Burundi (3), Costa Rica (3), Côte d'Ivoire (3), Gabon (3), The Gambia (3), Indonesia (3), Iraq (3), Kyrgyzstan (3), Mongolia (5), Mozambique (4), Oman (3), Saudi Arabia (3), Sierra Leone (3), Slovakia (3), Tajikistan (3), Venezuela (3) (relationships for each response by country in Appendix IX). None of these countries is in population decline; in fact, the mean dependency ratio and rate of population change for these countries are clustered around the mean value for African countries (Appendix IX, Fig. S63). In other words, these countries are fast-growing and young, so the negative relationships identified above are likely driven by other socio-economic and political conditions independent of population demography.

**Regional analysis**

The general relationships with the dependency ratio were retained in most regions for the principal socio-economic responses (Appendix X, Table S2). Domestic comprehensive wealth increased with dependency ratio in five of the six regions (adjusted $R^2$ = 0.08–0.85); the only exception was the Middle East, where the estimated relationship appeared negative but did not differ statistically from a slope =

0. Per-capita GDP had the same positive direction in Africa, Asia–Oceania, North America, and South America–Caribbean (adjusted $R^2 = 0.32–0.86$), whereas the slopes of the relationships in Europe and the Middle East were not statistically distinguishable from zero.

The productivity measures were also generally positive within regions. Research and development expenditure increased with the dependency ratio in Asia–Oceania, Europe, and North America, and per-capita patent applications increased in Africa, Asia–Oceania, and North America. The planetary pressure-adjusted Human Development Index increased in five regions, including the Middle East, with the strongest relationship in South America–Caribbean (Spearman's $\rho = 0.92$, adjusted $R^2 = 0.76$). Healthy life expectancy increased in four regions, with the steepest estimated slope in South America–Caribbean (7.46 years per unit logit dependency ratio). Freedom scores increased in five regions, and the corruption perception index indicated lower perceived corruption in four regions. There was little within-region evidence that income inequality varied with the dependency ratio: only North America had a negative Gini slope, consistent with greater income equality in older populations. The wellbeing rank was lower (better) only in Asia–Oceania. Thus, the positive cross-country relationships for wealth, development, healthy life expectancy, freedom, and lower corruption were generally not attributable solely to differences among continents, although the strength and precision of regional relationships varied with sample size and region.

**Lag analysis**

Best-scoring lag lengths differed among responses (Kruskal–Wallis $\chi^2 = 239.81$, df = 8, $p = 2.49 \times 10^{-47}$), but not among regions ($\chi^2 = 5.90$, df = 5, $p = 0.317$). The improvement in adjusted $R^2$ relative to contemporaneous models also differed among responses ($\chi^2 = 93.93$, df = 8, $P = 7.39 \times 10^{-17}$). Median selected lags were 0 years for freedom score, 1 year for the corruption perception index, 2 years for healthy life expectancy, 4 years for research and development expenditure, patent applications, Gini coefficient, and planetary pressure-adjusted Human Development Index, 5 years for GDP per capita, and 6 years for domestic comprehensive wealth (Appendix XI,

Table S3). The dispersion was substantial for most outcomes (interquartile range = 7–10 years), indicating heterogeneous country-level temporal patterns.

Support for a non-zero lag was most prevalent for the Gini coefficient (47 of 57 countries; 82.5%), research and development expenditure (56 of 78; 71.8%), per-capita GDP (131 of 197; 66.5%), patent applications (69 of 106; 65.1%), and domestic comprehensive wealth (95 of 151; 62.9%). It was least prevalent for freedom score (64 of 195; 32.8%), whose median best lag was zero. For the supported non-zero lag models, median lags ranged from 1 year for freedom score to 9 years for GDP per capita.

After alignment to each country's best-scoring lag, pooled annual and country-mean regressions showed the expected direction for every response (Appendix XI, Table S4). Older population structures were associated with higher domestic comprehensive wealth, research and development expenditure, patent applications, planetary pressure-adjusted Human Development Index, healthy life expectancy, freedom score, corruption perception index (less corruption), and per-capita GDP. Older populations were also associated with lower Gini coefficients, indicating greater income equality. Country-mean relationships were strong for patent applications (Spearman's $\rho = 0.76$; adjusted $R^2 = 0.57$), planetary pressure-adjusted Human Development Index ($\rho = 0.83$; adjusted $R^2 = 0.55$), and Gini coefficient ($\rho = -0.54$; adjusted $R^2 = 0.44$).

These lag-aligned associations should be interpreted cautiously. The selected lag maximises within-country descriptive fit across 11 candidate alignments, and the pooled regressions do not account for country-specific temporal autocorrelation, secular trends, common shocks, or other confounders. They nevertheless show that the broad relationships between older population structures and more favourable socio-economic outcomes are not limited to contemporaneous cross-sectional comparisons and that their temporal alignment varies markedly among outcomes and countries.

## Discussion

Our results provide a broad empirical counterpoint to deterministic claims that slower overall population growth or older age structures necessarily coincide with weaker national economic performance, income distribution, productivity, political stability, well-being, or health. Across the indicators we examined, countries with slower total population growth and older age structures often performed better on average. These associations should not be interpreted as evidence that population decline, lower fertility, or ageing causes better socio-economic outcomes. Rather, they suggest that demographic structure is intertwined with other factors, including human capital, productivity, institutions, migration, policy choices, and historical development trajectories. This distinction is important because our population-growth measure reflects net total population change and does not distinguish fertility, mortality, and migration.

Our results are also consistent with evidence that some forms of slower population growth can coincide with higher per-capita investment, lower environmental pressure, and improved individual outcomes under particular economic and institutional conditions (Casey and Galor, 2017; Lee and Mason, 2010; Prettner et al., 2013; Saraswati et al., 2024), but they do not show that smaller populations are inherently preferable or that reducing population growth would cause socio-economic improvements. Our findings therefore do not support pronatalist, antinatalist, anti-immigration, or coercive fertility policies. Instead, they indicate that societies can experience slower growth or ageing without inevitable socio-economic decline, particularly where institutions and policy systems adapt. Labour-market pressures in ageing societies can arise through multiple channels, including skills mismatches, sectoral demand, retirement patterns, labour-force participation, migration policy, and institutional constraints (Peeples, 2025). Because our analyses do not separate fertility, mortality, and migration, we do not estimate the contribution of immigration or refugee flows to the relationships we observe.

Not only did the simple dependency ratio we used have the strongest positive relationship to most of the socio-economic performance indicators we examined, the ratio is still necessarily conservative. Defined here as the ratio of the number of people aged ≥ 65 to the number people aged 15 to 64, the dependency ratio does not fully represent the number of 'dependants' relative to the working population. For example, workforce participation of people ≥ 65 years has been increasing in countries with an ageing population, especially in those with the highest number of years of education (Burtless and Aaron, 2013; Coulmas, 2007). Neither are people aged ≥ 65 necessarily economically dependent. In wealthier economies, improved development and stronger retirement systems mean many older adults can support themselves financially. In addition, volunteering in old age represents a substantial, although often overlooked, contribution to the economy. For example, even with conservative estimates of hourly wage, the volunteering sector in Canada was worth US$2 billion in 2008, or ~ 0.5% of that country's 2008 GDP (Gottlieb and Gillespie, 2008). Most importantly, the dependency ratio defined this way ignores the large cost savings associated with having fewer children < 15 years old in ageing populations (Bradshaw and Brook, 2014).

Nonetheless, changing age structures will inevitably present economic and fiscal challenges (Bloom et al., 2015; Faruqee and Mühleisen, 2003; Peeples, 2025). But there are solutions to these challenges, such as greater investment in education to enlarge the effective workforce, delaying retirement age to promote higher income taxes, and redesigning pension financing (Bloom et al., 2015; Campante et al., 2021; Myrskylä et al., 2025; Skirbekk, 2023). Even the few economic analyses on the topic that purport "profound social and economic implications" concede that a transition to older societies in a few countries (e.g., Japan) is "manageable" with structural reforms, technological advances, and debt stabilisation (Faruqee and Mühleisen, 2003). Such adjustments are entirely realistic in the low-corruption, high rule-of-law countries where populations are ageing the fastest.

Several interpretive limits follow directly from our design. First, population change in our analyses is total net population change, not fertility-specific change. Countries can have similar growth rates for different reasons, including differences in births, deaths, immigration, emigration, conflict-related displacement, or age composition; this follows from the demographic balancing framework and from recent evidence that fertility, mortality, migration, and historical age structure can contribute differently to observed population-growth rates (Abel and Cohen, 2019; Canudas-Romo et al., 2022; Vollset et al., 2020). Second, age structure and population growth are themselves outcomes of development, health, education, labour markets, gender equity, and migration systems, so reverse causality and bidirectional relationships are plausible (Bloom et al., 2024; Doepke et al., 2023; Lee and Mason, 2010; Myrskylä et al., 2025). Third, national-level associations can mask within-country inequalities and differences among demographic groups, and aggregate associations should not necessarily be interpreted as individual-level effects or mechanisms (Robinson, 1950; Shih et al., 2023). Fourth, our analyses identify broad global, regional, and temporal patterns, but they do not identify the policy mechanisms that would allow particular countries to adapt to demographic change. Future research should therefore examine which combinations of age-specific skills, labour-force participation, migration, public finance, health, education, productivity, and institutions explain why some countries age or grow slowly without socio-economic decline (Bloom et al., 2015; Maestas et al., 2023; Myrskylä et al., 2025; Skirbekk, 2023).

In conclusion, although we do not take a normative position on whether population should or should not decline, our analysis finds no systematic evidence that slower or negative population growth coincides with weaker economic performance, productivity, income distribution, or the well-being and health of citizens. In many cases, declining or slow-growing populations have on average more robust economies, and wealthier, happier, and healthier citizens. Indeed, the direction of causality might in fact be the converse of that suggested in popular narratives (or even bi-directional), requiring more research into the complex, multi-dimensional relationships at play. There is evidence to suggest that reduced fertility (a precursor to declining population growth) might itself be an *effect* of improved economic development, related to rising female participation in the labour force, education, health, wages, and income (Doepke et al., 2023; Lee and Mason, 2010; Organisation for Economic Co-operation and Development, 2024), rather than a definitive *cause* of reduced economic development. Moreover, recent research is challenging the view that these relationships are strictly monotonic (Doepke et al., 2023; Luci and Thévenon, 2011). Our results suggest that demographic change should not be treated as an inevitable socio-economic threat, nor as evidence in favour of any single population policy. A more appropriate interpretation is that countries can follow different demographic trajectories without necessarily compromising socio-economic performance, and that the central policy question is how institutions, labour markets, migration systems, education, health, public finance, and productivity adapt to those trajectories..

## Declaration of Generative AI and AI-assisted technologies in the writing process

The authors used some AI-assisted technologies to code some of the analysis in R.

## Declaration of competing interest

The authors declare that they have no known competing financial interests or personal relationships that could have appeared to influence the work reported in this paper.

## Supplementary information

Supplementary Appendices I–XI; Figures S1–S63; Table S1–S4.

# Tables

**Table 1 |** Eleven different measures of productivity we considered in the analysis.

| indicator | units | source | reference |
|---|---|---|---|
| labour productivity | 2018 GDP per employment, in 2010 $US | World Bank | Dieppe (2021) |
| value-added manufacturing | % of GDP | World Bank | data.worldbank.org |
| gross capital formation | % of GDP | World Bank | data.worldbank.org |
| per-capita capital stock | constant 2018 national prices | Penn World Table version 10.01 | Feenstra et al. (2015) |
| per-capita output-side real GDP | $US at current purchasing power parities | Penn World Table version 10.01 | Feenstra et al. (2015) |
| per-capita capital services levels | $US at current purchasing power parities | Penn World Table version 10.01 | Feenstra et al. (2015) |
| total factor productivity | log difference, % | World Bank | Dieppe (2021) |
| gross savings | % of GDP | World Bank | data.worldbank.org |
| gross domestic expenditures on research and development | % of GDP | World Bank | data.worldbank.org |
| per-capita patent applications | patents/person | World Bank | data.worldbank.org |
| human capital index | unitless (0 to 1) | Penn World Table version 10.01 | Feenstra et al. (2015) |

## Figures

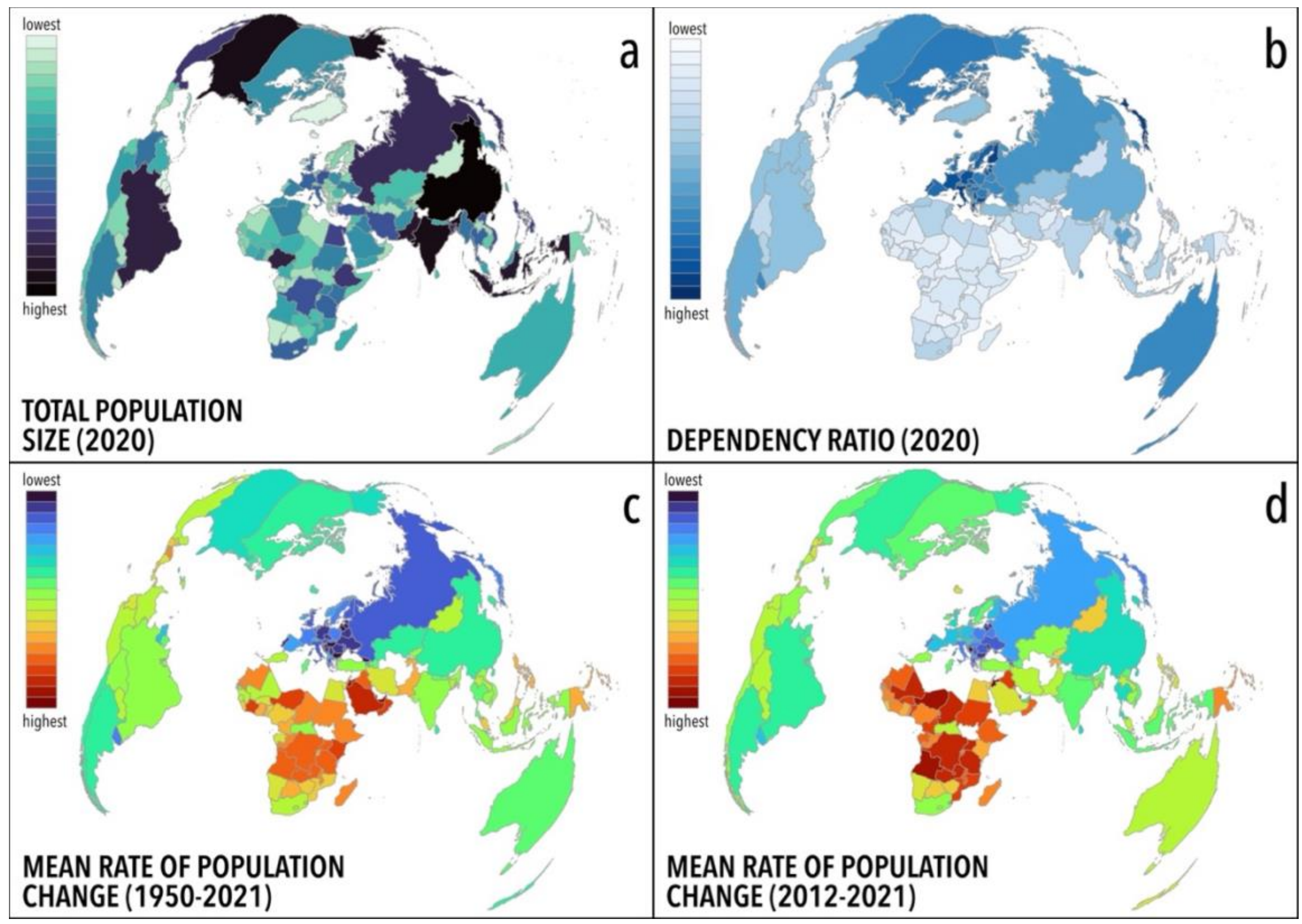

**Fig. 1 |** Global distribution of hypothesised national-level drivers of wealth and well-being: (**a**) total population size (2020) (United Nations Population Division); (**b**) dependency ratio ($N_{>65} \div N_{16\text{–}65}$) (United Nations Population Division); (**c**) mean rate of population change $r$ from 1950–2021; (**d**) mean rate of population change $r$ from 2012–2021.

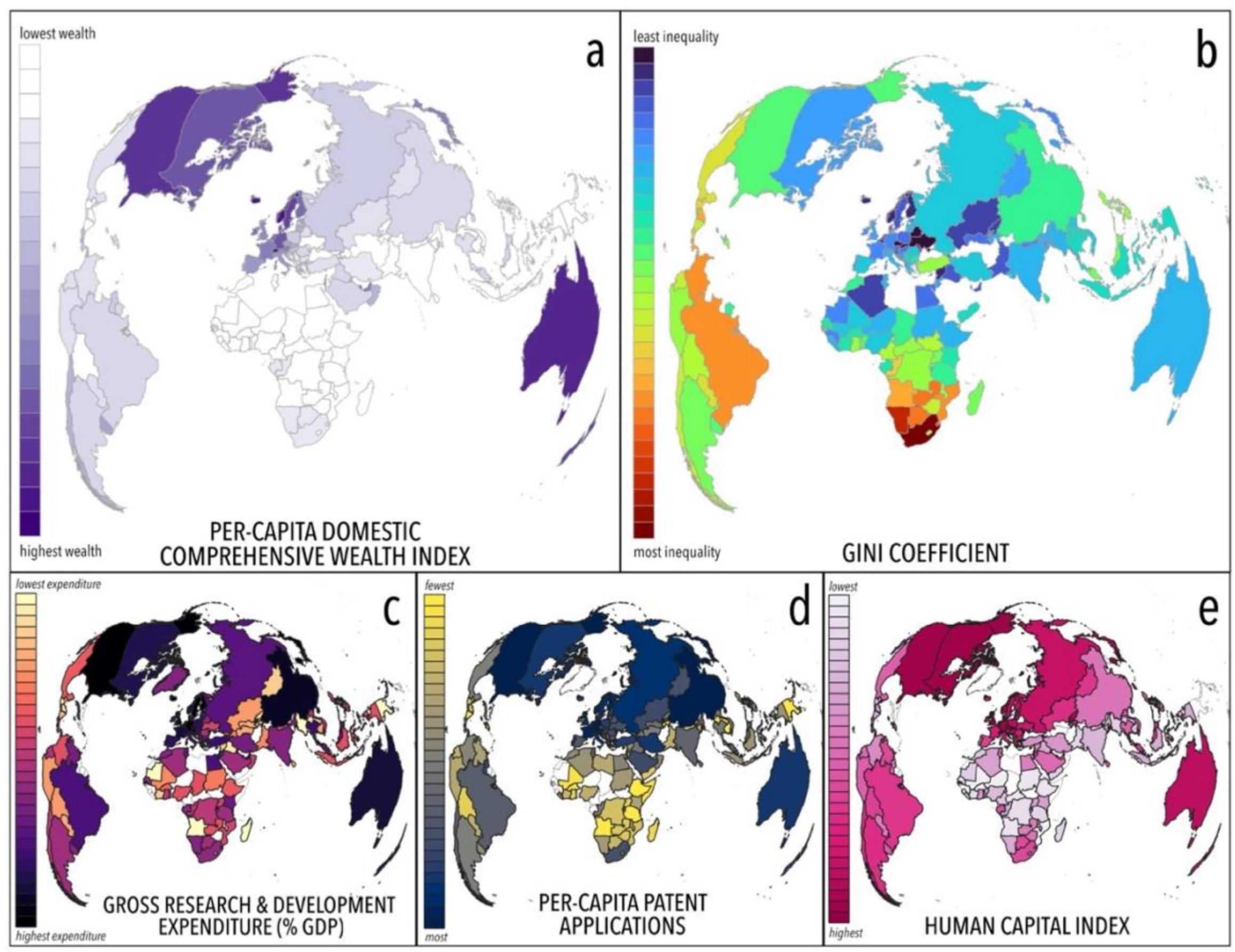

**Fig. 2 |** Global distribution of national-level wealth, income inequality, and productivity measures: (**a**) per-capita domestic comprehensive wealth index (data.worldbank.org); (**b**) Gini coefficient of income inequality (data.worldbank.org); (**c**) gross expenditure on research and development as a percentage of GDP (data.worldbank.org; most recent year); (**d**) per-capita patent applications (most recent year; data.worldbank.org); (**e**) human capital index (Feenstra et al., 2015).

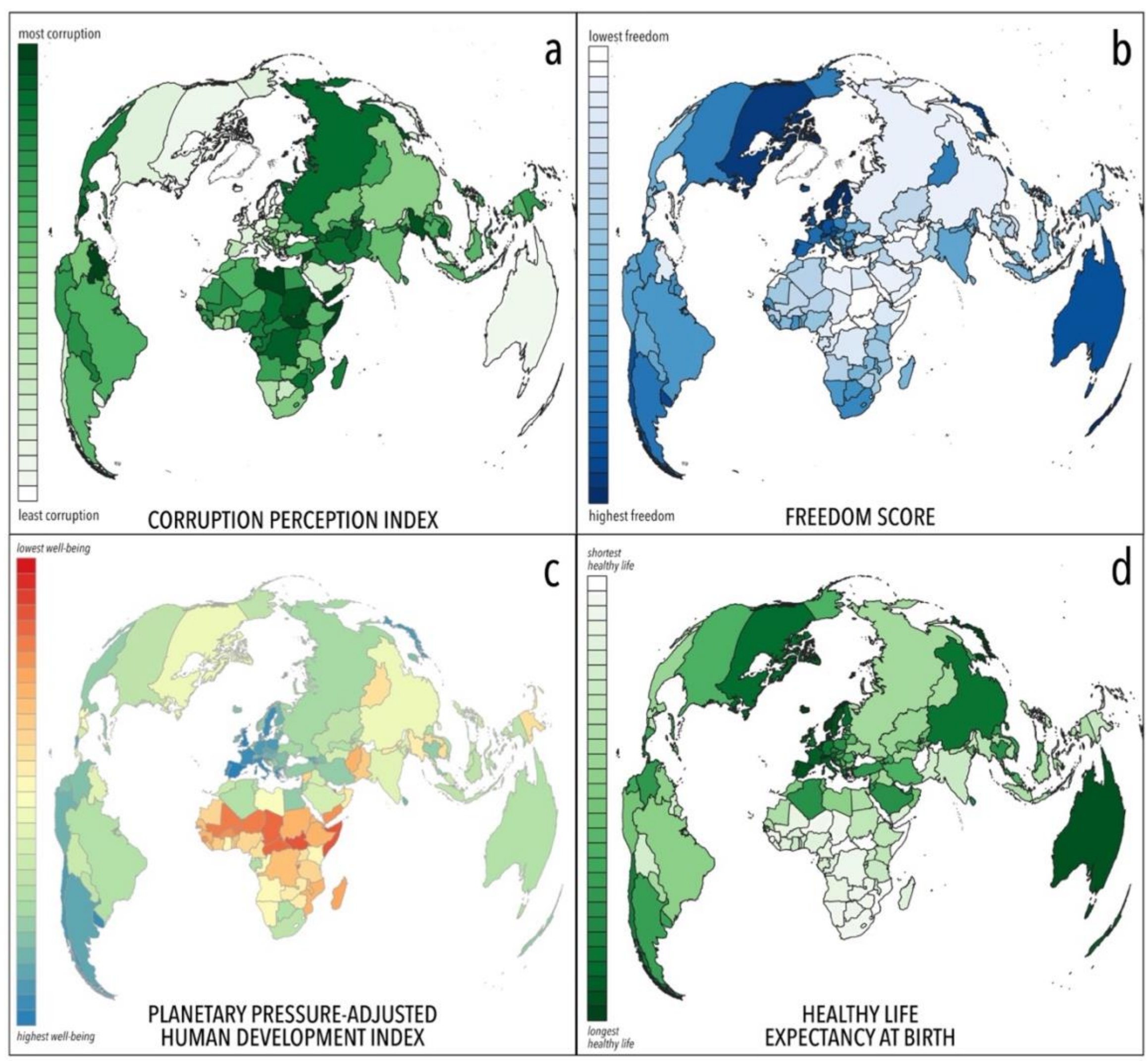

**Fig. 3 |** Global distribution of national-level corruption, freedom, well-being, and healthy life expectancy indices: (**a**) 2024 corruption perception index (Transparency International); (**b**) freedom score (Freedom House *Freedom in the World*); (**c**) planetary pressure-adjusted Human Development Index (United Nations Development Programme, 2024b); (**d**) healthy life expectancy at birth (World Health Organization; data.who.int).

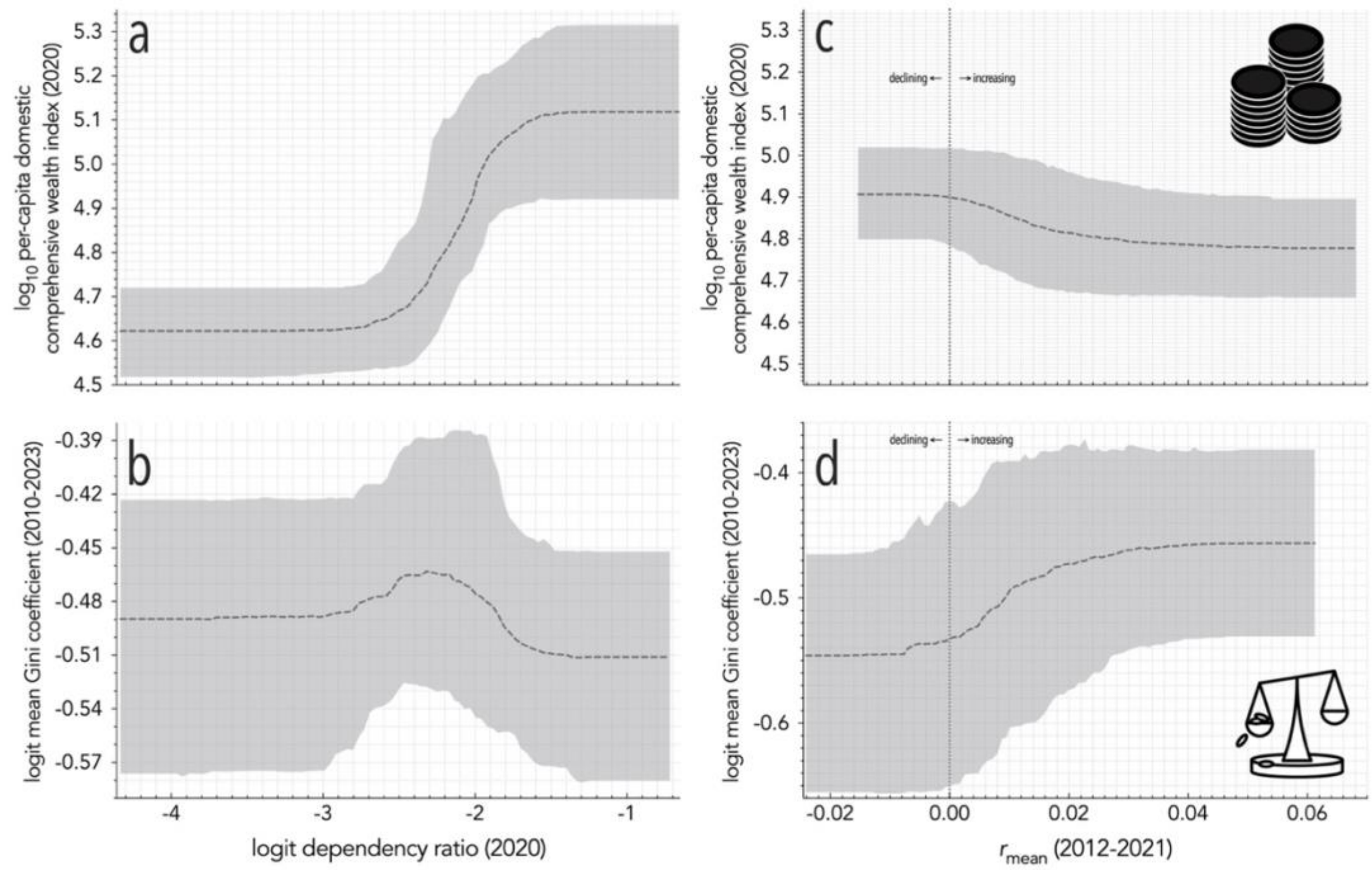

**Fig. 4 |** Resampled boosted regression tree results showing the relationships between (top panels) $\log_{10}$ per-capita domestic comprehensive wealth index and (**a**) logit dependency ratio 2020 and (**c**) mean rate of population change ($r_{\mathrm{mean}}$ for 2012–2021). Full results for the wealth index including $\log_{10}$ total population size in 2020 ($N_{2020}$) and relative influence in Appendix II (Fig. S5). Bottom panels show the relationship between the Gini coefficient (income inequality) and (**b**) the dependency ratio and (**d**) $r_{\mathrm{mean}}$. Full results for the Gini coefficient including $\log_{10}$ total population size in 2020 ($N_{2020}$) and relative influence in Appendix III (Fig. S8).

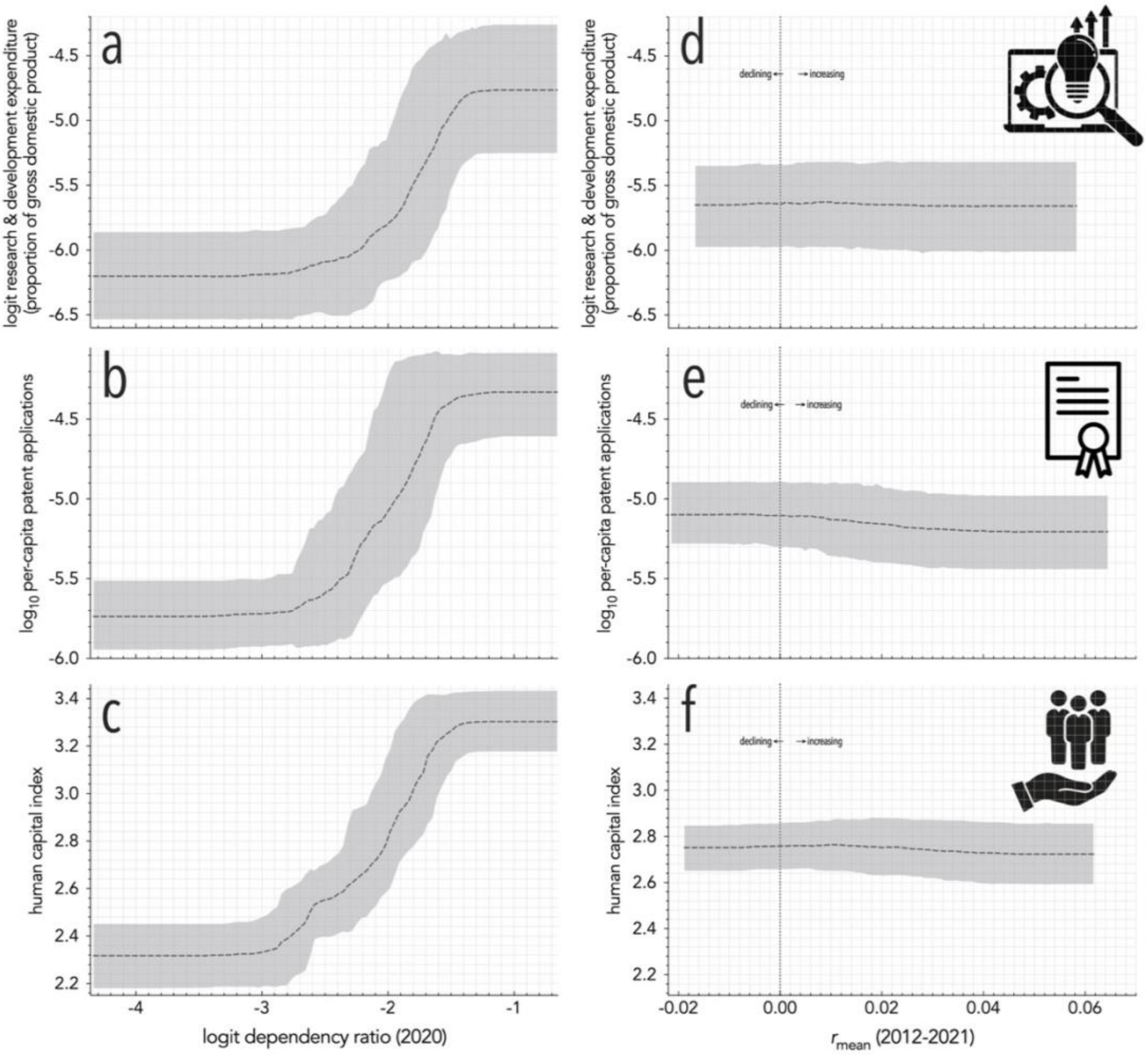

**Fig. 5 |** Resampled boosted regression tree results showing the relationships between the productivity responses (**a**, **d**) logit research and development expenditure as a proportion of GDP, (**b**, **e**) $\log_{10}$ per-capita patent applications, and (**c**, **f**) the human capital index and two predictors: logit dependency ratio and mean rate of population change ($r_{mean}$ for 2012–2021). Full results for each productivity response including $\log_{10}$ total population size in 2020 ($N_{2020}$) in Appendix IV.

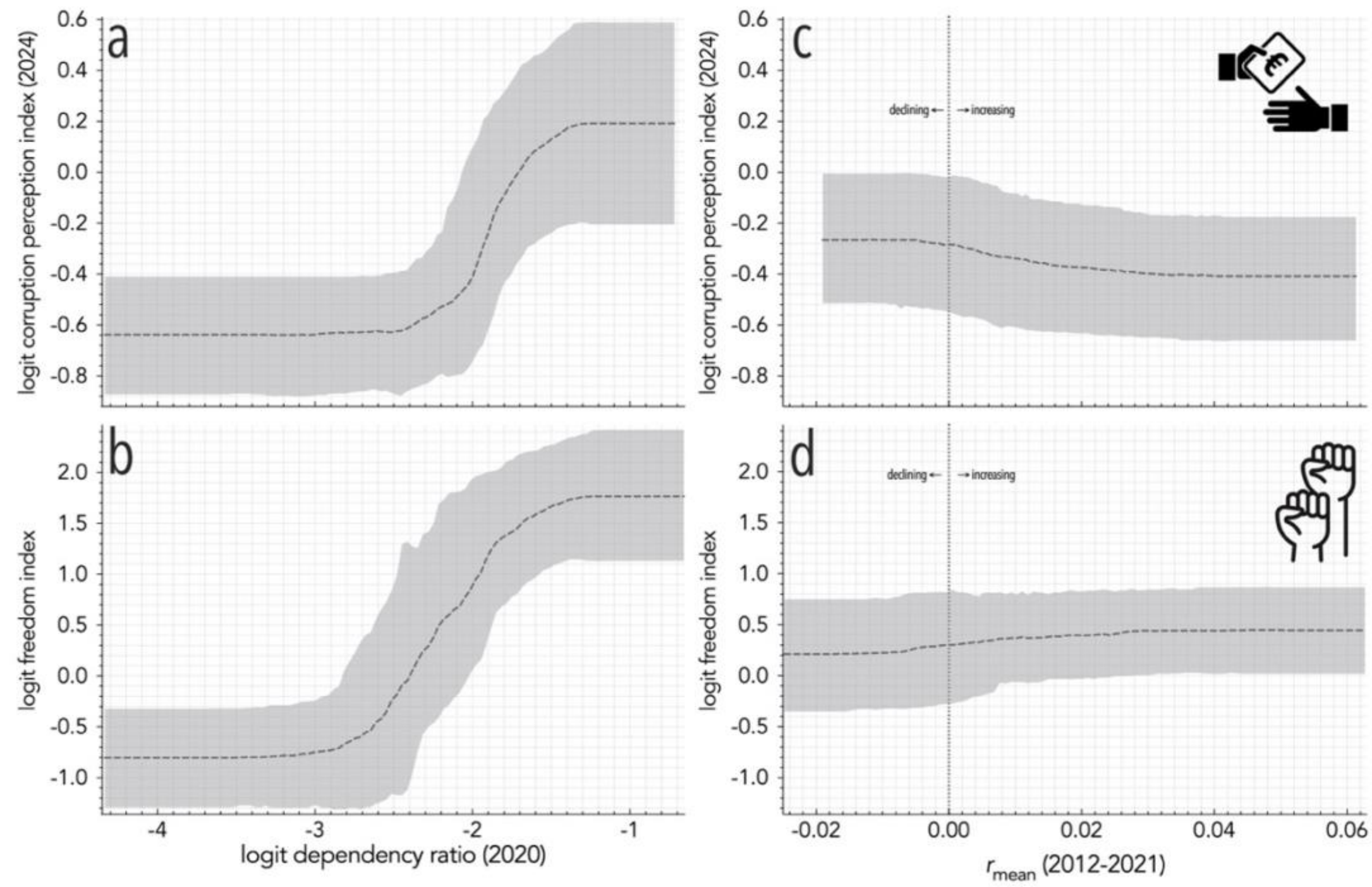

**Fig. 6 |** Resampled boosted regression tree results showing the relationships between (**a**, **c**) logit corruption perception index (higher values = less corruption) and (**b**, **d**) logit freedom score (higher values = more freedom) and two predictors: logit dependency ratio and mean rate of population change ($r_{mean}$ for 2012–2021). Full results for each response including $\log_{10}$ total population size in 2020 ($N_{2020}$) in Appendix V.

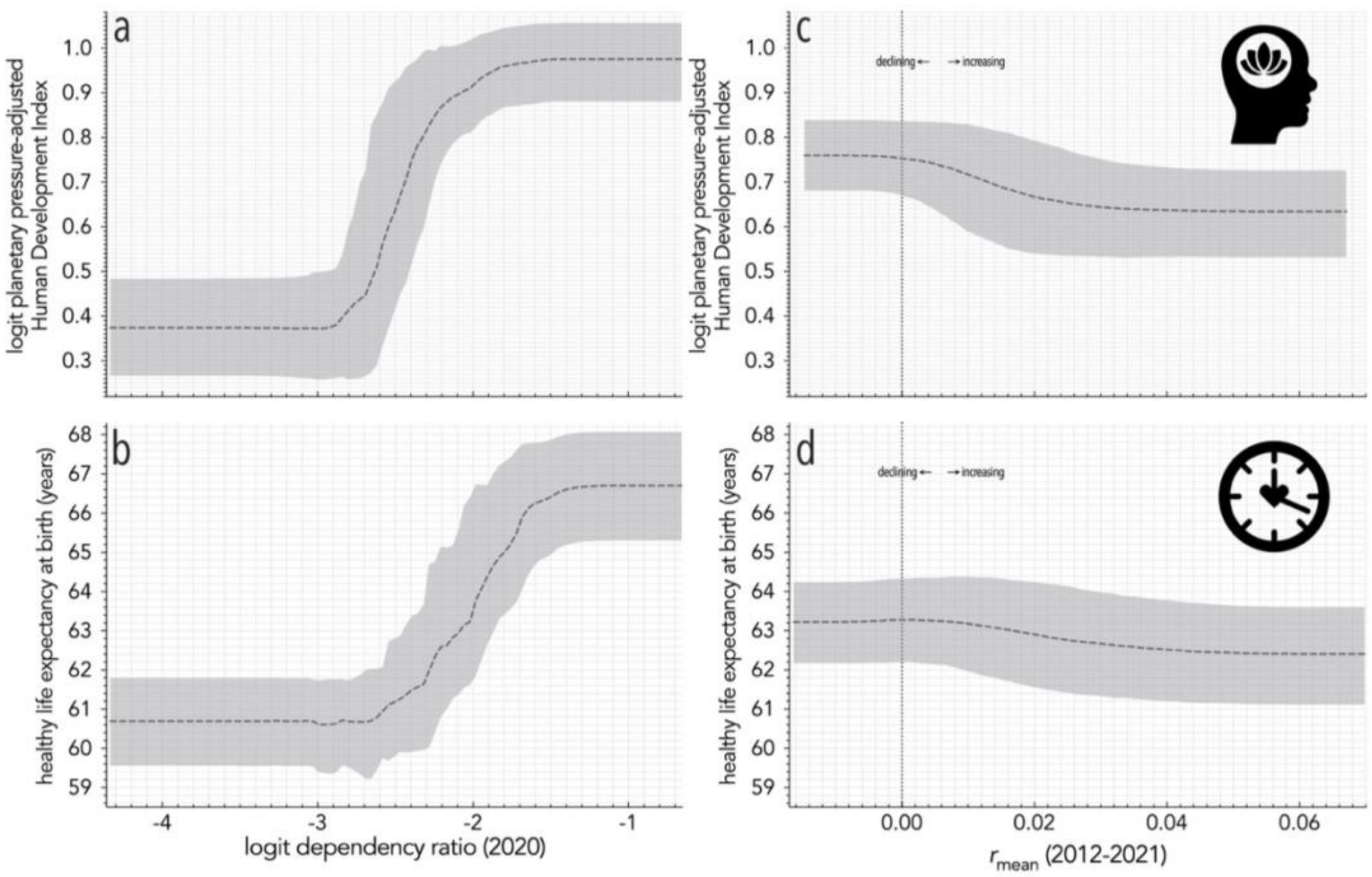

**Fig. 7 |** Resampled boosted regression tree results showing the relationships between (**a**, **c**) logit planetary pressure-adjusted Human Development Index and (**b**, **c**) healthy life expectancy at birth and two predictors: logit dependency ratio and mean rate of population change ($r_{mean}$ for 2012–2021). Full results for each response including $log_{10}$ total population size in 2020 ($N_{2020}$) in Appendices V and VII.

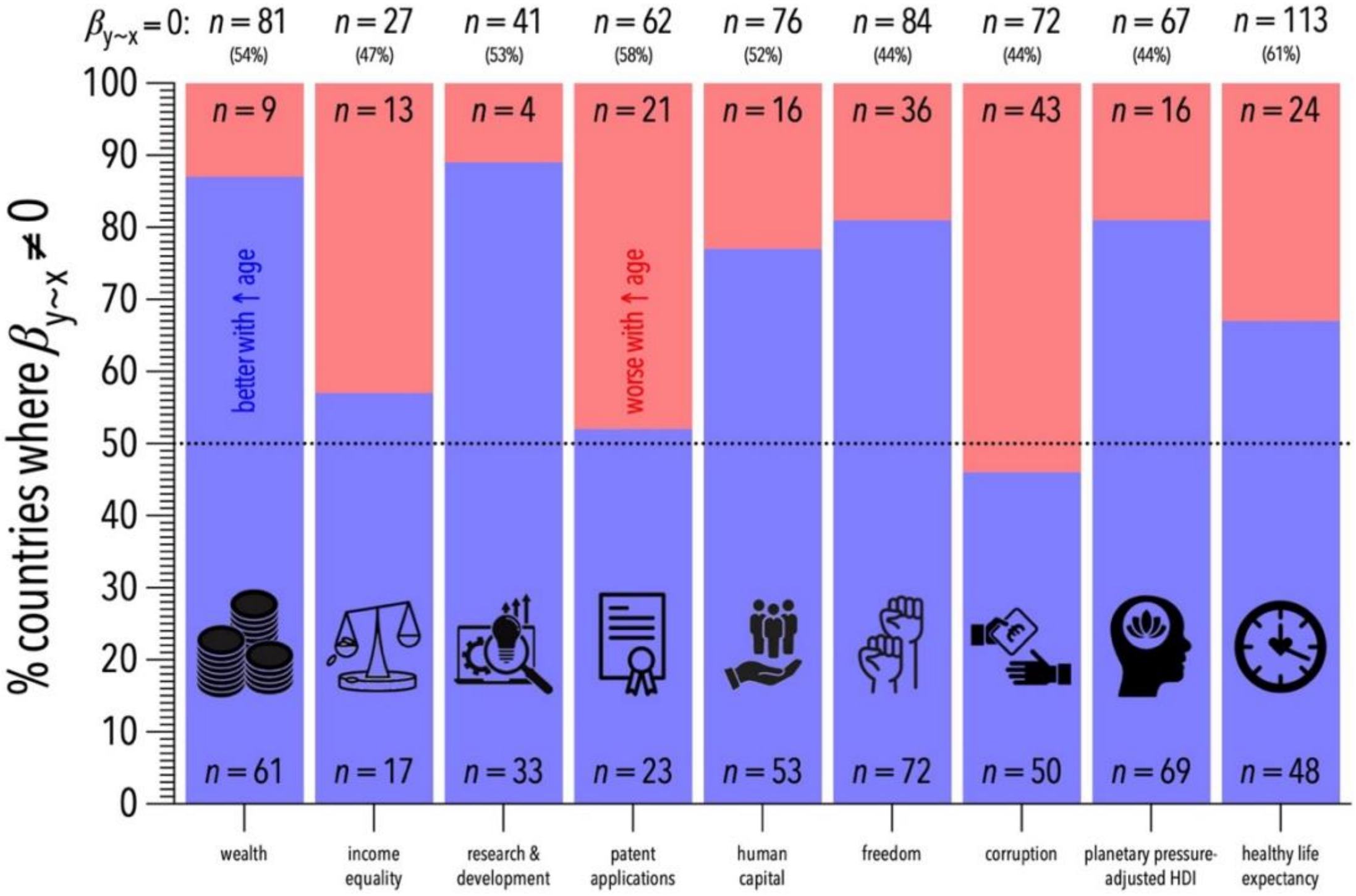

**Fig. 8 |** Percentage of countries where the yearly time series indicated either a better (blue) or worse (red) response outcome with an increasing dependency ratio (i.e., older population) after accounting for temporal autocorrelation (slope $\beta_{y\sim x}$ indicates evidence for a non-zero slope for the relationship between response *y* and dependency ratio *x*). Results for nine different responses shown: per-capita domestic comprehensive wealth index, income inequality (Gini coefficient), gross research and development expenditure (proportion of GDP), per-capita patent applications, human capital index, freedom score, corruption perception index, planetary pressure-adjusted human development index [HDI], healthy life expectancy at birth. Samples sizes (*n*) across the bottom of the figure (in blue) indicate number of countries with temporal autocorrelation-corrected evidence for a 'positive' relationship, and *n* across the top of the figure (in red) indicate number of countries with temporal autocorrelation-corrected evidence for a 'negative' relationship (note: Gini coefficient and corruption perception index use higher values for higher income inequality and more corruption, respectively, so the sign of the slopes is reversed for these responses). Sample sizes (*n*) in black at the top of the figure indicate the number of countries (and % of total countries assessed) for which the temporal autocorrelation-corrected confidence interval of the slope ($\beta$) included zero.

# Population change, age structure, and socio-economic performance

# Supplementary information

### Appendix Ia. Limitations of gross domestic product as an economic indicator

One of the most cited indicators of the size of a nation's economy is its gross domestic product (GDP), which measures the market value of all final goods and services produced within a defined period (usually quarterly and annually)[1]. Note the adjective 'final', which denotes the exclusion of intermediate goods from the calculation to avoid double counting, because their value is already incorporated in final goods. Put simply, GDP is the sum of a nation's government expenditures (public spending on the provision of goods and services, infrastructure, etc.), personal consumption expenditures (household payments by for goods and services), net exports (value of all exports minus value of all imports), and gross investment (total spending on new capital goods and changes in inventories)[1]. Because total GDP reflects both productivity and population size (both over time and among countries; Fig. S1), it is often expressed on a per-capita basis. When comparing GDP across time within a country, economists typically use 'real' GDP, which adjusts for inflation to reflect changes in actual output rather than price values. Economic 'growth' is measured as the percentage increase in real GDP from one time period to the next, while a 'recession' is typically defined as a sustained decline in real GDP. And to compare across countries, it is more meaningful to adjust GDP for purchasing power parity (PPP), which accounts for differences in local prices of equivalent goods and services.

**Fig. S1** | (a) Total (sum of all countries) annual global gross domestic product (GDP; source: World Bank; data.worldbank.org) corrected for purchasing power parity (PPP) and expressed in 2021 international dollars relative to the size of the global human population (source: United Nations Population Division) from 1990 to 2023 (least-squares regression $R^2 = 0.982$; black dashed line and grey-shaded prediction 95% confidence interval). Various years are indicated as point labels. (b) Power-law ($\log_{10}$-$\log_{10}$) relationship between global gross domestic product (GDP; source: World Bank) corrected for purchasing power parity (PPP) for the most recent year (2023) of data availability (source: World Bank) and population size in 2023 (source: United Nations Population Division). Least-squares power-law regression $R^2 = 0.772$ (grey dashed line with light grey-shaded prediction 95% confidence interval). Lines of best fit for two contrasting regions ($\text{EUROPE}_{x\sim y}$ and $\text{AFRICA}_{x\sim y}$; dashed blue and green lines, respectively) shown to demonstrate regional differences in GDP accumulation with population size. Several example countries are labelled according to their ISO 3-character country code: TUV = Tuvalu; DMA = Dominica; LUX = Luxembourg; QAT = Qatar; BDI = Burundi; NLD = Netherlands AUS = Australia; CAN = Canada; COD = Democratic Republic of Congo; ETH = Ethiopia; NGA = Nigeria; DEU = Germany; IDN = Indonesia; BRA = Brazil; USA = United States; IND = India; CHN = China. Points are colour-coded by major region (**AFR** = Africa; **ASIOC** = Asia + Oceania; **EUR** = Eurasia; **ME** = Middle East; **NAM** = North America; **SACAR** = South America + Caribbean).

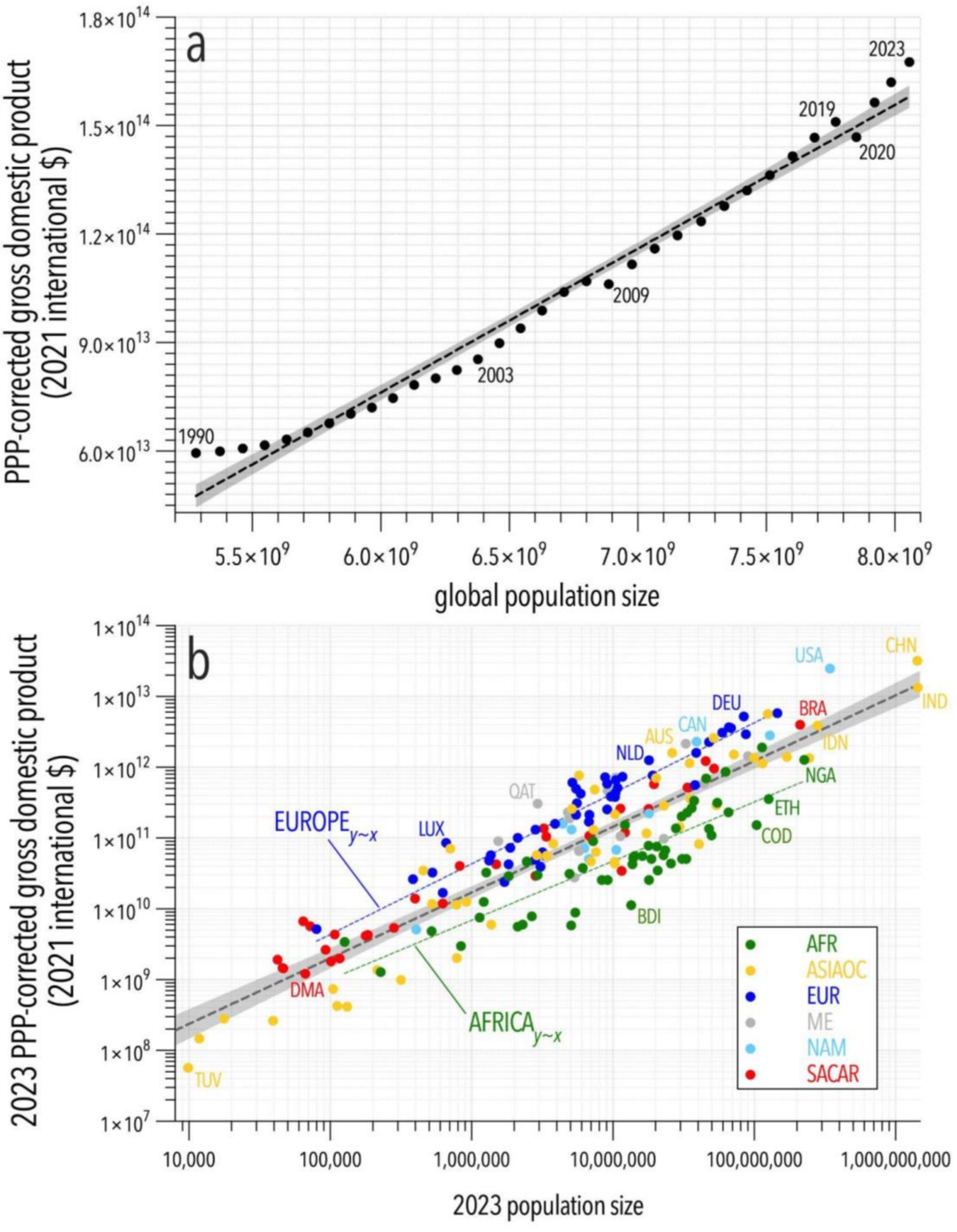

However, the picture that GDP provides is still incomplete. In practice, GDP captures only ‘market-based economic activity’, although some ‘nonmarket’ production is imputed and included, such as nonprofit spending on emergency housing and health care, and national defence spending[1]. And although GDP is a common way to measure economic activity, it is not a perfect measure of prosperity. For example, it treats activities that make people better off and worse off alike, while leaving out important considerations such as inequality, environmental costs, and impacts on future generations. More specifically, every expense included in the calculation of GDP is taken as a positive contribution, regardless of whether it enhances or diminishes well-being[2,3]. GDP also ignores non-monetary welfare-enhancing transactions that fall outside the market, as well as certain cross-border flows such as net factor income from abroad. GDP can also misrepresent economic activity by failing to capture the size of the shadow economy (e.g., unpaid household labour, volunteer work, black market activity), thereby undervaluing total production relative to economic well-being it is assumed to measure[4]. Importantly, GDP does not account for income distribution, so nations with high inequality among their citizens (e.g., a small minority of extremely wealthy individuals alongside a majority who are poor) can still have a large GDP[4]. Neither does GDP account for long-term environmental degradation, because it does not treat the depreciation of natural and community capital as an economic cost[5]. GDP therefore fails to capture intergenerational equity because it overlooks these future economic costs[6].

Because of these well-known limitations, GDP was developed to measure economic output, not necessarily the living standards or overall welfare of a nation's citizens[5]. So, even a quantitative increase of the economy (either in gross or per capita terms) often does not affect many individual components of human well-being, such as social inequality, safety, health, and civil rights[1]. For example, per-capita GDP is only weakly correlated with the most common development indices[7,8]. This is why most economists note that GDP was designed to measure economic output, not overall well-being. Nevertheless, GDP per capita is still often used as a rough proxy for living standards, and this nuance is frequently overlooked by politicians, policymakers, and journalists[6].

In conclusion, although GDP measures the scale of a nation's economy, it does not capture overall economic performance or reflect the average social and economic well-being of either current or future citizens. Per-capita GDP is often used as a rougher proxy for individual living standards, but even this has clear limitations[6].

## Appendix Ib. Relationship between gross domestic product and dependency ratio/population rate of change

Despite the limitations of GDP as an indicator of 'wealth' (Appendix Ia), we repeated the main analyses using per-capita gross domestic product adjusted for purchasing power parity (PPP), which better reflects average living standards across countries. Given the correlation between per-capita PPP-GDP and national per-capita domestic comprehensive wealth index (Appendix II), we expect similar relationships to dependency ratio, rate of population change, and population size.

*Boosted regression trees*

The resampled boosted regression tree analysis revealed that of the three predictors we considered to explain variation in per-capita PPP-GDP among countries, the dependency ratio explained the highest proportion of the total variance in the response for the models including either $r_{\text{mean}}$ from 1950–2021 or 2012–2021 (Fig. S2).

**Fig. S2** | Boosted regression tree results testing the relationships between $\log_{10}$ per-capita purchasing power parity-adjusted gross domestic product and three hypothesised predictors: logit dependency ratio (2020), $\log_{10}$ total population size in 2020 ($N_{2020}$), and mean rate of population change ($r_{\mathrm{mean}}$ for both 1950–2021 in black, and 2012–2021 in grey). Panel **a** shows the mean (± 95% confidence interval) relative contribution (%) of each predictor after 10,000 iterations of the model. The model explained 41.6–73.3% and 40.8–72.8% of the variation ($\beta_{\mathrm{CV}}$) in the response for those including $r$ calculated from 1950–2021 and 2012–2021, respectively. Panels **b**–**d**: mean and 95% prediction confidence intervals for each predictor after Gaussian resampling $r_{\mathrm{mean}}$ with its associated standard deviation (from 1950–2021 in black and 2012–2021 in grey) and resampling countries to reduce the effects of spatial autocorrelation.

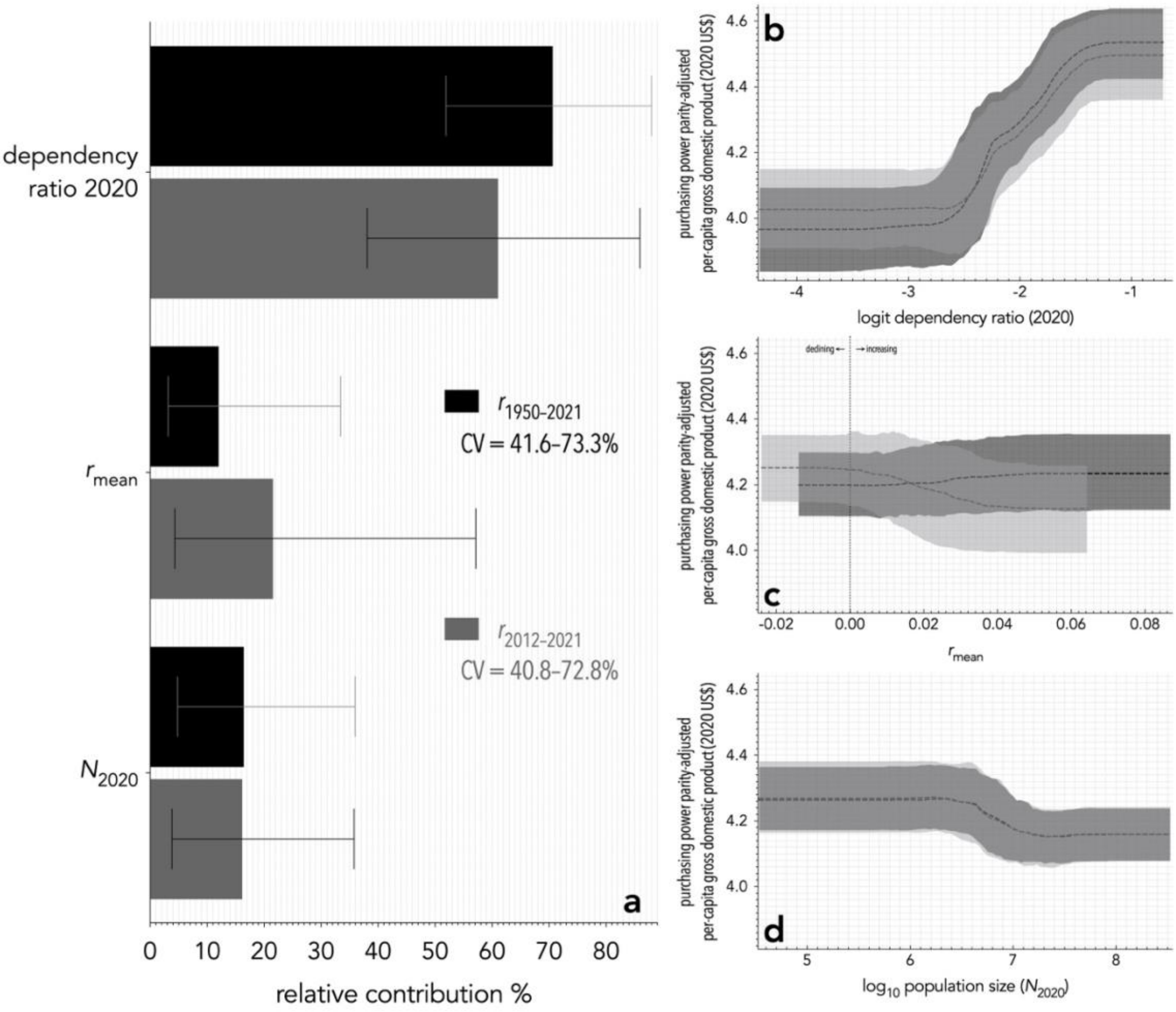

## Appendix II. Domestic comprehensive wealth index

**Fig. S3** | Power-law ($\log_{10}$-$\log_{10}$) relationship between national per-capita domestic comprehensive wealth index for the year 2020 (source: World Bank and purchase power parity (PPP)-adjusted per-capita gross domestic product (source: World Bank). Least-squares regression gives $R^2 = 0.846$. Bubble size indicates relative total population size of each country in 2020. Bubbles are colour-coded by major region (**AFR** = Africa; **ASIOC** = Asia + Oceania; **EUR** = Eurasia; **ME** = Middle East; **NAM** = North America; **SACAR** = South America + Caribbean). Labels for a sample of the points are ISO 3-character country codes.

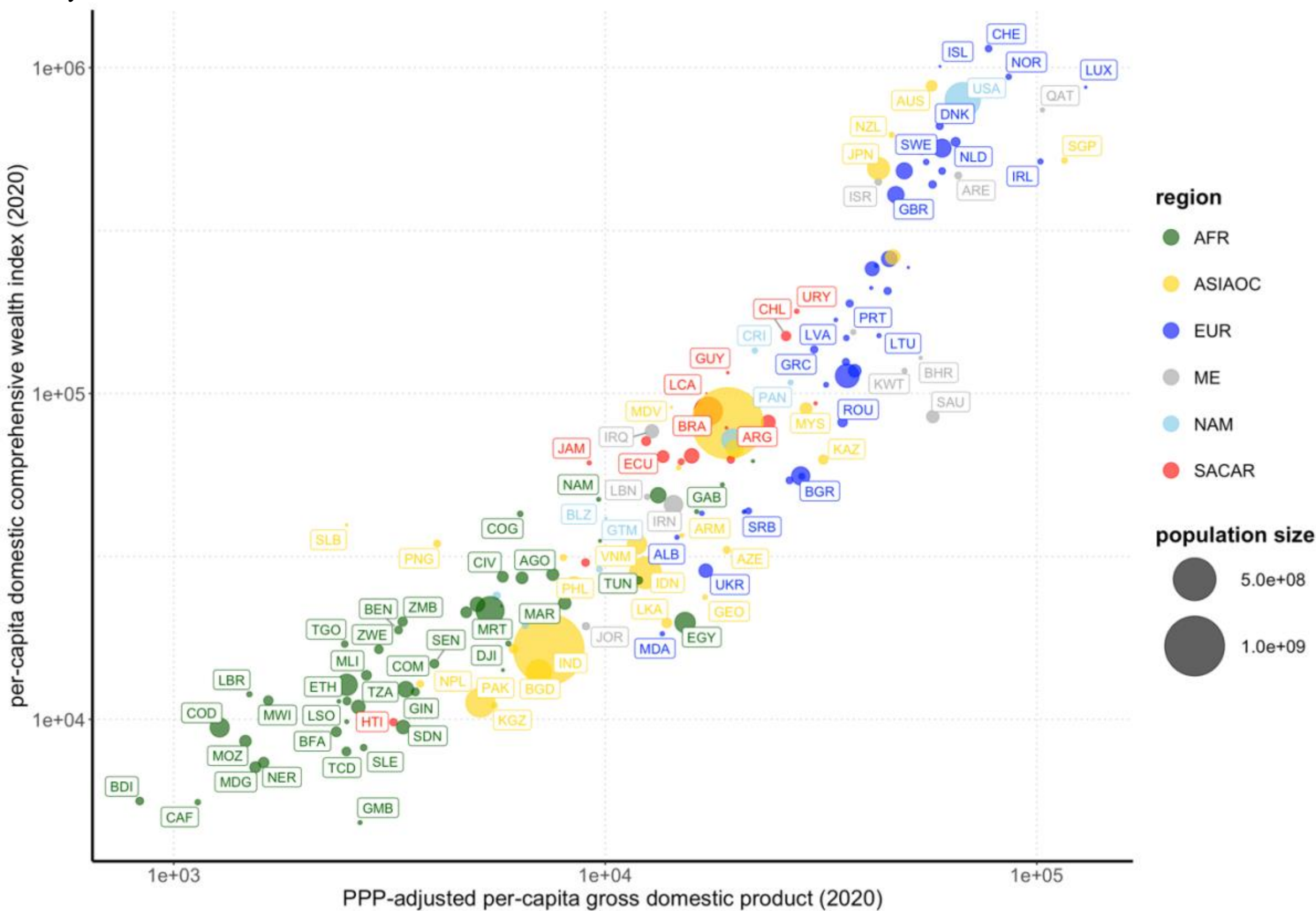

**Fig. S4** | Relationship between $\log_{10}$ national per-capita domestic comprehensive wealth index for the year 2020 (source: World Bank) and logit dependency ratio (2020) (source: United Nations Population Division). Bubble size indicates relative total population size of each country in 2020. Bubbles are colour-coded by major region: **AFR (green)** = Africa; **ASIOC (gold)** = Asia + Oceania; **EUR (blue)** = Eurasia; **ME** = Middle East (silver); **NAM (light blue)** = North America; **SACAR (red)** = South America + Caribbean. Labels for a sample of the points are ISO 3-character country codes.

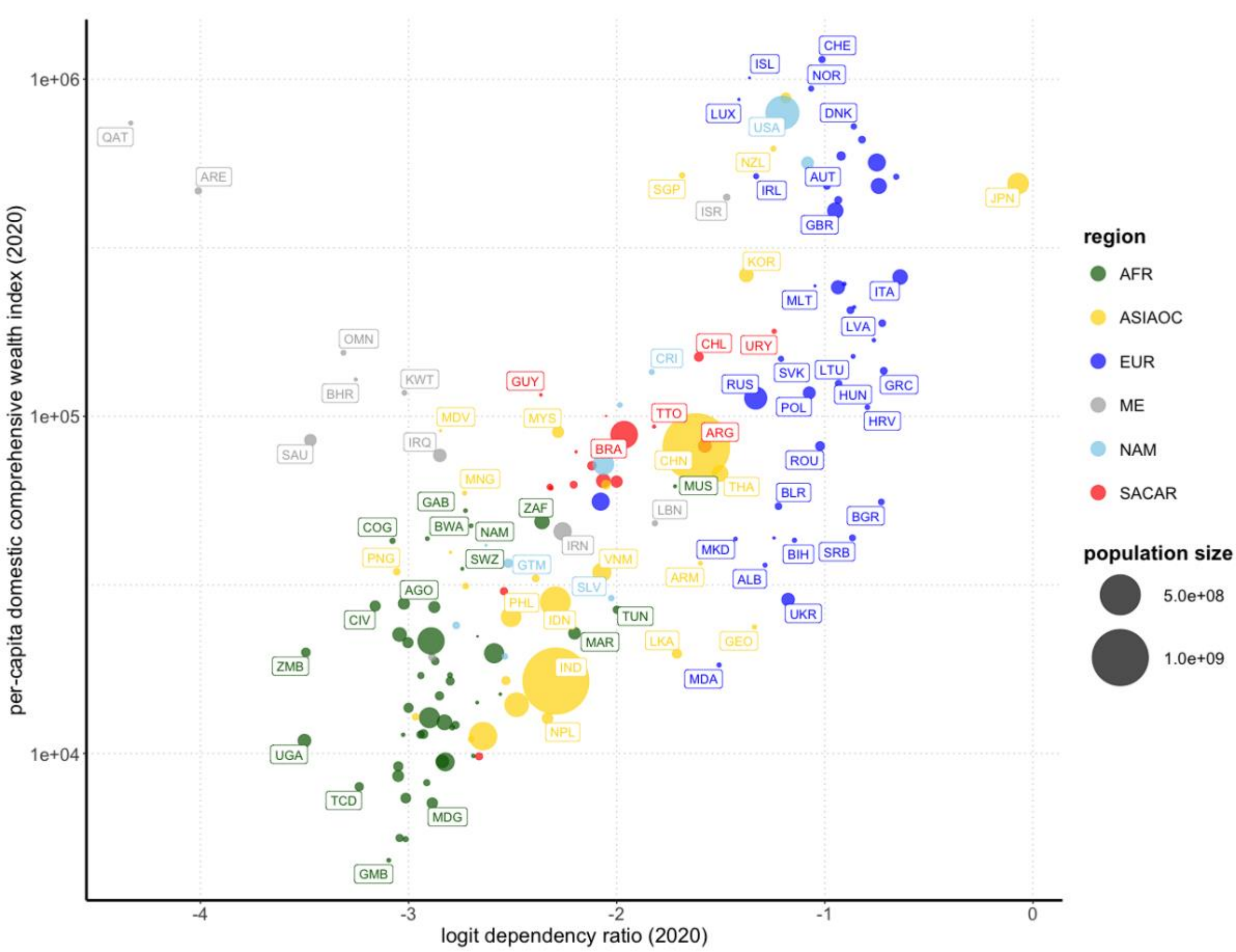

*Boosted regression trees*

The resampled boosted regression tree analysis revealed that of the three predictors we considered to explain variation in the domestic comprehensive wealth index among countries, the dependency ratio explained the highest proportion of the total variance in the response for the models including either $r_{\text{mean}}$ from 1950–2021 or 2012–2021 (Fig. S5). The predicted relationship revealed a quasi-threshold effect where per-capita domestic comprehensive wealth declined rapidly below a dependency ratio of ~ 0.16 (-1.65 on the logit scale) to stabilise at ~ 0.09 (-2.3 on the logit scale) (Fig. S5, S6).

**Fig. S5** | Boosted regression tree results testing the relationships between $\log_{10}$ per-capita domestic comprehensive wealth index and three hypothesised predictors: logit dependency ratio (2020), $\log_{10}$ total population size in 2020 ($N_{2020}$), and mean rate of population change ($r_{mean}$ for both 1950–2021 in black, and 2012–2021 in grey). Panel **a** shows the mean (± 95% confidence interval) relative contribution (%) of each predictor after 10,000 iterations of the model. The model explained 35.1–69.0% and 35.6–68.3% of the variation ($\beta_{CV}$) in the response for those including $r$ calculated from 1950–2021 and 2012–2021, respectively. Panels **b**–**d**: mean and 95% prediction confidence intervals for each predictor after Gaussian resampling $r_{mean}$ with its associated standard deviation (from 1950–2021 in black and 2012–2021 in grey) and resampling countries to reduce the effects of spatial autocorrelation. Prediction curve for panel b ($r_{mean}$ 1950–2021 only) on the untransformed scale for both $x$ and $y$ axes shown in Fig. S6).

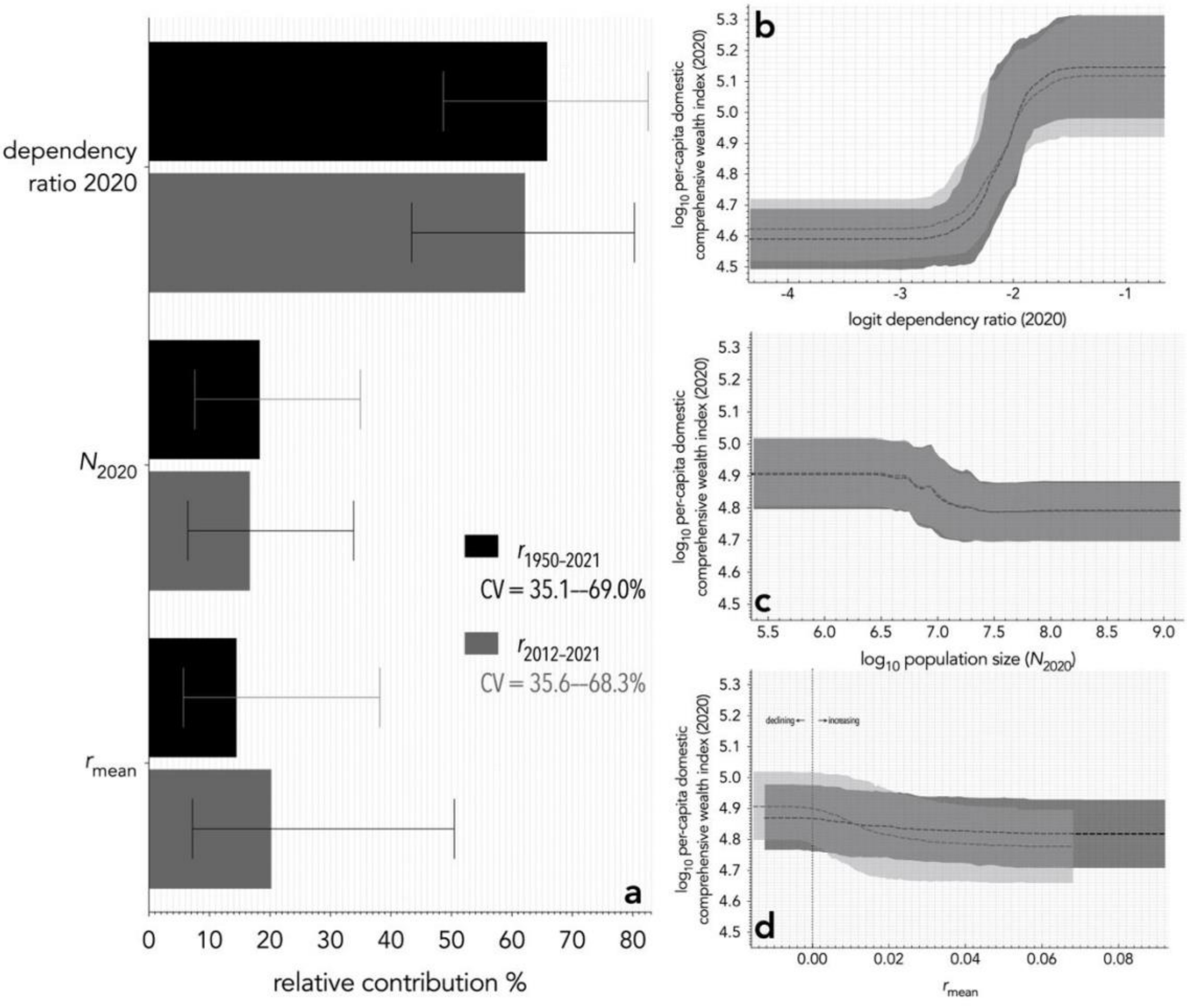

**Fig. S6** | Predicted boosted regression tree relationship (± 95% confidence interval calculated from 10,000 iterations of the model) between per-capita domestic comprehensive wealth index and the dependency ratio (2020) on the linear scale after accounting for total population size in 2020 and mean rate of population change ($r_{mean}$) from 1950–2021. There were 26 countries in this dataset with a dependency ratio between ≥ 0.09 and ≤ 0.16 (i.e., where the relationship dipped precipitously from high to low wealth; vertical dotted lines):

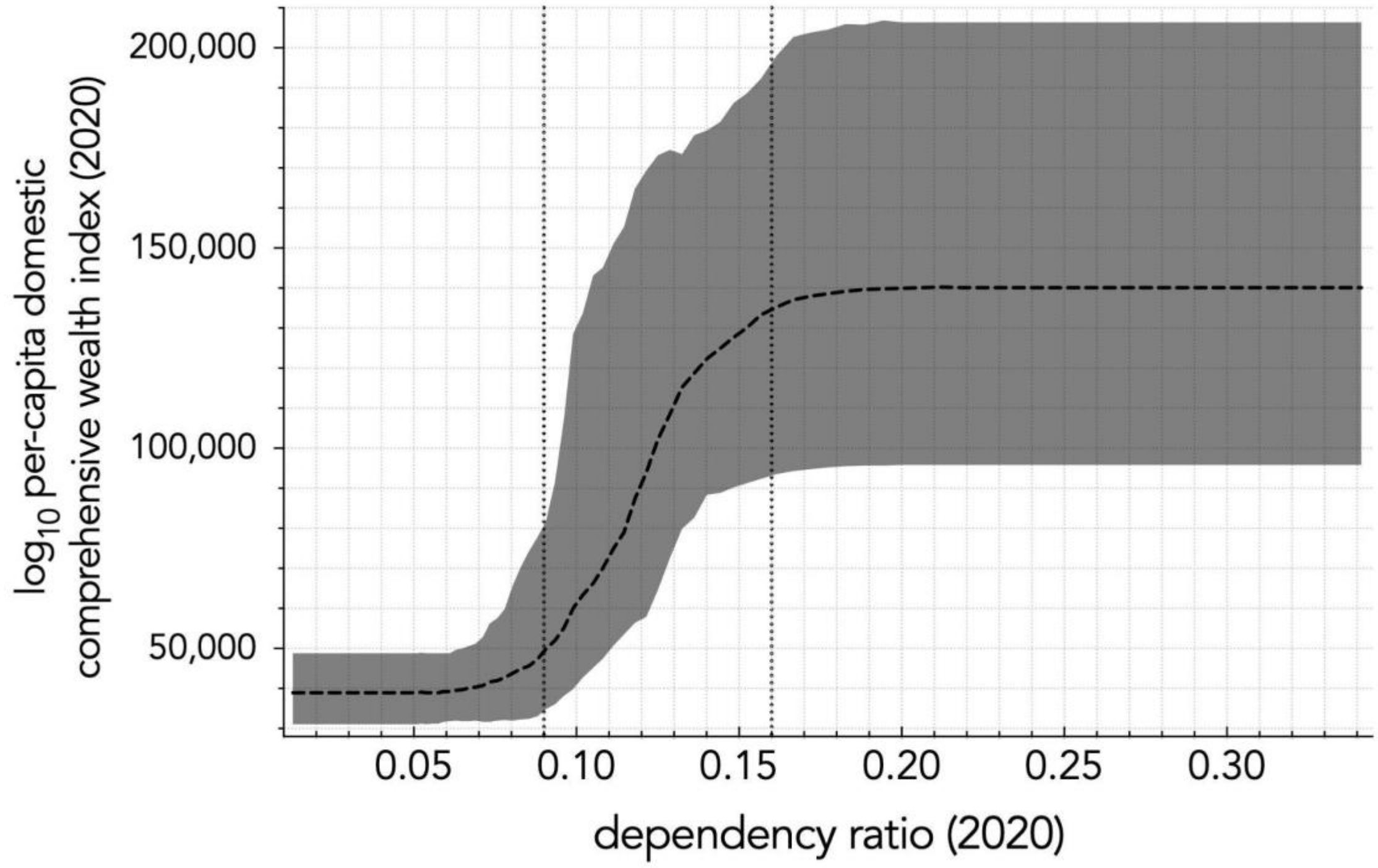

Despite the response already correcting for total population size (i.e., *per-capita* domestic comprehensive wealth index), there was still a weak effect of population size on the wealth response; here, larger populations had slightly lower domestic wealth (Fig. S5). There was also a weak relationship between rate of population change and the wealth response (Fig. S5d), but in the opposite direction to the hypothesis that declining or slow-growing populations have compromised economies. Instead, declining/slow-growing populations had *higher* mean wealth than faster-growing nations (Fig. S5).

## Appendix III. Income inequality (Gini coefficient)

**Fig. S7** | Relationship between the mean Gini coefficient (income inequality) (0 = perfect income equality; 100 = perfect income inequality) (source: World Bank) and the logit dependency ratio ($\Sigma N_{\geq 65} \div \Sigma N_{16\text{–}65}$; see Methods). Bubble size indicates relative total population size of each country in 2020. Bubbles are colour-coded by major region (**AFR** = Africa; **ASIOC** = Asia + Oceania; **EUR** = Eurasia; **ME** = Middle East; **NAM** = North America; **SACAR** = South America + Caribbean). Labels for a sample of the points are ISO 3-character country codes.

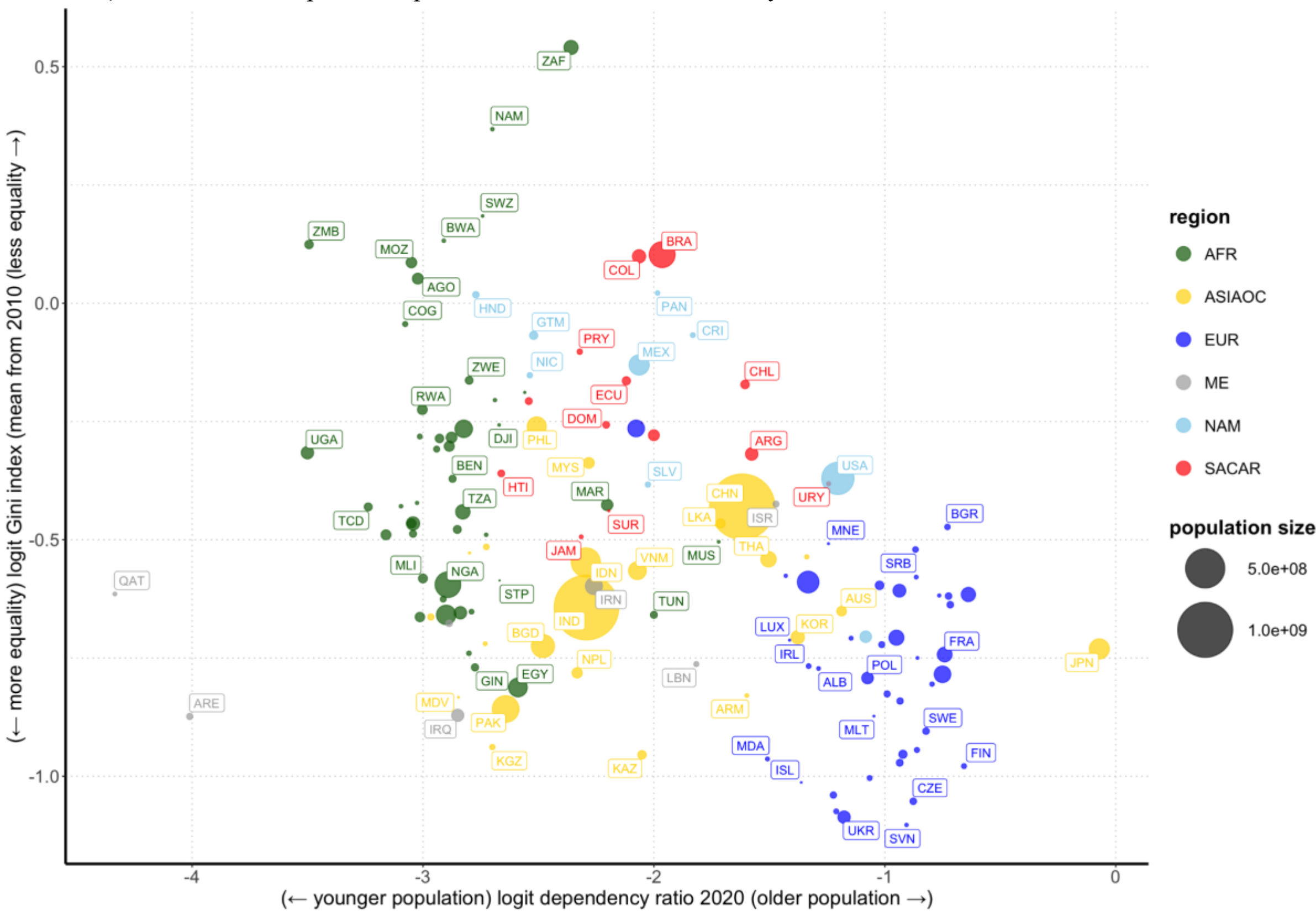

### *Boosted regression trees*

The resampled boosted regression tree analysis revealed that of the three predictors considered, $r_{\text{mean}}$ accounted for the largest share of variance in income inequality (either $r_{\text{mean}}$ from 1950–2021 or 2012–2021), but with only slightly lower share from the dependency ratio (Fig. S8). Although overall variance explained was lower due to higher uncertainty, the predicted patterns suggested that as $r_{\text{mean}}$ increased, so too did income inequality (Fig. S8). There was also a slight decrease in income inequality as the dependency ratio increased (Fig. S8). There was no clear signal from total population size (Fig. S8d). Overall, declining/slow-growing and older populations had *lower* mean income inequality than faster-growing/younger nations (Fig. S8).

**Fig. S8** | Boosted regression tree results testing the relationships between logit Gini coefficient (income inequality) and three hypothesised predictors: logit dependency ratio (2020), $\log_{10}$ total population size in 2020 ($N_{2020}$), and mean rate of population change ($r_{mean}$ for both 1950–2021 in black, and 2012–2021 in grey). Panel **a** shows the mean (± 95% confidence interval) relative contribution (%) of each predictor after 10,000 iterations of the model. The model explained 2.9–46.4% and 0–51.2% of the variation ($\beta_{CV}$) in the response for those including $r$ calculated from 1950–2021 and 2012–2021, respectively. Panels **b**–**d**: mean and 95% prediction confidence intervals for each predictor after Gaussian resampling $r_{mean}$ with its associated standard deviation (from 1950–2021 in black and 2012–2021 in grey) and resampling countries to reduce the effects of spatial autocorrelation.

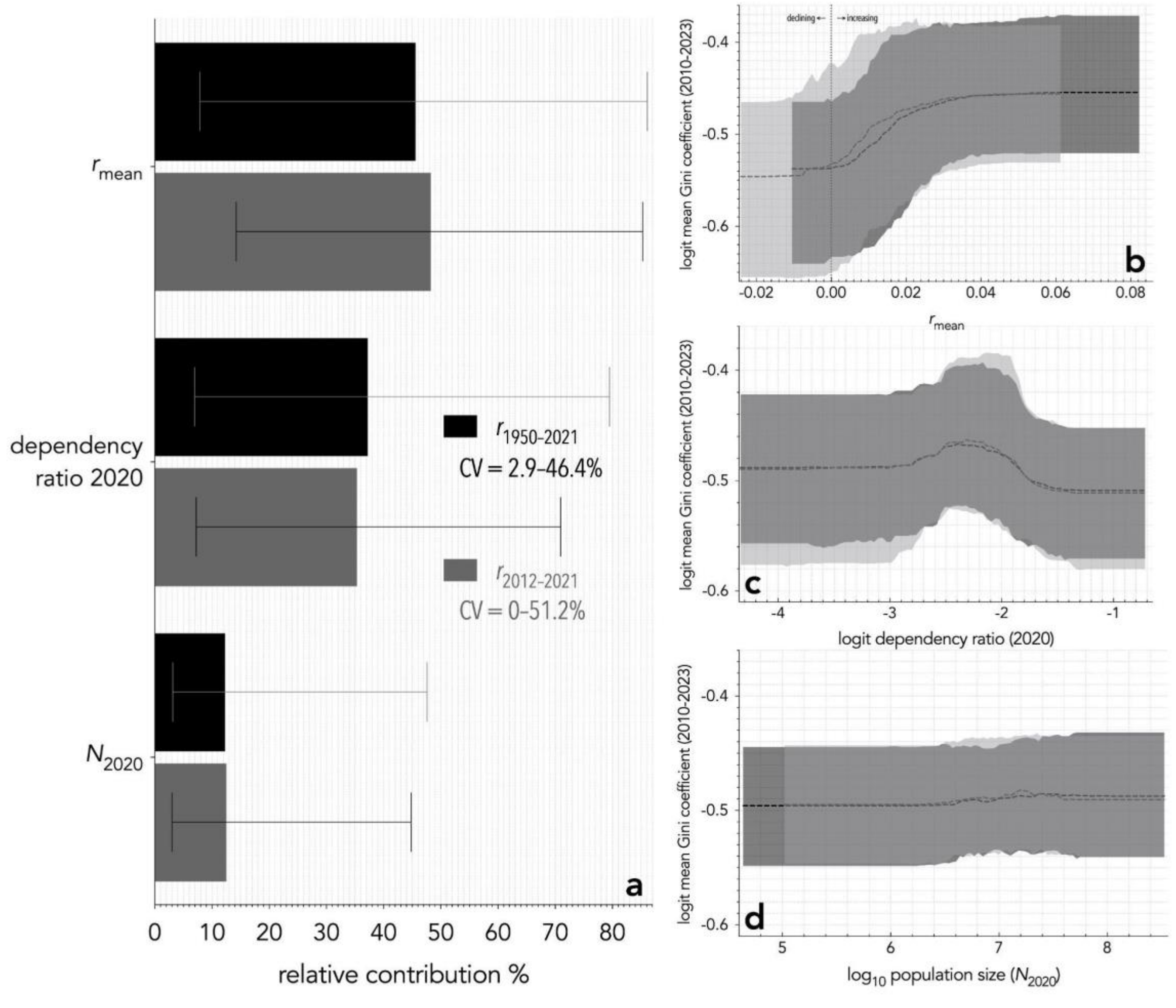

## Appendix IV. Productivity

(*i*) **Labour productivity** (2018 gross domestic product per employment, in 2010 \$US[9]; World Bank).
We examined the relationship between labour productivity; however, this metric is strongly correlated with per-capita gross domestic product (power-law $R^2$ = 0.906), because both measure output relative to individuals in the workforce or total population.

**Fig. S9** | Power-law ($\log_{10}$-$\log_{10}$) relationship between labour productivity in terms of gross domestic product per employment (\$US 2010) for the year 2018 (source: World Bank) and national purchase power parity (PPP)-adjusted per-capita gross domestic product (source: World Bank). Least-squares regression gives $R^2$ = 0.906. Bubble size indicates relative total population size of each country in 2020. Bubbles are colour-coded by major region (**AFR** = Africa; **ASIOC** = Asia + Oceania; **EUR** = Eurasia; **ME** = Middle East; **NAM** = North America; **SACAR** = South America + Caribbean). Labels for a sample of the points are ISO 3-character country codes.

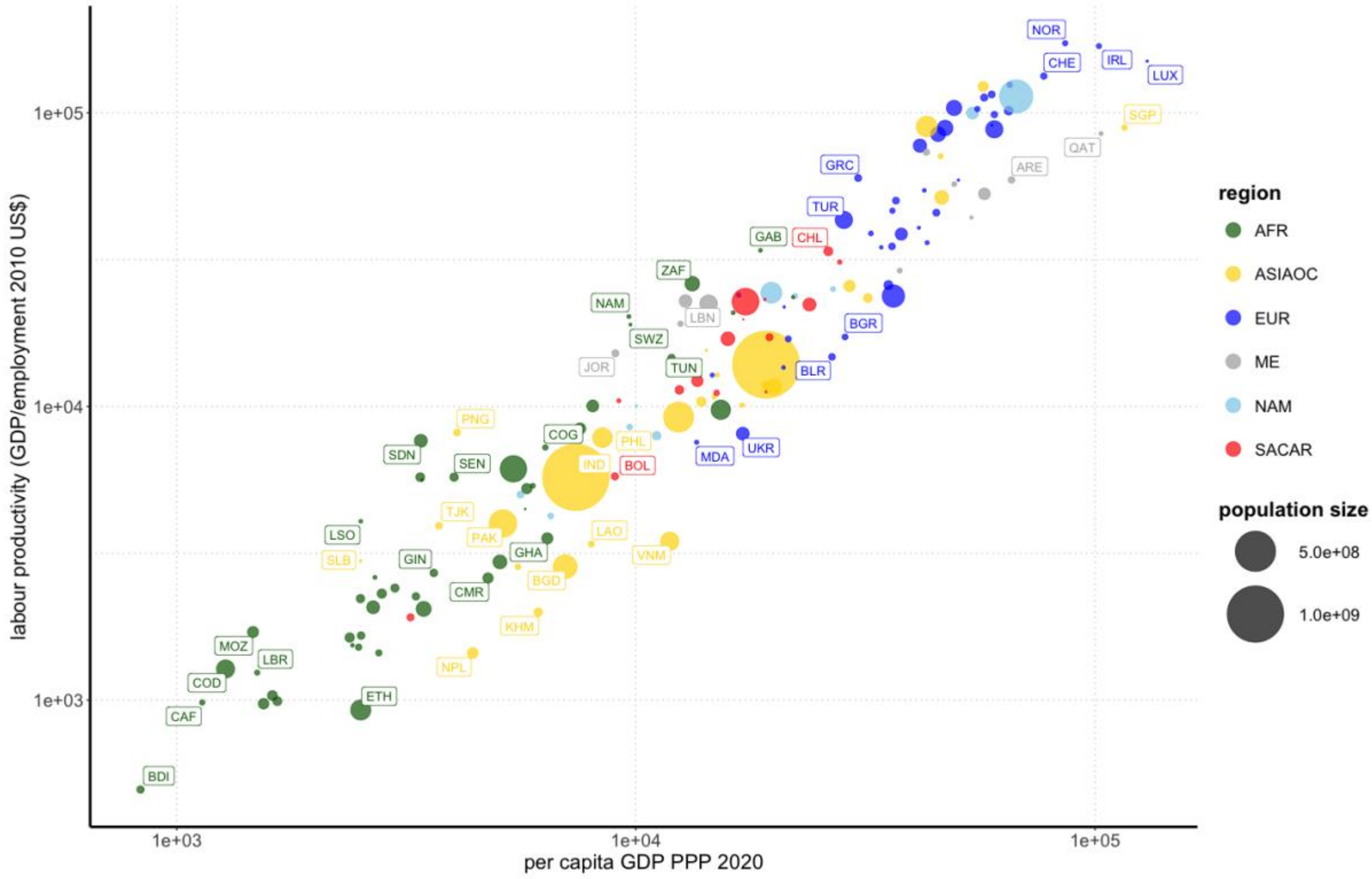

(*ii*) **Value-added manufacturing** (% of gross domestic product; World Bank).
This is defined as the physical or chemical transformation of materials or components into new products. Value added is the contribution to the economy by a producer or an industry or an institutional sector, estimated by the total value of output produced and deducting the total value of intermediate consumption of goods and services used to produce that output. There was no indication of a relationship of the most recent values for each country to the dependency ratio (Fig. S10) or rate of population change (Fig. S11), so we disregarded this variable for testing.

**Fig. S10** | Relationship between value-added manufacturing (% of gross domestic product; source: World Bank) and the logit of the dependency ratio. Bubble size indicates relative total population size of each country in 2020. Bubbles are colour-coded by major region (**AFR** = Africa; **ASIOC** = Asia + Oceania; **EUR** = Eurasia; **ME** = Middle East; **NAM** = North America; **SACAR** = South America + Caribbean). Labels for a sample of the points are ISO 3-character country codes.

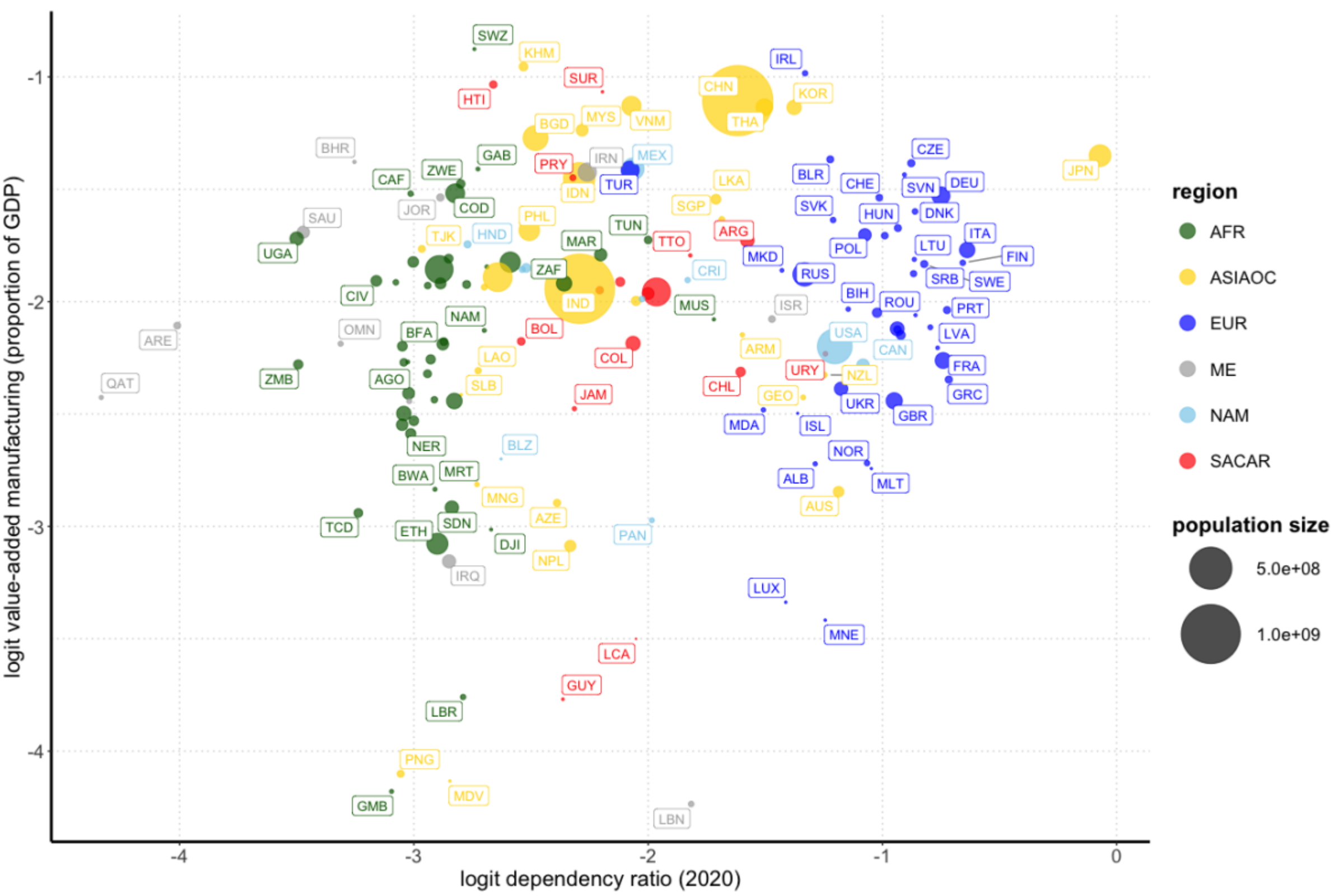

**Fig. S11** | Relationship between value-added manufacturing (% of gross domestic product; source: World Bank) and the mean rate of population change from 2012–2021. Bubble size indicates relative total population size of each country in 2020. Bubbles are colour-coded by major region (**AFR** = Africa; **ASIOC** = Asia + Oceania; **EUR** = Eurasia; **ME** = Middle East; **NAM** = North America; **SACAR** = South America + Caribbean). Labels for a sample of the points are ISO 3-character country codes.

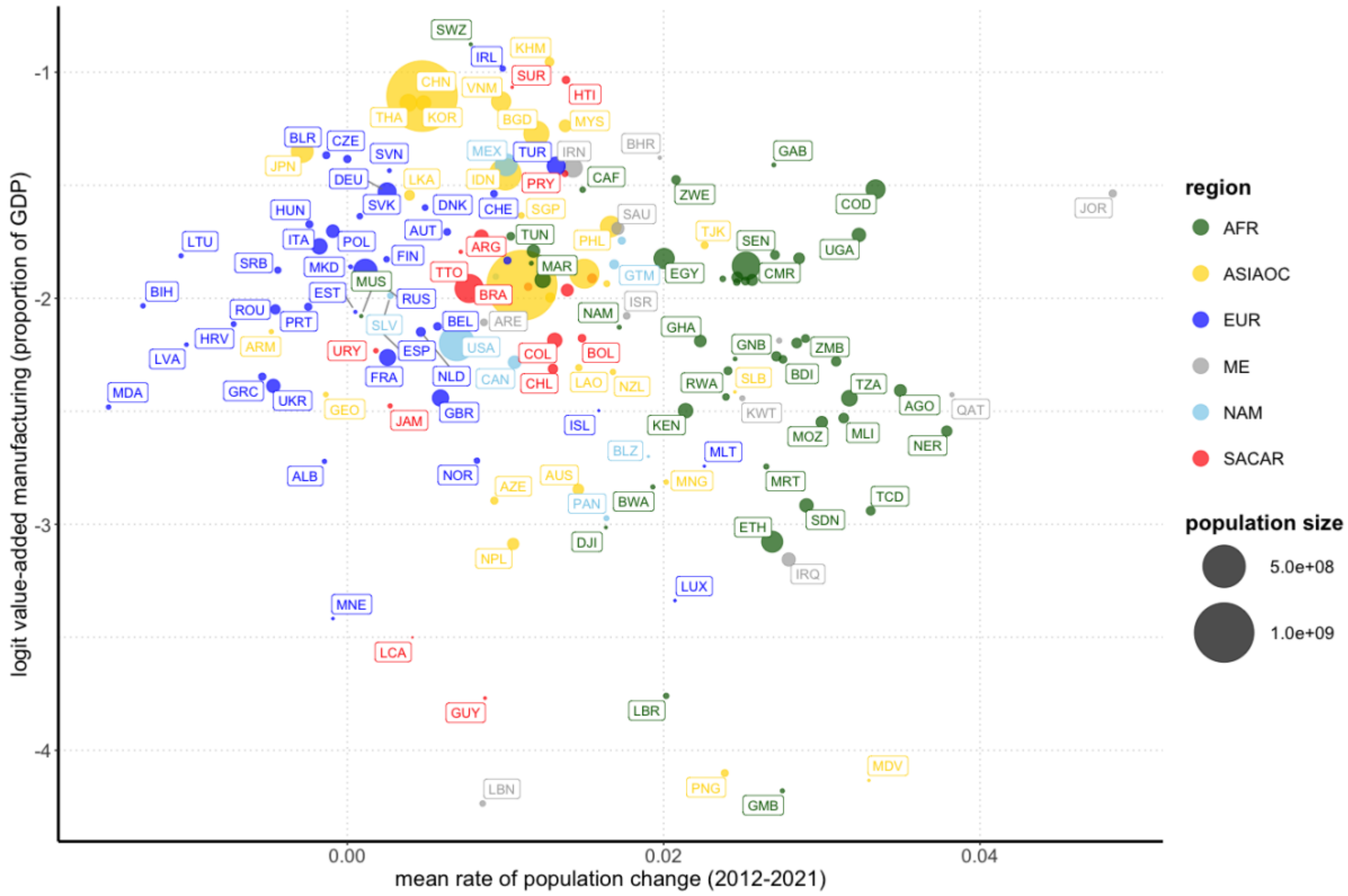

(*iii*) **Gross capital formation** (% of gross domestic product; World Bank).

This variable includes acquisitions less disposals of produced assets for purposes of fixed capital formation, inventories or valuables. As for value-added manufacturing, there was no indication of a relationship between the most recent value per country and the dependency ratio (Fig. S12) or rate of population change (Fig. S13), so we disregarded this variable for testing.

**Fig. S12** | Relationship between gross capital formation (% of gross domestic product; source: World Bank) and the logit of the dependency ratio. Bubble size indicates relative total population size of each country in 2020. Bubbles are colour-coded by major region (**AFR** = Africa; **ASIOC** = Asia + Oceania; **EUR** = Eurasia; **ME** = Middle East; **NAM** = North America; **SACAR** = South America + Caribbean). Labels for a sample of the points are ISO 3-character country codes.

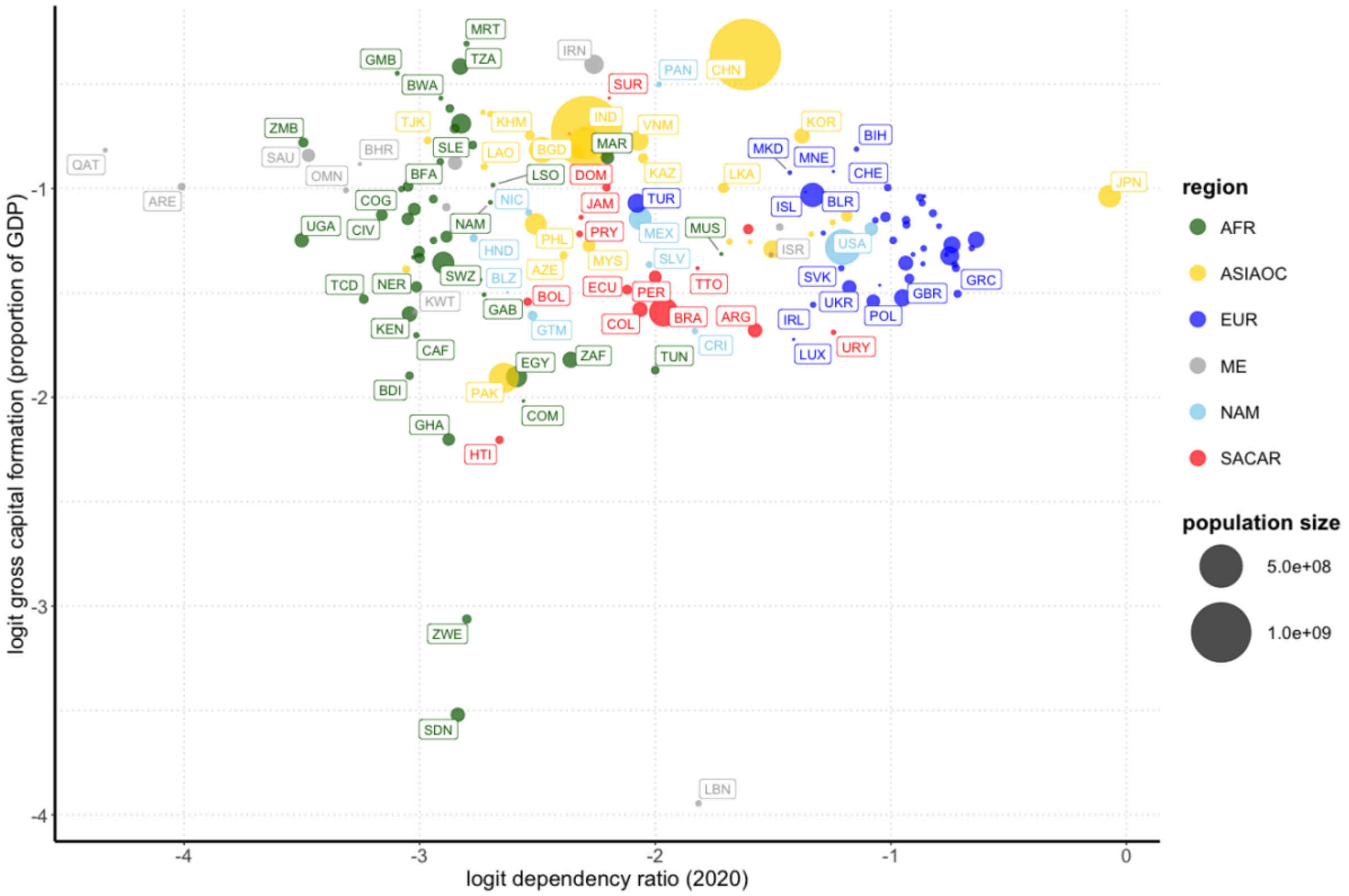

**Fig. S13** | Relationship between gross capital formation (% of gross domestic product; source: World Bank) and the mean rate of population change from 2012–2021. Bubble size indicates relative total population size of each country in 2020. Bubbles are colour-coded by major region (**AFR** = Africa; **ASIOC** = Asia + Oceania; **EUR** = Eurasia; **ME** = Middle East; **NAM** = North America; **SACAR** = South America + Caribbean). Labels for a sample of the points are ISO 3-character country codes.

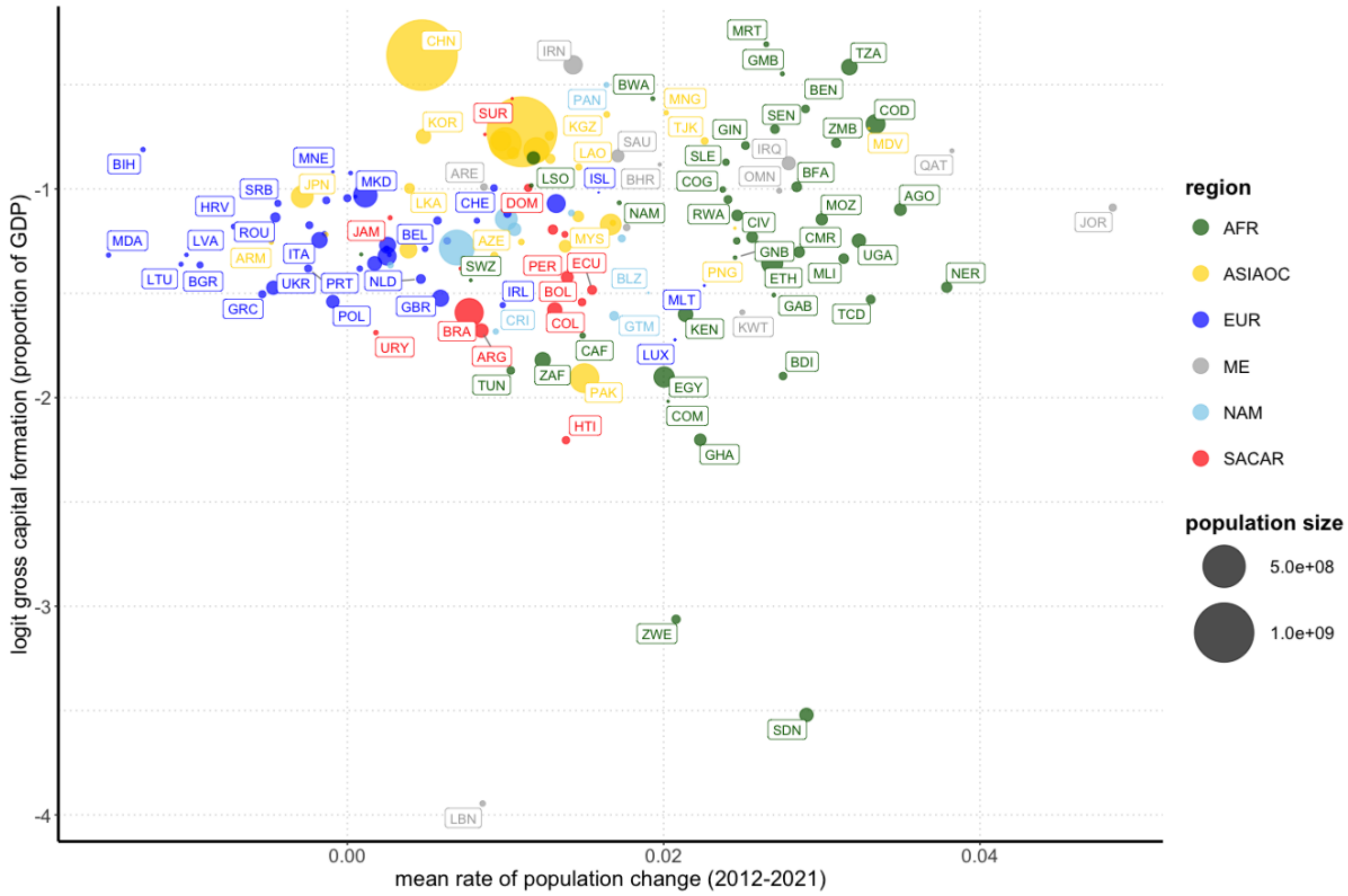

(*iv*) **Per-capita capital stock at constant 2018 national prices** (Penn World Table version 10.01, Groningen Growth and Development Centre, Faculty of Economics and Business; rug.nl/ggdc/productivity/pwt)[10]. Capital stocks are estimated based on cumulating and depreciation past investments using the perpetual inventory method. However, this metric (2019 values) is strongly correlated with per-capita gross domestic product (power-law $R^2 = 0.912$; Fig. S14), so we disregarded it for analysis.

**Fig. S14** | Power-law ($\log_{10}$-$\log_{10}$) relationship between national per-capita capital stock at constant 2018 national prices (source: Penn World Table version 10.01, Groningen Growth and Development Centre, Faculty of Economics and Business)[10] and purchase power parity (PPP)-adjusted per-capita gross domestic product (source: World Bank). Least-squares regression gives $R^2 = 0.906$. Bubble size indicates relative total population size of each country in 2020. Bubbles are colour-coded by major region (**AFR** = Africa; **ASIOC** = Asia + Oceania; **EUR** = Eurasia; **ME** = Middle East; **NAM** = North America; **SACAR** = South America + Caribbean). Labels for a sample of the points are ISO 3-character country codes.

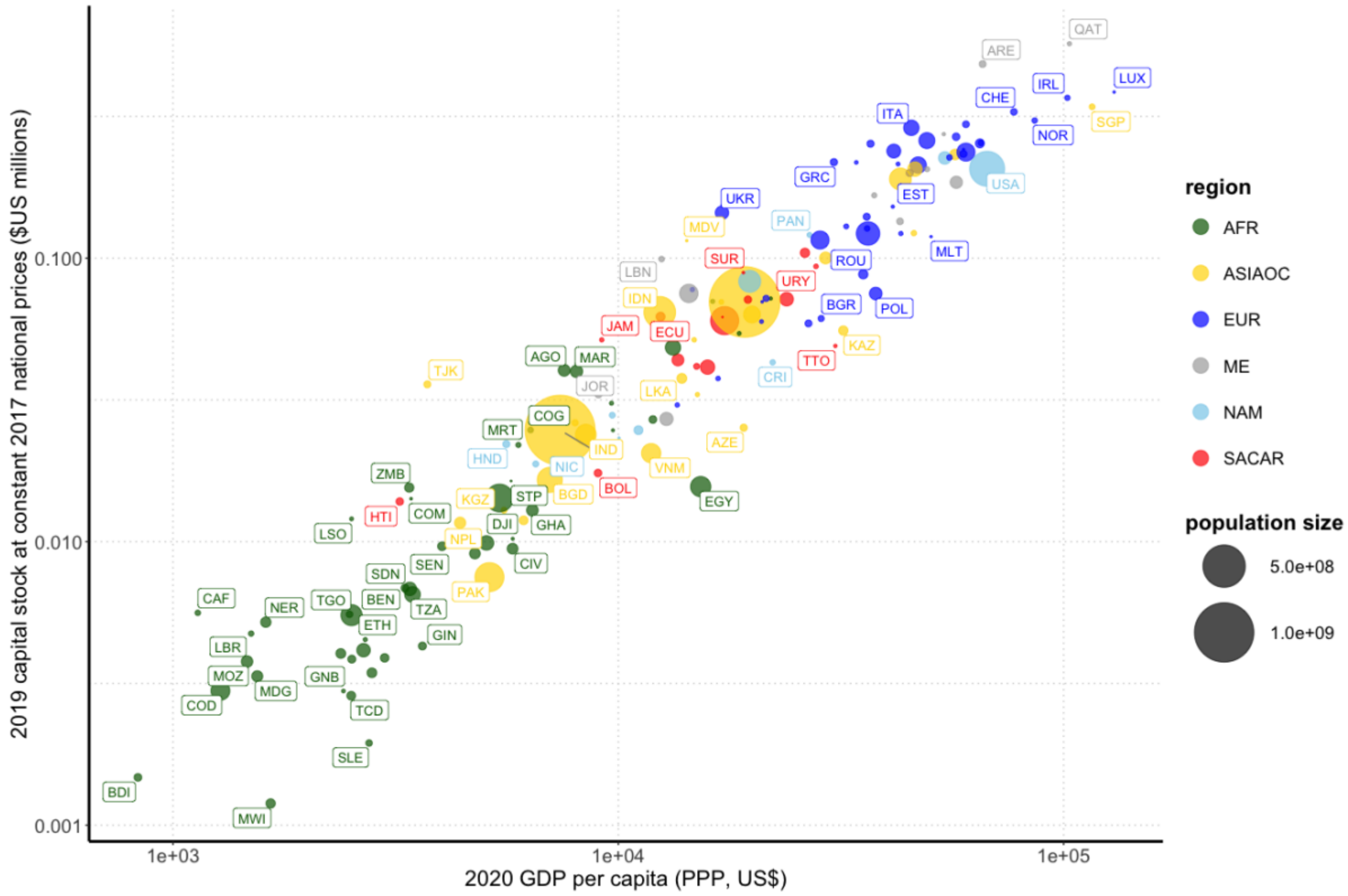

(*v*) **Per-capita output-side real gross domestic product at current purchasing power parities** (Penn World Table version 10.01, Groningen Growth and Development Centre, Faculty of Economics and Business; rug.nl/ggdc/productivity/pwt)[10].

This metric compares relative productive capacity across countries at a single point in time, and it too is strongly correlated with gross domestic product (power-law $R^2$ = 0.983; Fig. S15), so we disregarded it for analysis.

**Fig. S15** | Power-law ($\log_{10}$-$\log_{10}$) relationship between per-capita output-side real gross domestic product at current purchasing power parities (source: Penn World Table version 10.01, Groningen Growth and Development Centre, Faculty of Economics and Business)[10] and purchase power parity (PPP)-adjusted per-capita gross domestic product (source: World Bank). Least-squares regression gives $R^2$ = 0.906. Bubble size indicates relative total population size of each country in 2020. Bubbles are colour-coded by major region (**AFR** = Africa; **ASIOC** = Asia + Oceania; **EUR** = Eurasia; **ME** = Middle East; **NAM** = North America; **SACAR** = South America + Caribbean). Labels for a sample of the points are ISO 3-character country codes.

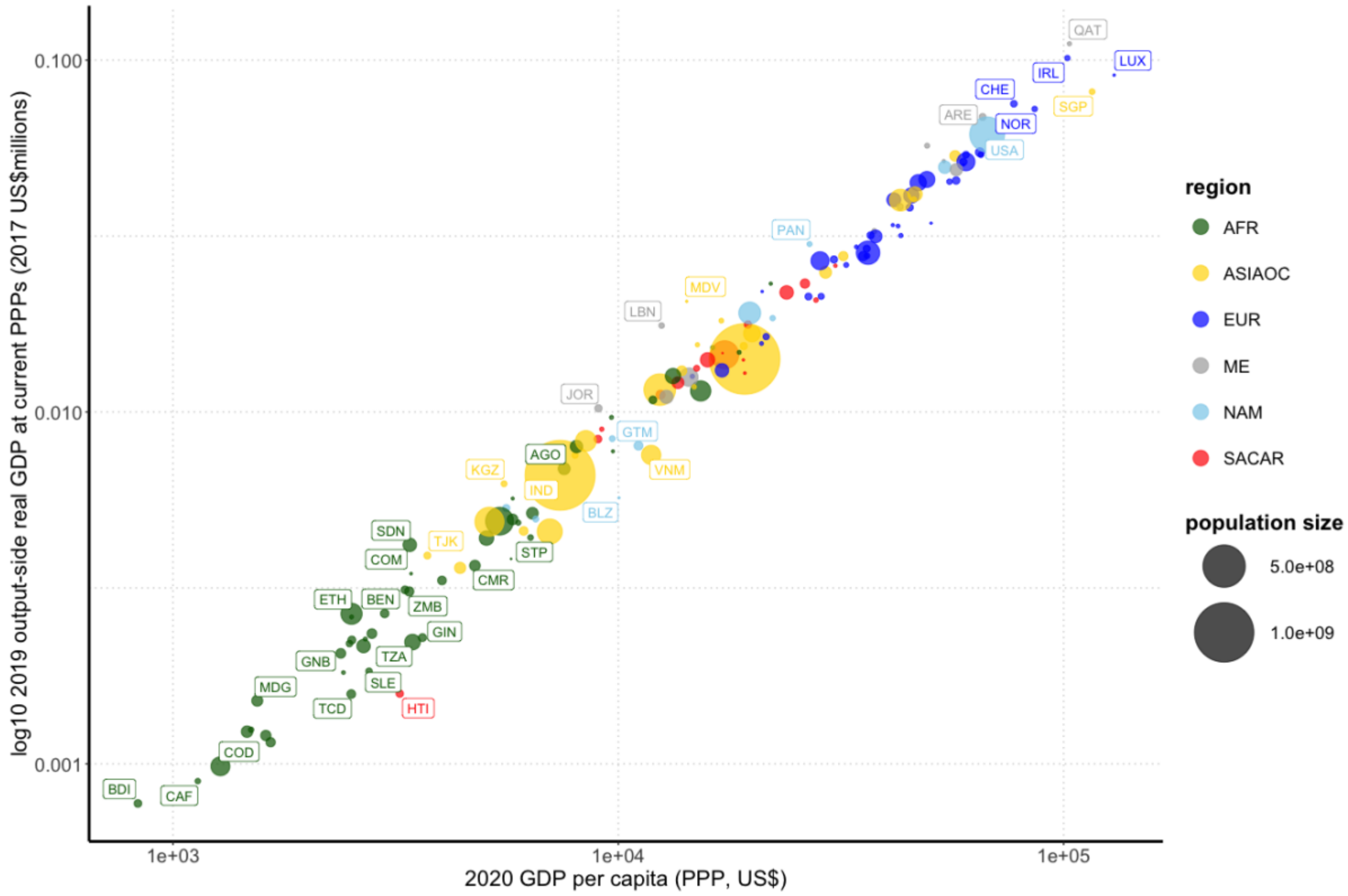

(*vi*) **Total factor productivity**[9] (World Bank).

Human capital-adjusted total factor productivity growth rates are the residual of labour productivity growth calculated by subtracting the contribution of human capital and capital deepening from labour productivity growth. The human capital contribution uses data from the Penn World Table 9.0 (Groningen Growth and Development Centre, Faculty of Economics and Business)[10], augmented to 2018 using the Barro and Lee[11] database and estimates from Cohen and Leker[12] of average years of schooling. These estimates follow the methodology of Penn World Table 9.1 and use the same rates of returns to schooling from Psacharopoulos[13].

**Fig. S16** | Relationship between total factor productivity (log difference, %) (source: World Bank)[9] and purchase power parity (PPP)-adjusted per-capita gross domestic product (source: World Bank). Least-squares regression gives $R^2 = 0.032$. Bubble size indicates relative total population size of each country in 2020. Bubbles are colour-coded by major region (**AFR** = Africa; **ASIOC** = Asia + Oceania; **EUR** = Eurasia; **ME** = Middle East; **NAM** = North America; **SACAR** = South America + Caribbean). Labels for a sample of the points are ISO 3-character country codes.

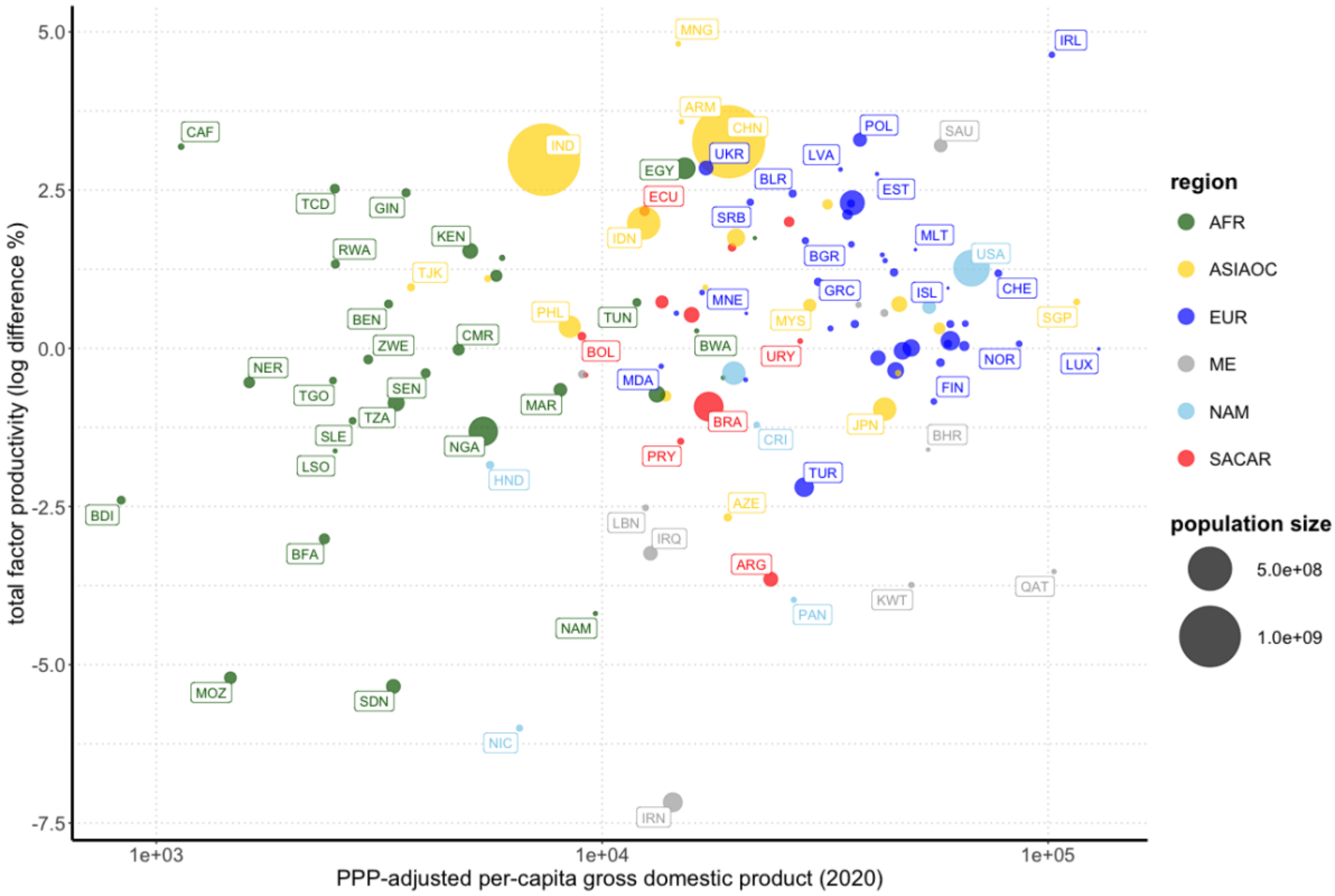

*Boosted regression trees*

**Fig. S17** | Boosted regression tree results testing the relationships between total factor productivity (log difference, %) and three hypothesised predictors: logit dependency ratio (2020), $\log_{10}$ total population size in 2020 ($N_{2020}$), and mean rate of population change ($r_{\mathrm{mean}}$ for both 1950–2021 in black, and 2012–2021 in grey). Panel **a** shows the mean (± 95% confidence interval) relative contribution (%) of each predictor after 10,000 iterations of the model. The model explained 1.4–48.0% and 2.7–44.7% of the variation ($\beta_{\mathrm{CV}}$) in the response for those including $r$ calculated from 1950–2021 and 2012–2021, respectively. Panels **b**–**d**: mean and 95% prediction confidence intervals for each predictor after Gaussian resampling $r_{\mathrm{mean}}$ with its associated standard deviation (from 1950–2021 in black and 2012–2021 in grey) and resampling countries to reduce the effects of spatial autocorrelation.

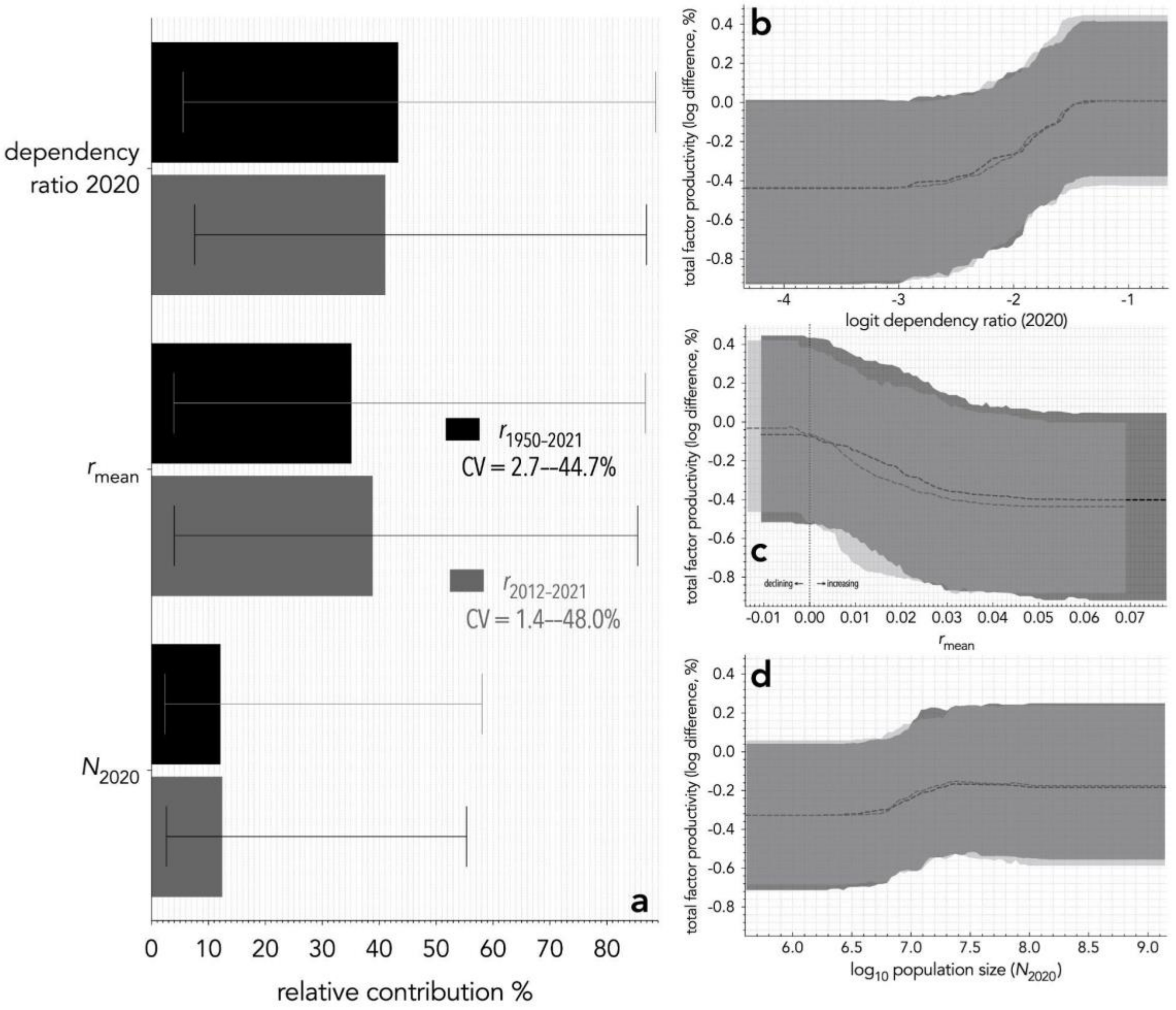

(***vii***) **Per-capita capital services levels at current purchasing power parities** (Penn World Table version 10.01, Groningen Growth and Development Centre, Faculty of Economics and Business; rug.nl/ggdc/productivity/pwt)[10].

This variable estimates capital stock using prices for structures and equipment that are constant across countries, and is also highly correlated with gross domestic product (power-law $R^2 = 0.909$; Fig. S18), so we disregarded it for analysis.

**Fig. S18** | Power-law ($\log_{10}$-$\log_{10}$) relationship between per-capita capital services levels at current purchasing power parities (source: Penn World Table version 10.01, Groningen Growth and Development Centre, Faculty of Economics and Business)[10] and purchase power parity (PPP)-adjusted per-capita gross domestic product (source: World Bank). Least-squares regression gives $R^2 = 0.906$. Bubble size indicates relative total population size of each country in 2020. Bubbles are colour-coded by major region (**AFR** = Africa; **ASIOC** = Asia + Oceania; **EUR** = Eurasia; **ME** = Middle East; **NAM** = North America; **SACAR** = South America + Caribbean). Labels for a sample of the points are ISO 3-character country codes.

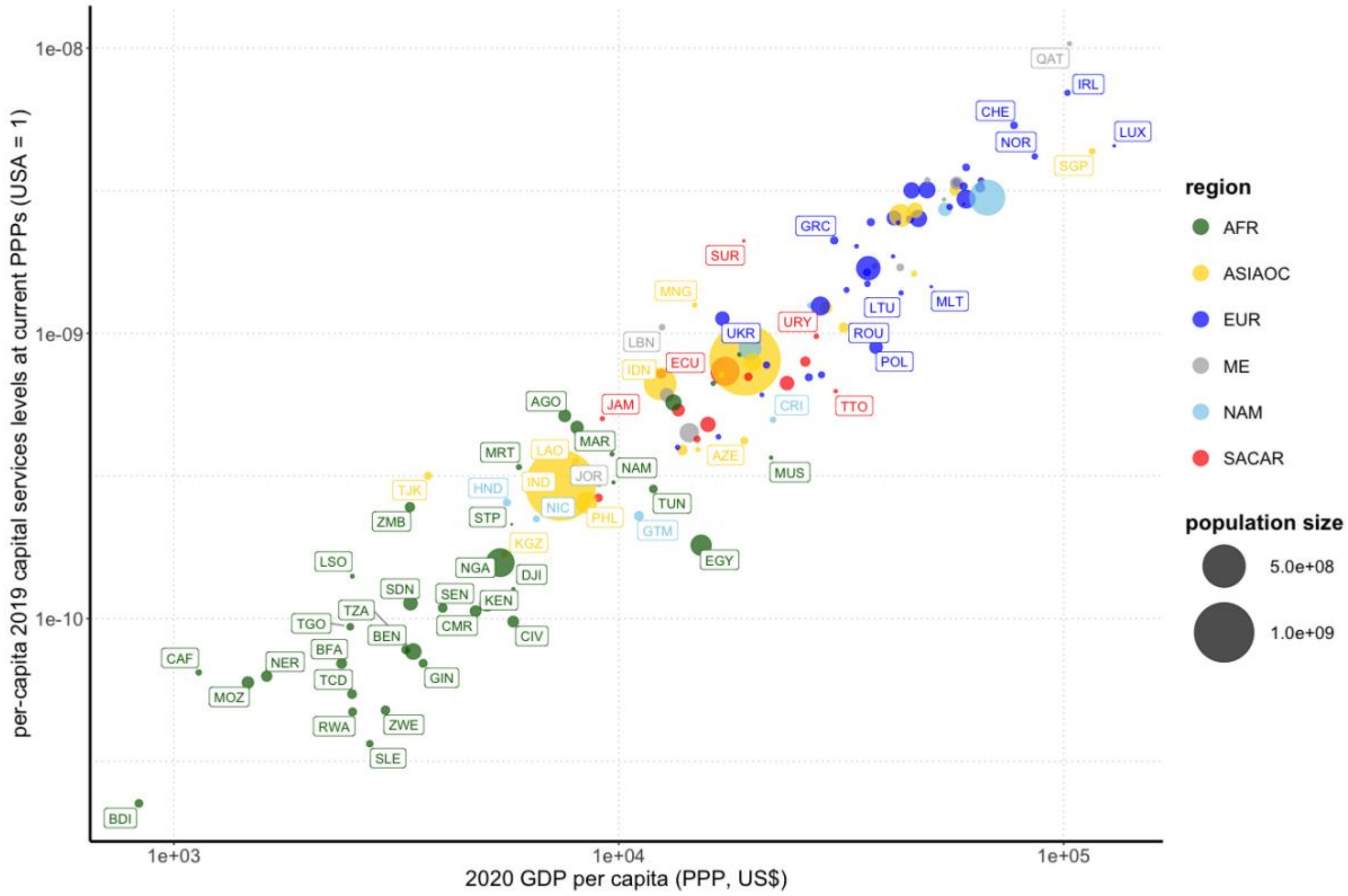

(*viii*) **Gross savings** (% of gross domestic product; World Bank).

This is an amount representing the part of disposable income (adjusted for the change in pension entitlements) that is not spent on final consumption (expressed as a percentage of gross domestic product). Gross savings are calculated as gross national income less total consumption, plus net transfers.

**Fig. S19** | Relationship between gross savings as a percentage of gross domestic product (source: World Bank) and purchase power parity (PPP)-adjusted per-capita gross domestic product (source: World Bank). Least-squares regression gives $R^2$ = 0.083. Bubble size indicates relative total population size of each country in 2020. Bubbles are colour-coded by major region (**AFR** = Africa; **ASIOC** = Asia + Oceania; **EUR** = Eurasia; **ME** = Middle East; **NAM** = North America; **SACAR** = South America + Caribbean). Labels for a sample of the points are ISO 3-character country codes.

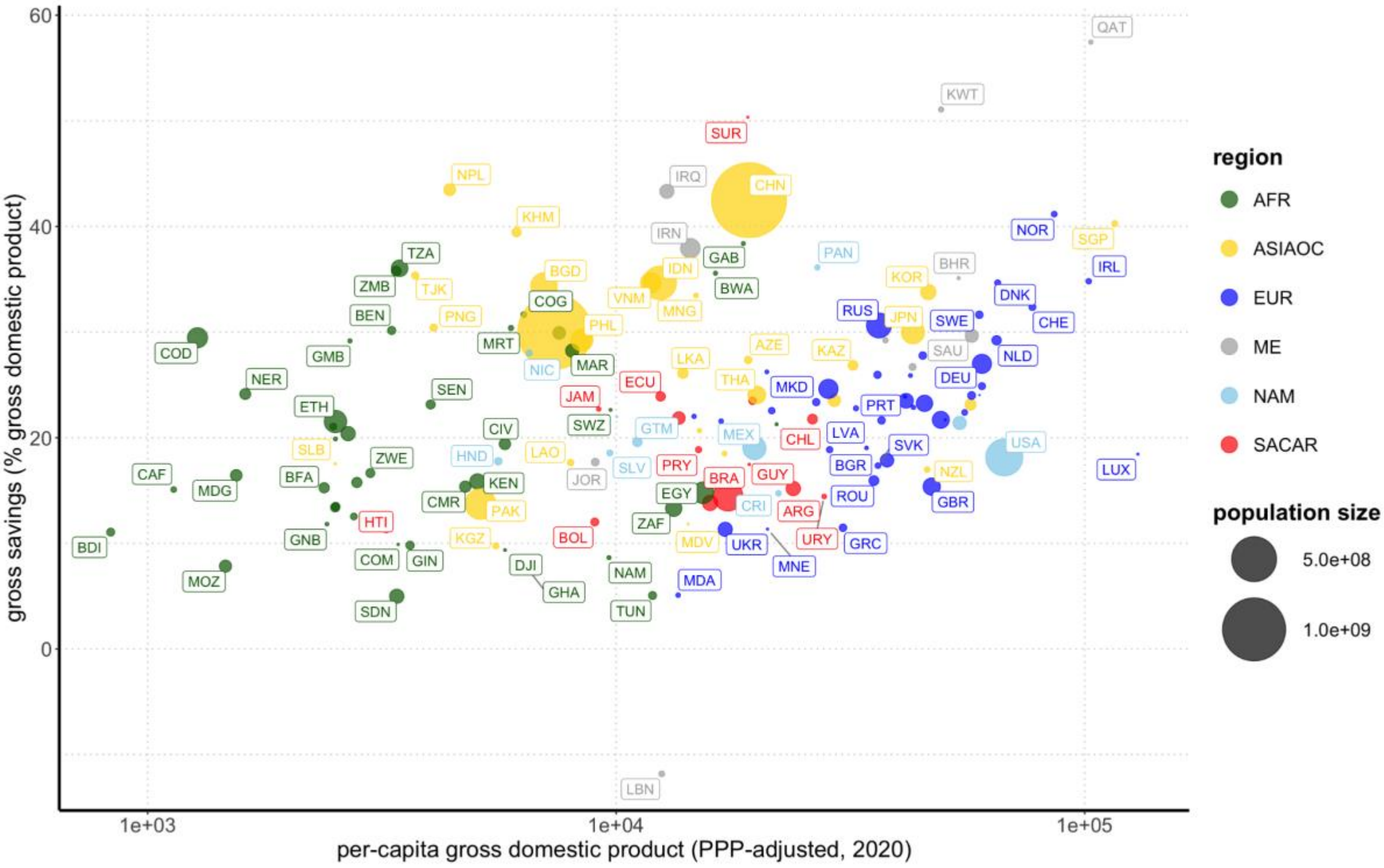

**Fig. S20** | Relationship between gross savings as a percentage of gross domestic product (source: World Bank) and the logit of the dependency ratio. Bubble size indicates relative total population size of each country in 2020. Bubbles are colour-coded by major region (**AFR** = Africa; **ASIOC** = Asia + Oceania; **EUR** = Eurasia; **ME** = Middle East; **NAM** = North America; **SACAR** = South America + Caribbean). Labels for a sample of the points are ISO 3-character country codes.

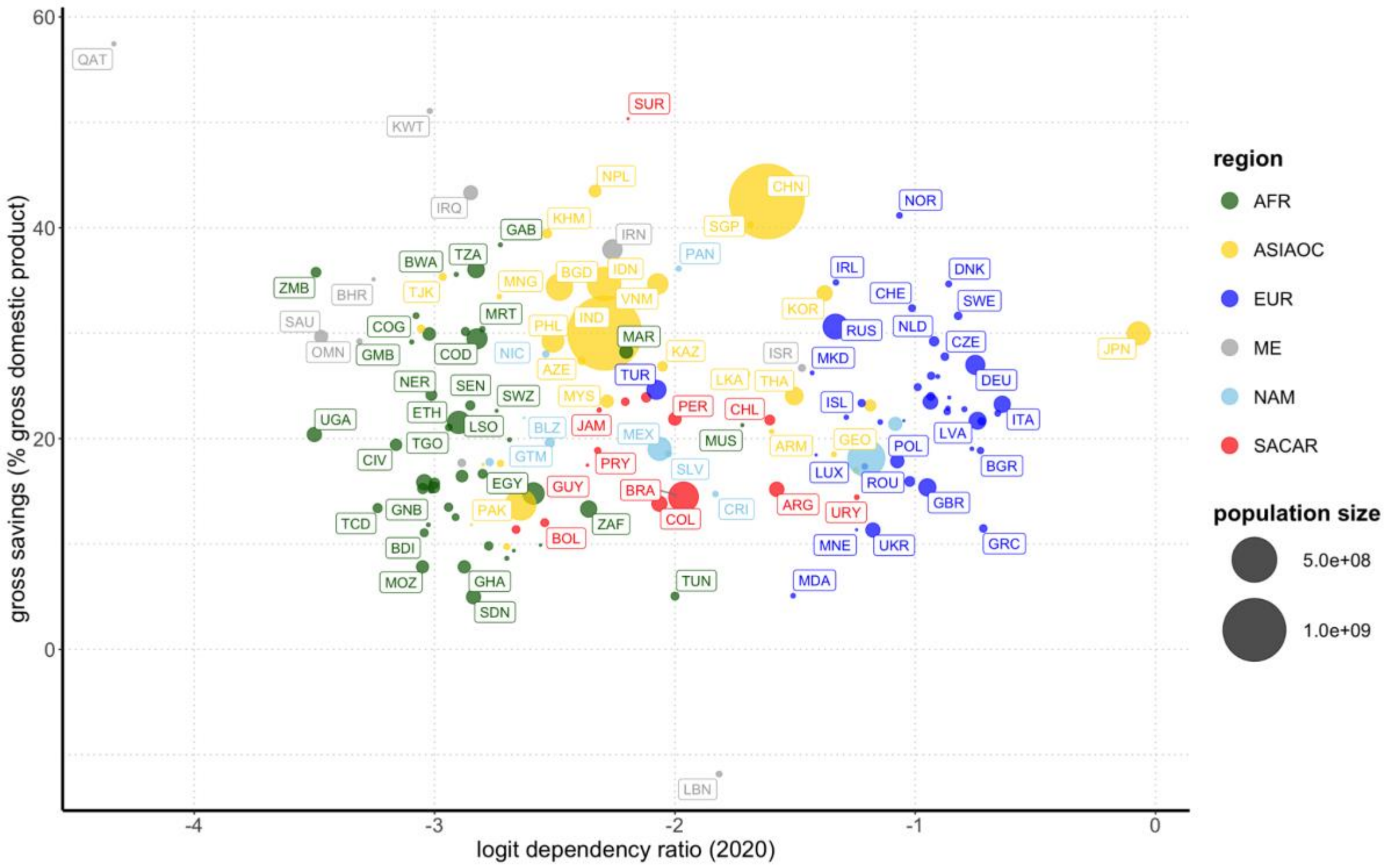

**Fig. S21** | Relationship between gross savings as a percentage of gross domestic product (source: World Bank) and the mean rate of population change from 2012–2021. Bubble size indicates relative total population size of each country in 2020. Bubbles are colour-coded by major region (**AFR** = Africa; **ASIOC** = Asia + Oceania; **EUR** = Eurasia; **ME** = Middle East; **NAM** = North America; **SACAR** = South America + Caribbean). Labels for a sample of the points are ISO 3-character country codes.

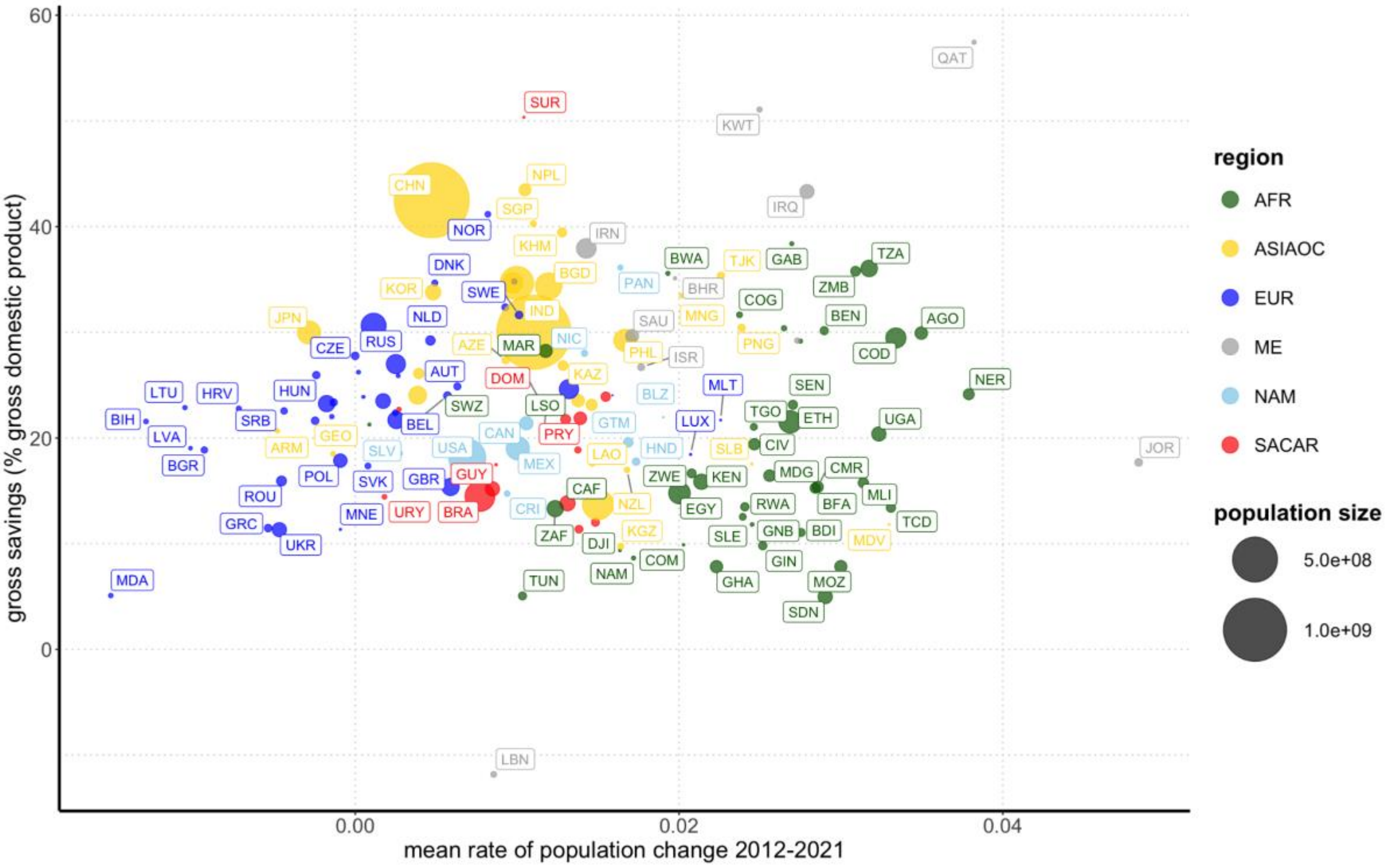

(*ix*) **Gross domestic expenditures on research and development** (% of gross domestic product; World Bank).

**Fig. S22** | Relationship between logit gross domestic expenditures on research and development as a percentage of gross domestic product (source: World Bank) and $\log_{10}$ purchase power parity (PPP)-adjusted per-capita gross domestic product (source: World Bank). Least-squares regression gives $R^2 = 0.379$. Bubble size indicates relative total population size of each country in 2020. Bubbles are colour-coded by major region (**AFR** = Africa; **ASIOC** = Asia + Oceania; **EUR** = Eurasia; **ME** = Middle East; **NAM** = North America; **SACAR** = South America + Caribbean). Labels for a sample of the points are ISO 3-character country codes.

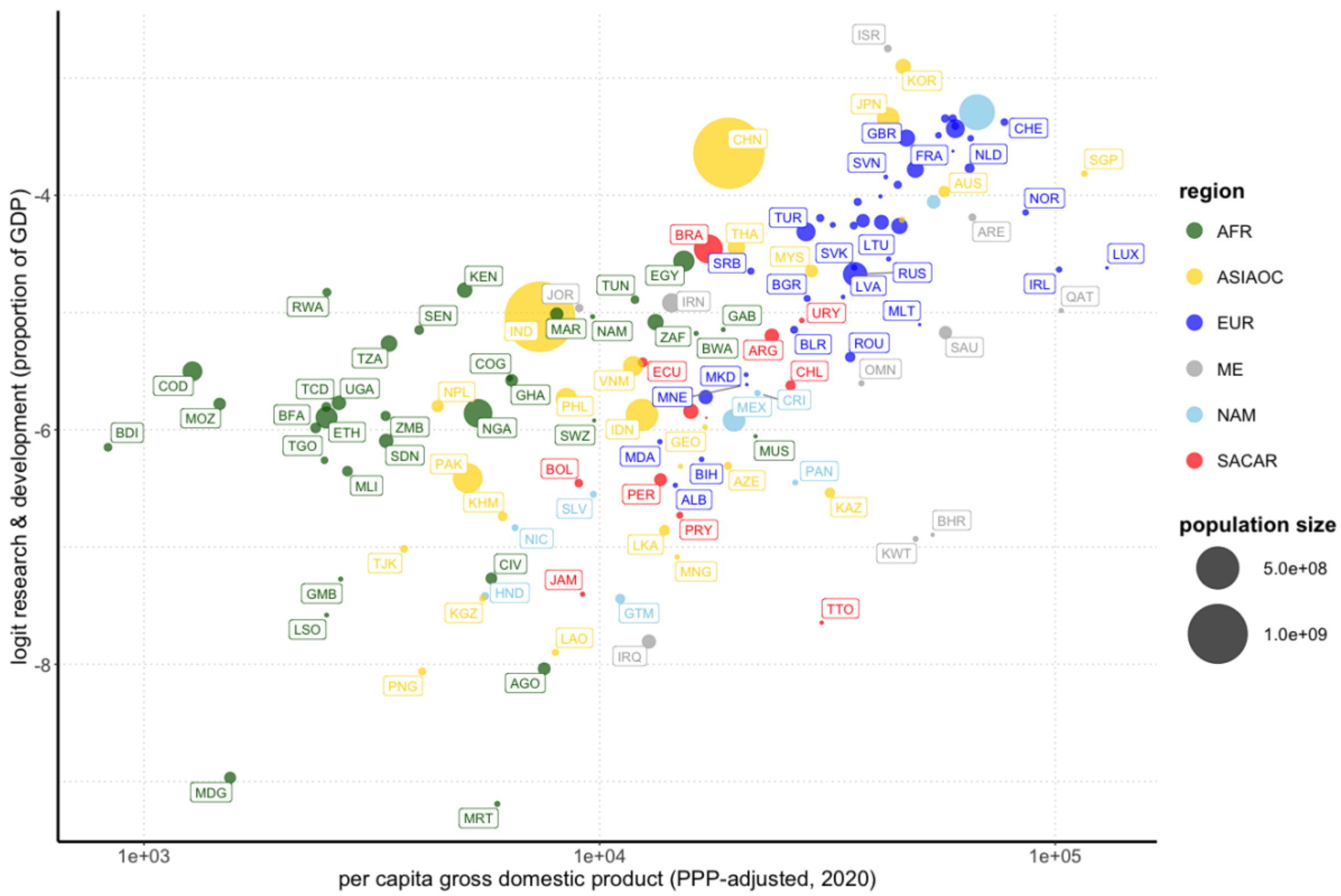

*Boosted regression trees*

**Fig. S23** | Boosted regression tree results testing the relationships between the logit expenditure on research and development as a proportion of gross domestic product and three hypothesised predictors: logit dependency ratio (2020), $\log_{10}$ total population size in 2020 ($N_{2020}$), and mean rate of population change ($r_{mean}$ for both 1950–2021 in black, and 2012–2021 in grey). Panel **a** shows the mean (± 95% confidence interval) relative contribution (%) of each predictor after 10,000 iterations of the model. The model explained 26.2–69.4% and 27.7–69.4% of the variation ($\beta_{CV}$) in the response for those including $r$ calculated from 1950–2021 and 2012–2021, respectively. Panels **b**–**d**: mean and 95% prediction confidence intervals for each predictor after Gaussian resampling $r_{mean}$ with its associated standard deviation (from 1950–2021 in black and 2012–2021 in grey) and resampling countries to reduce the effects of spatial autocorrelation.

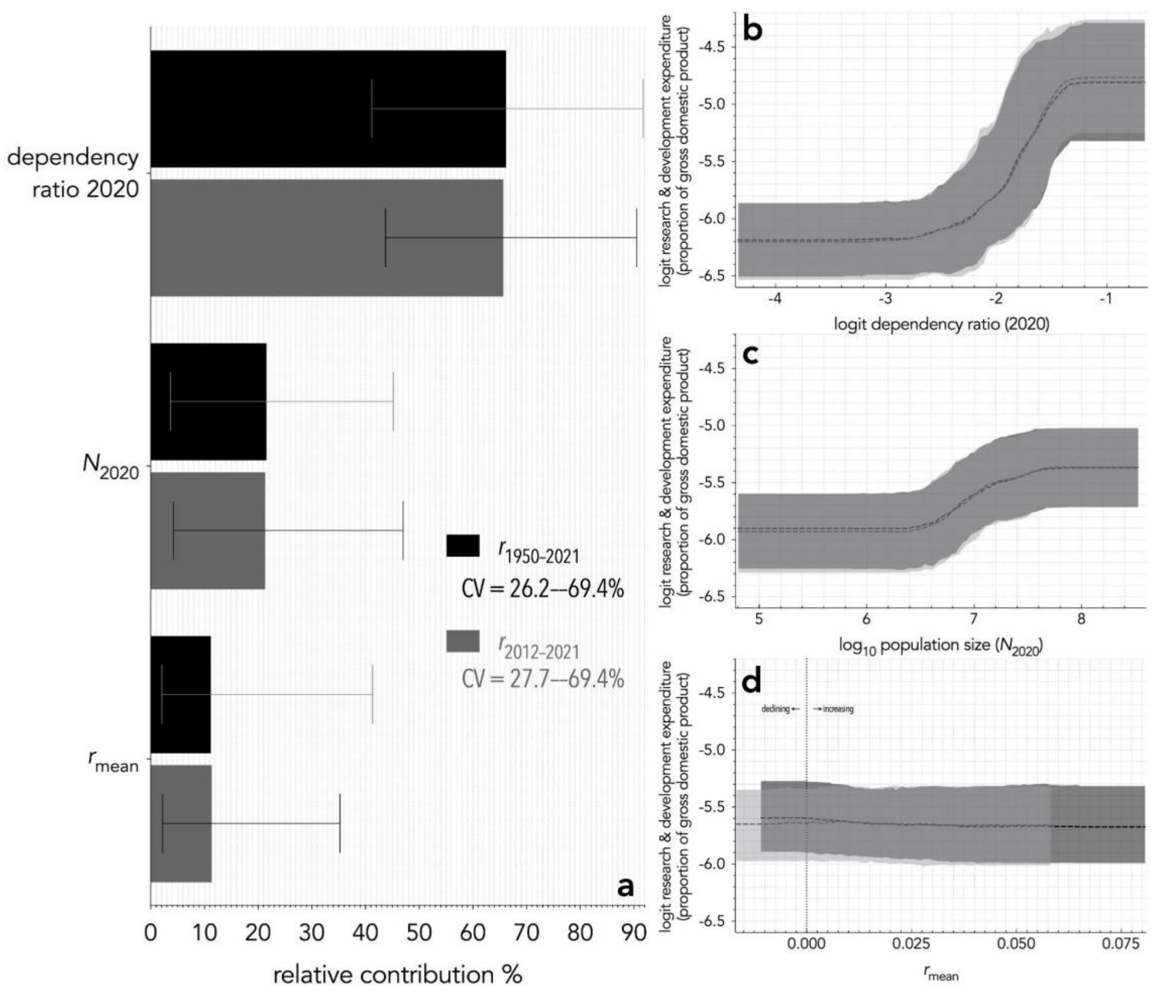

($x$) **Per-capita patent applications** (World Bank).

Worldwide patent applications filed through the Patent Cooperation Treaty procedure or with a national patent office for exclusive rights for an invention — a product or process that provides a new way of doing something or offers a new technical solution to a problem. Given its moderate correlation with gross domestic product (power-law $R^2 = 0.581$; Fig. S24), we included this variable in the analysis.

**Fig. S24** | Power-law ($\log_{10}$-$\log_{10}$) relationship between per-capita patent applications (source: World Bank) and purchase power parity (PPP)-adjusted per-capita gross domestic product (source: World Bank). Least-squares regression gives $R^2$ = 0.581. Bubble size indicates relative total population size of each country in 2020. Bubbles are colour-coded by major region (**AFR** = Africa; **ASIOC** = Asia + Oceania; **EUR** = Eurasia; **ME** = Middle East; **NAM** = North America; **SACAR** = South America + Caribbean). Labels for a sample of the points are ISO 3-character country codes.

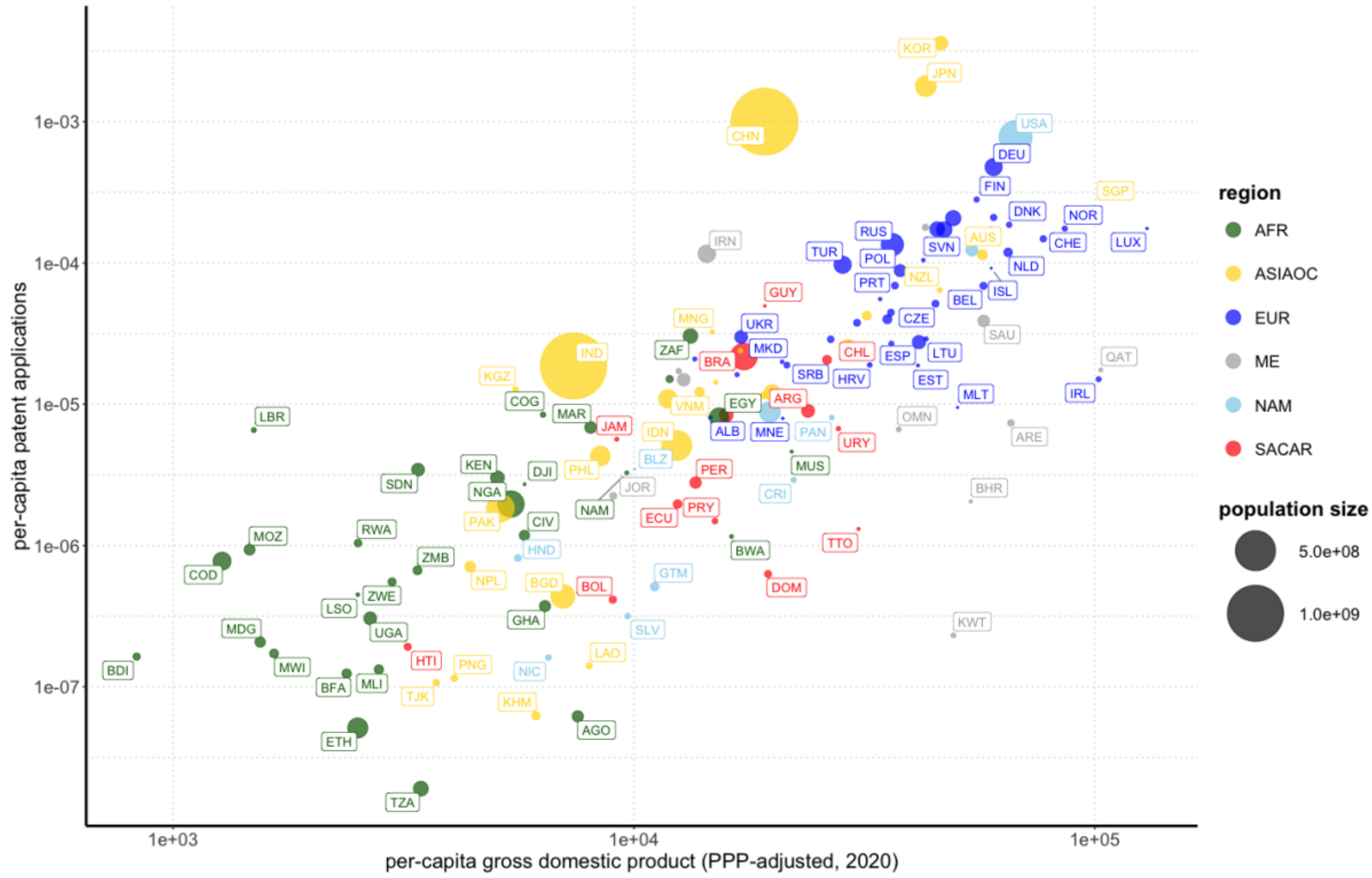

*Boosted regression trees*

**Fig. S25**. Boosted regression tree results testing the relationships between the $\log_{10}$ per-capita patent applications and three hypothesised predictors: logit dependency ratio (2020), $\log_{10}$ total population size in 2020 ($N_{2020}$), and mean rate of population change ($r_{\mathrm{mean}}$ for both 1950–2021 in black, and 2012–2021 in grey). Panel **a** shows the mean (± 95% confidence interval) relative contribution (%) of each predictor after 10,000 iterations of the model. The model explained 48.3–78.6% and 49.3–78.3% of the variation ($\beta_{\mathrm{CV}}$) in the response for those including $r$ calculated from 1950–2021 and 2012–2021, respectively. Panels **b**–**d**: mean and 95% prediction confidence intervals for each predictor after Gaussian resampling $r_{\mathrm{mean}}$ with its associated standard deviation (from 1950–2021 in black and 2012–2021 in grey) and resampling countries to reduce the effects of spatial autocorrelation.

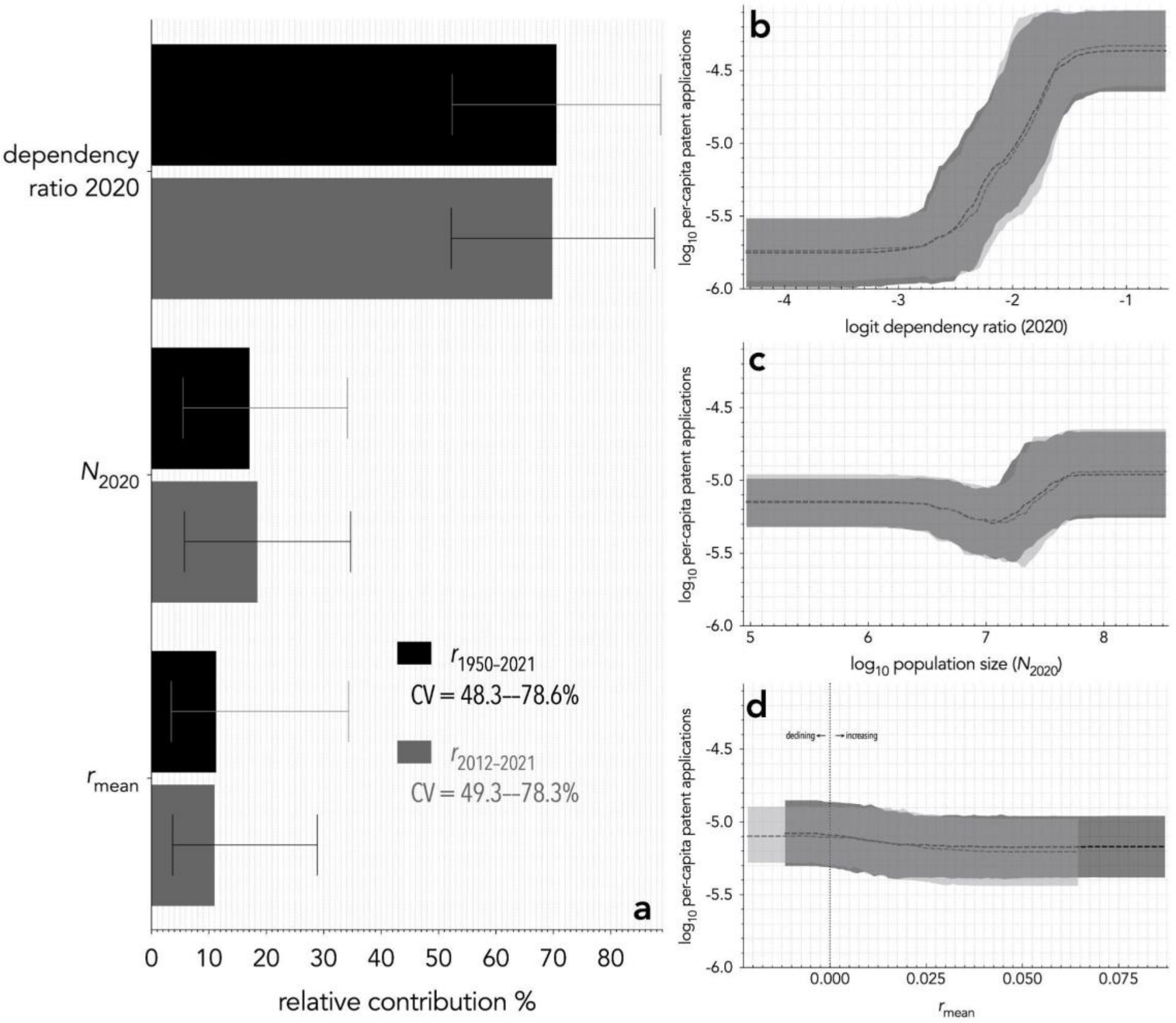

(*xi*) **Human capital index** (Penn World Table version 10.01, Groningen Growth and Development Centre, Faculty of Economics and Business; rug.nl/ggdc/productivity/pwt)[10].

This variable is based on mean years of schooling and returns to education; given its moderate correlation with gross domestic product (log-linear $R^2 = 0.685$; Fig. S26), we included this variable in the analysis.

**Fig. S26** | Relationship between the human capital index (source: Penn World Table version 10.01, Groningen Growth and Development Centre, Faculty of Economics and Business)[10] and purchase power parity (PPP)-adjusted per-capita gross domestic product (source: World Bank). Least-squares regression gives $R^2 = 0.906$. Bubble size indicates relative total population size of each country in 2020. Bubbles are colour-coded by major region (**AFR** = Africa; **ASIOC** = Asia + Oceania; **EUR** = Eurasia; **ME** = Middle East; **NAM** = North America; **SACAR** = South America + Caribbean). Labels for a sample of the points are ISO 3-character country codes.

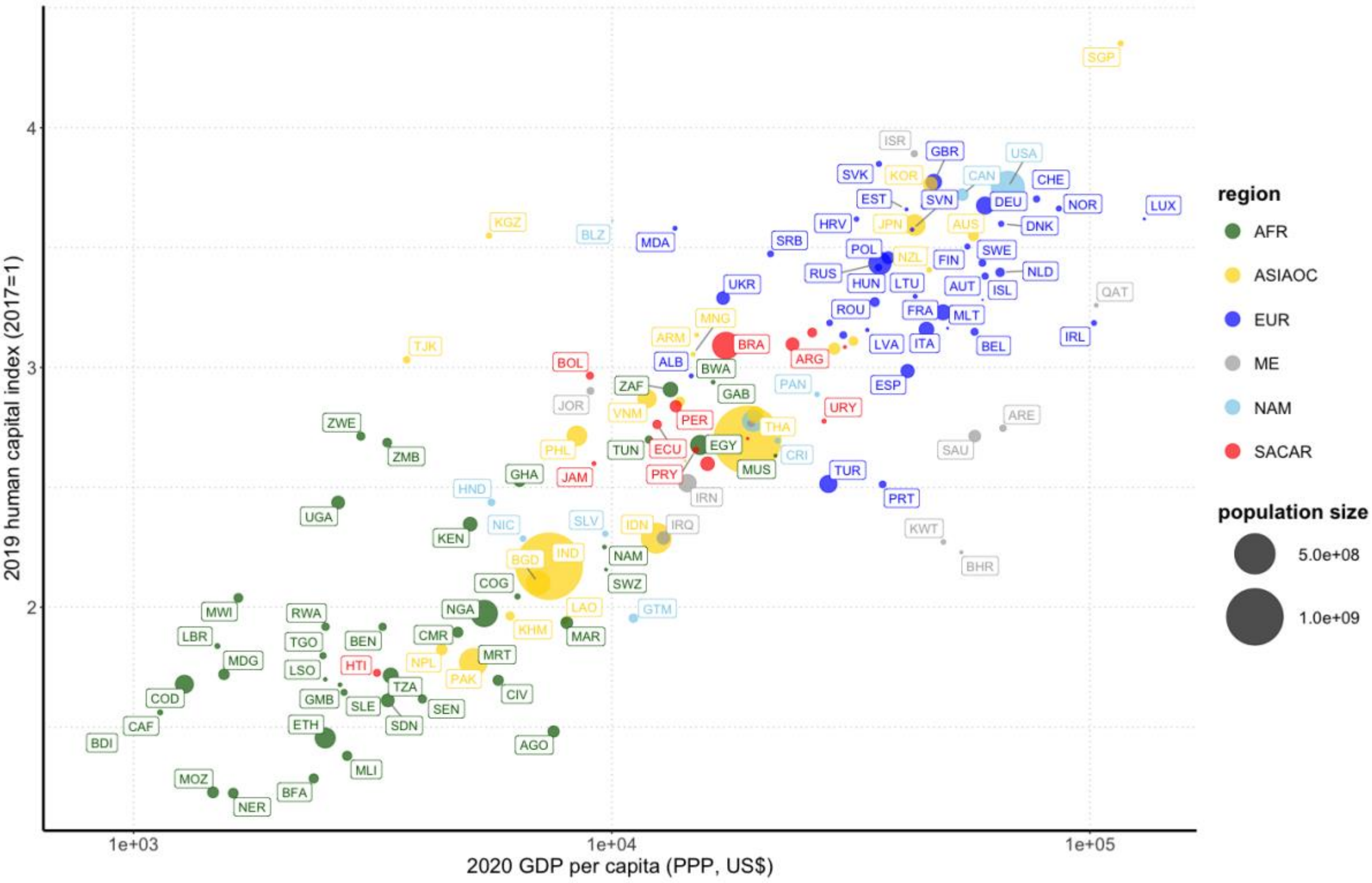

## *Boosted regression trees*

**Fig. S27** | Boosted regression tree results testing the relationships between the 2019 human capital index and three hypothesised predictors: logit dependency ratio (2020), $\log_{10}$ total population size in 2020 ($N_{2020}$), and mean rate of population change ($r_{mean}$ for both 1950–2021 in black, and 2012–2021 in grey). Panel **a** shows the mean (± 95% confidence interval) relative contribution (%) of each predictor after 10,000 iterations of the model. The model explained 53.6–79.3% and 54.1–79.7% of the variation ($\beta_{CV}$) in the response for those including *r* calculated from 1950–2021 and 2012–2021, respectively. Panels **b**–**d**: mean and 95% prediction confidence intervals for each predictor after Gaussian resampling $r_{mean}$ with its associated standard deviation (from 1950–2021 in black and 2012–2021 in grey) and resampling countries to reduce the effects of spatial autocorrelation.

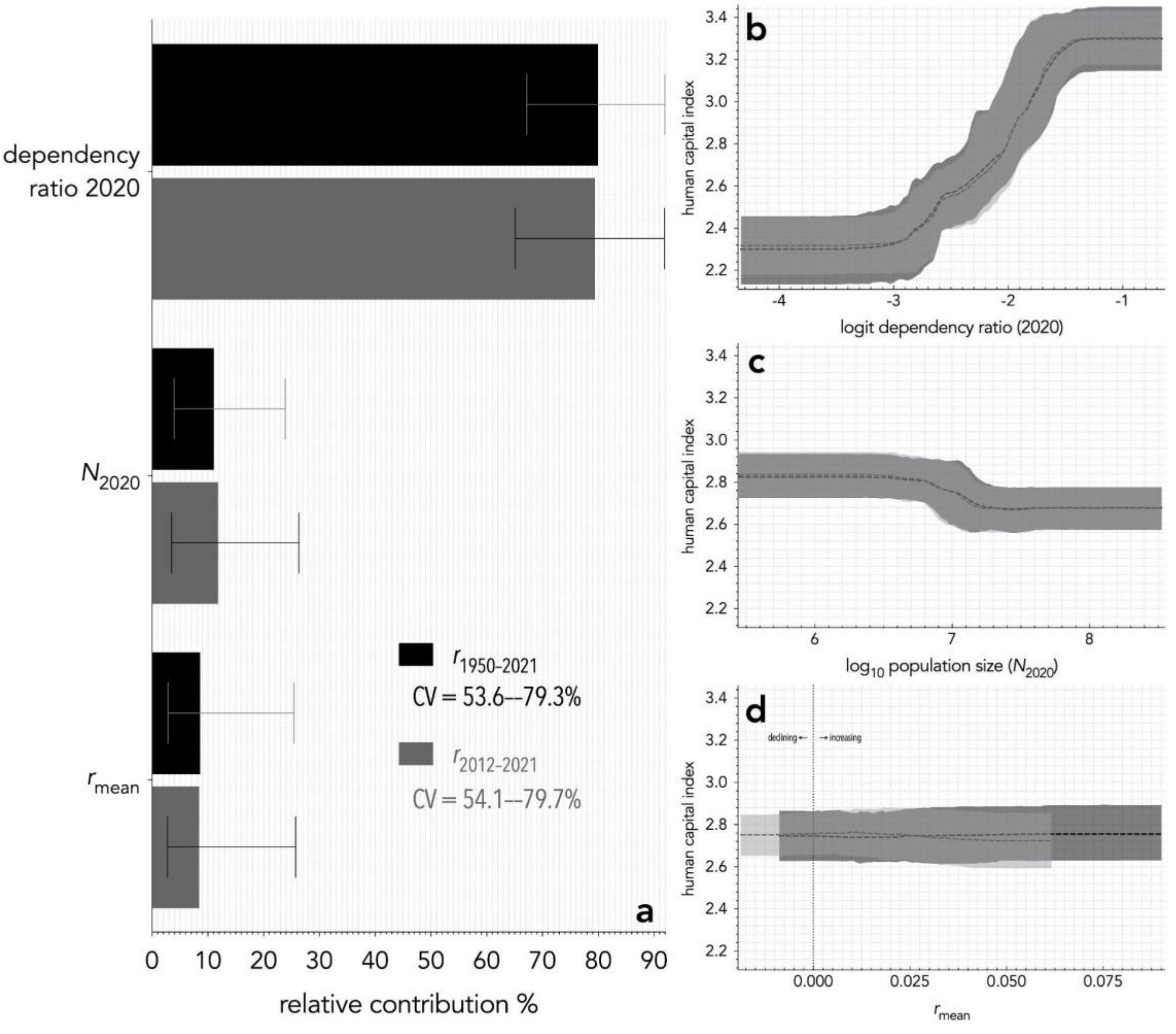

## Appendix V. Corruption and freedom

### *Corruption perception index*

Transparency International's (transparency.org) corruption perception index ranks countries and territories based on the perceived degree of public sector corruption, as judged by experts and business leaders. The index combines data from up to 13 independent sources, including expert assessments and business surveys from trusted institutions. Each source is converted to a scale from 0 to 100, where 0 means very corrupt and 100 means very clean, with a country's score being the average of the available sources for that year.

**Fig. S28** | Relationship between the logit corruption perception index (source: Transparency International) and purchase power parity (PPP)-adjusted per-capita gross domestic product (source: World Bank). Least-squares regression gives $R^2$ = 0.527. Bubble size indicates relative total population size of each country in 2020. Bubbles are colour-coded by major region (**AFR** = Africa; **ASIOC** = Asia + Oceania; **EUR** = Eurasia; **ME** = Middle East; **NAM** = North America; **SACAR** = South America + Caribbean). Labels for a sample of the points are ISO 3-character country codes.

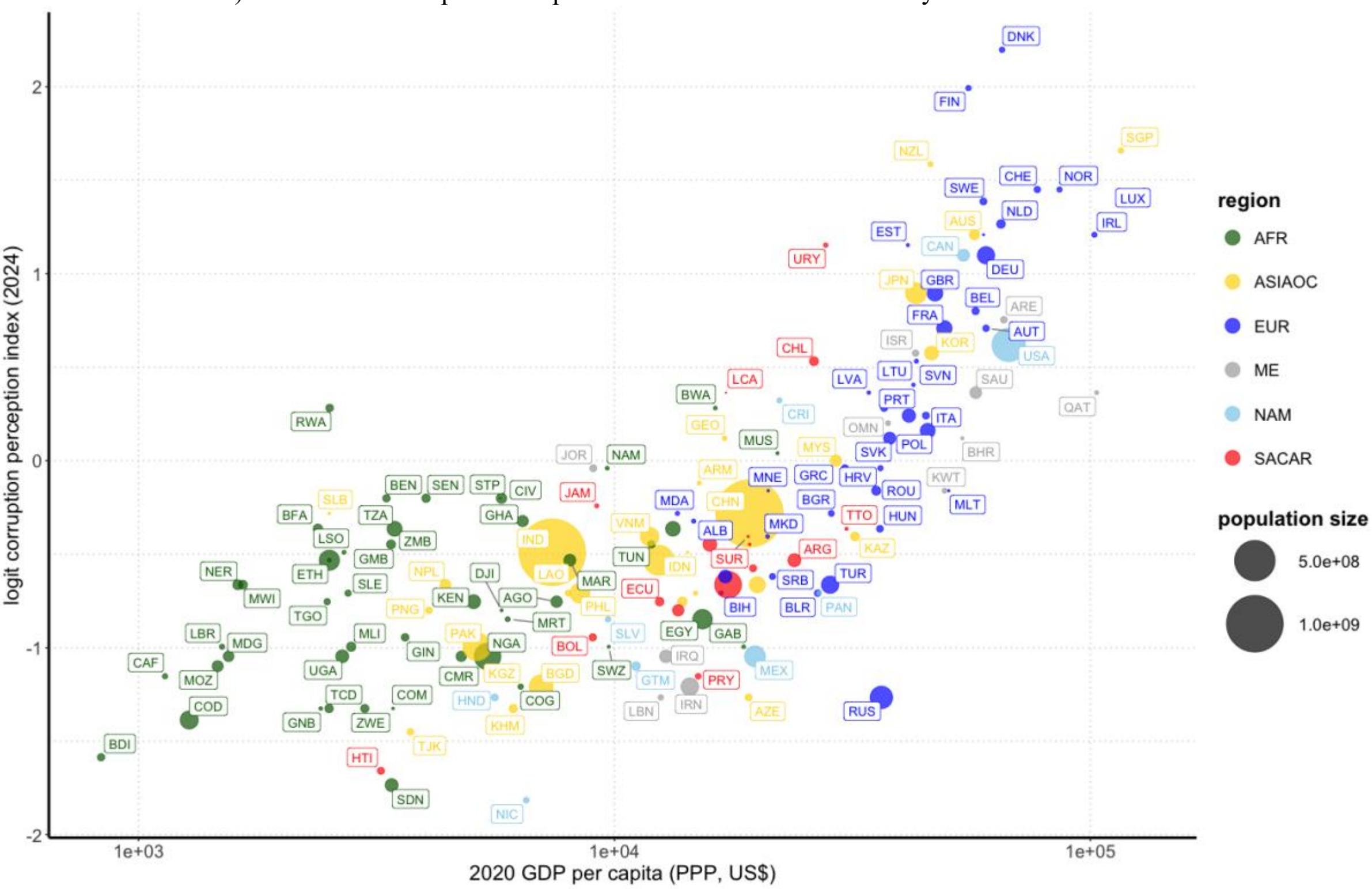

### *Freedom*

This is the aggregate (total) score of all categories assessed in the Freedom House *Freedom in the World* report (freedomhouse.org) measuring dimensions of a country's electoral process, political pluralism and participation, functioning of government, freedom of expression and of belief, associational and organisational rights, rule of law, and personal autonomy and individual rights.

**Fig. S29** | Relationship between the logit freedom score (Source: Freedom House) and purchase power parity (PPP)-adjusted per-capita gross domestic product (source: World Bank). Least-squares regression gives $R^2 = 0.301$. Bubble size indicates relative total population size of each country in 2020. Bubbles are colour-coded by major region (**AFR** = Africa; **ASIOC** = Asia + Oceania; **EUR** = Eurasia; **ME** = Middle East; **NAM** = North America; **SACAR** = South America + Caribbean). Labels for a sample of the points are ISO 3-character country codes.

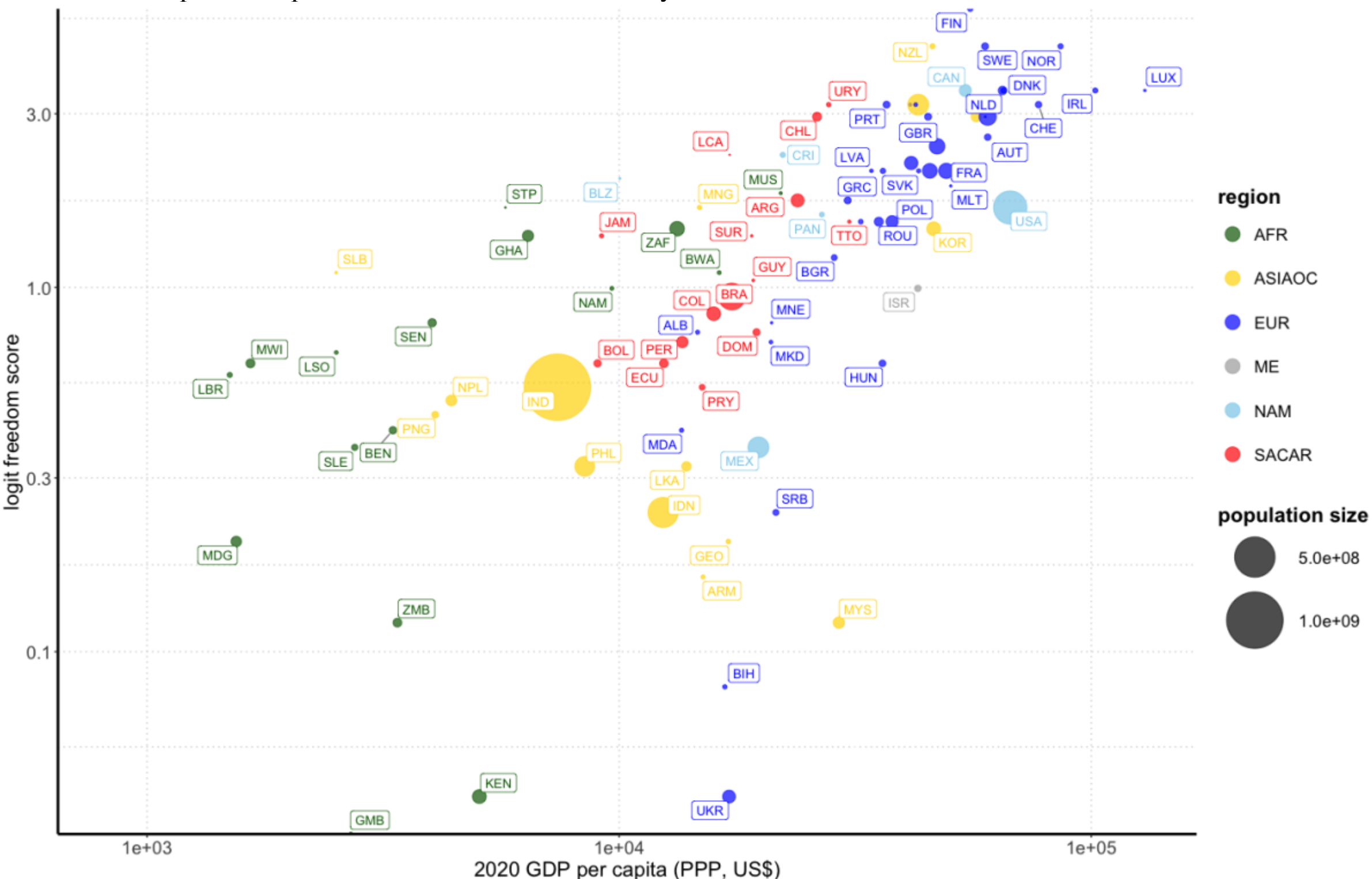

**Fig. S30** | Relationship between the logit 2025 freedom score (Source: Freedom House) and the logit corruption perception index (source: Transparency International). Least-squares regression gives $R^2 = 0.556$. Bubble size indicates relative total population size of each country in 2020. Bubbles are colour-coded by major region (**AFR** = Africa; **ASIOC** = Asia + Oceania; **EUR** = Eurasia; **ME** = Middle East; **NAM** = North America; **SACAR** = South America + Caribbean). Labels for a sample of the points are ISO 3-character country codes.

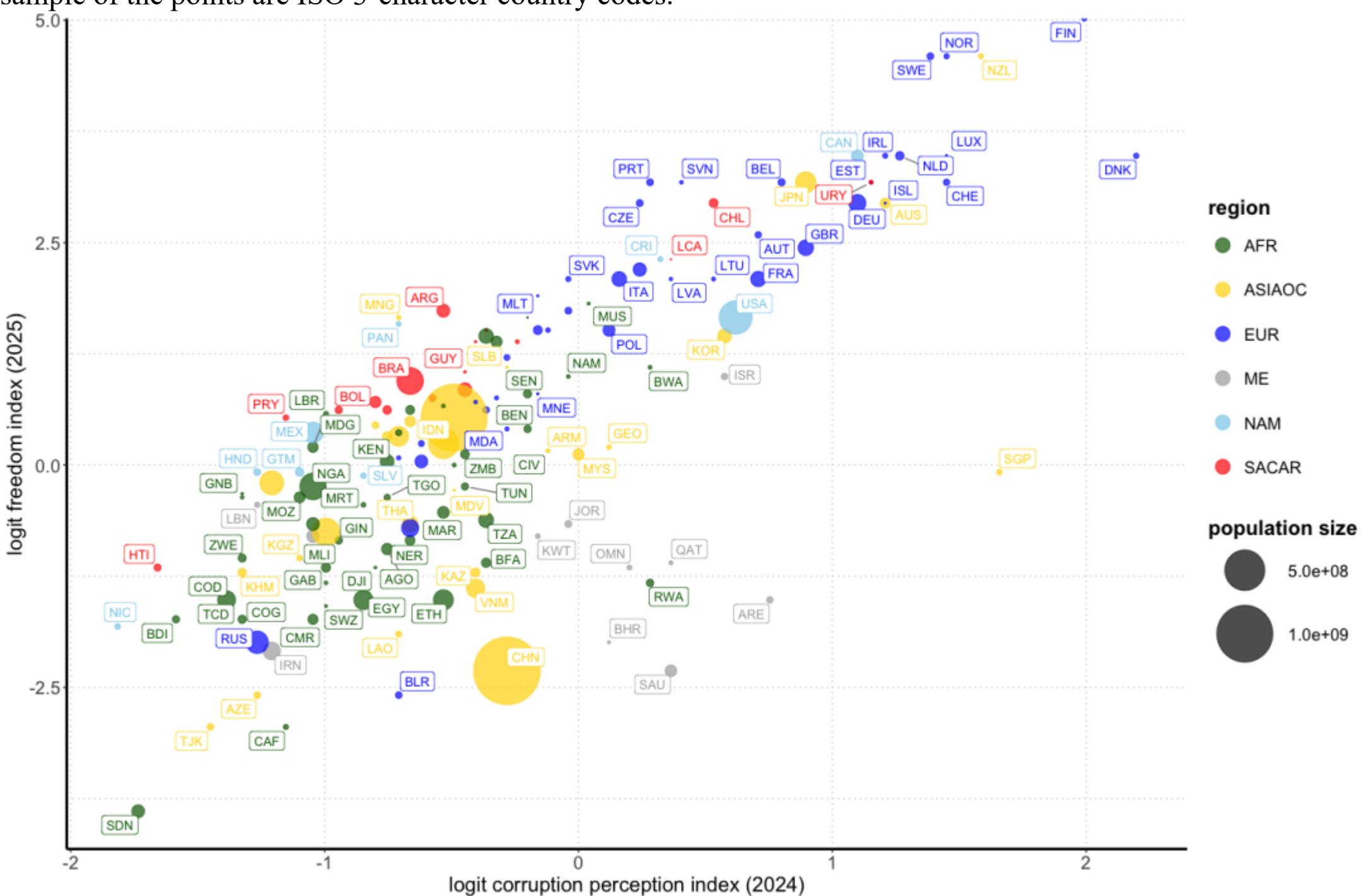

***Global parliament index* (2025)**

The global parliament index (freedomhouse.org) analyses the political leanings of each government and places countries into one of ten categories based upon multiple factors; including the governing party's policies, manifestos, and record in power (-3: populist/authoritarian left; -2: left wing; -1: centre-left; 0: centrist; 1: centrist-authoritarian; 2: centre-right; 3: right wing; 4: populist/authoritarian right; 5: dictatorship; 6: powerful monarchic system). When compared to the freedom score (Fig. S31), more left-wing governments today tend to have a higher freedom score (slope 95% confidence interval: -0.891 to -0.522; $R^2 = 0.273$).

**Fig. S31** | Relationship between the logit 2025 freedom score (Source: Freedom House) and the 2025 global parliament index (political spectrum) (source: Arden Strategies). Least-squares regression gives $R^2 = 0.273$. Bubble size indicates relative total population size of each country in 2020. Bubbles are colour-coded by major region (**AFR** = Africa; **ASIOC** = Asia + Oceania; **EUR** = Eurasia; **ME** = Middle East; **NAM** = North America; **SACAR** = South America + Caribbean). Labels for a sample of the points are ISO 3-character country codes.

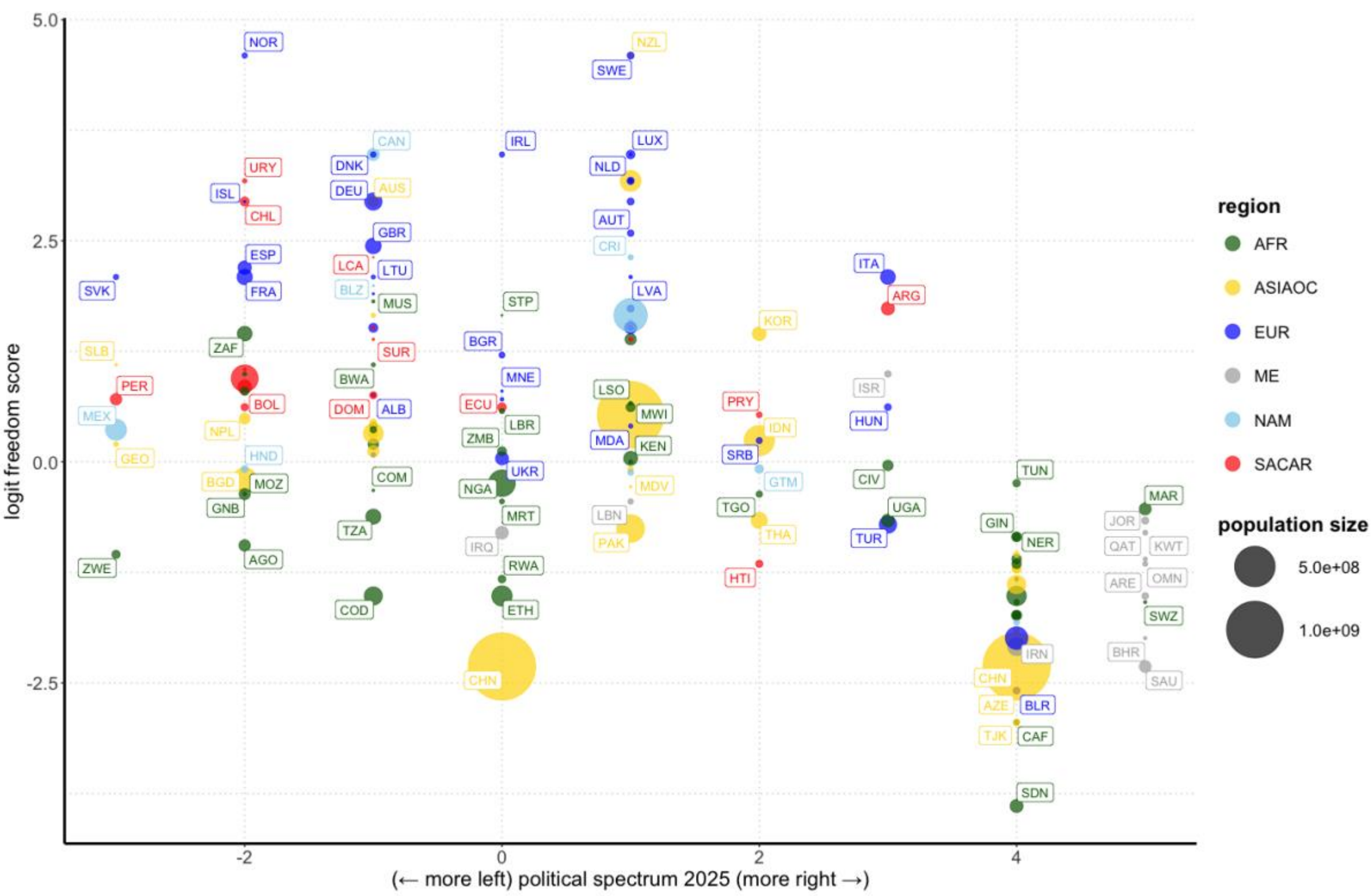

## Appendix VI. Human Development

*Human development index*

The Human Development Index is a composite measure of three dimensions of human development: health (measured as life expectancy), education (measured as mean years of schooling), and standard of living (measured as per-capita gross national income)[14]. The composite index itself is a geometric mean of the normalised indices across the three dimensions. The index captures only part of the multidimensional definition of 'human development', because it does not reflect inequalities, poverty, security, empowerment, or environmental integrity[14].

**Fig. S32** | Relationship between the logit of the Human Development Index (2023) (source: United Nations Environment Programme) and national purchase power parity (PPP)-adjusted per-capita gross domestic product (source: World Bank) and. Least-squares regression gives $R^2 = 0.883$. Bubble size indicates relative total population size of each country in 2020. Bubbles are colour-coded by major region (**AFR** = Africa; **ASIOC** = Asia + Oceania; **EUR** = Eurasia; **ME** = Middle East; **NAM** = North America; **SACAR** = South America + Caribbean). Labels for a sample of the points are ISO 3-character country codes.

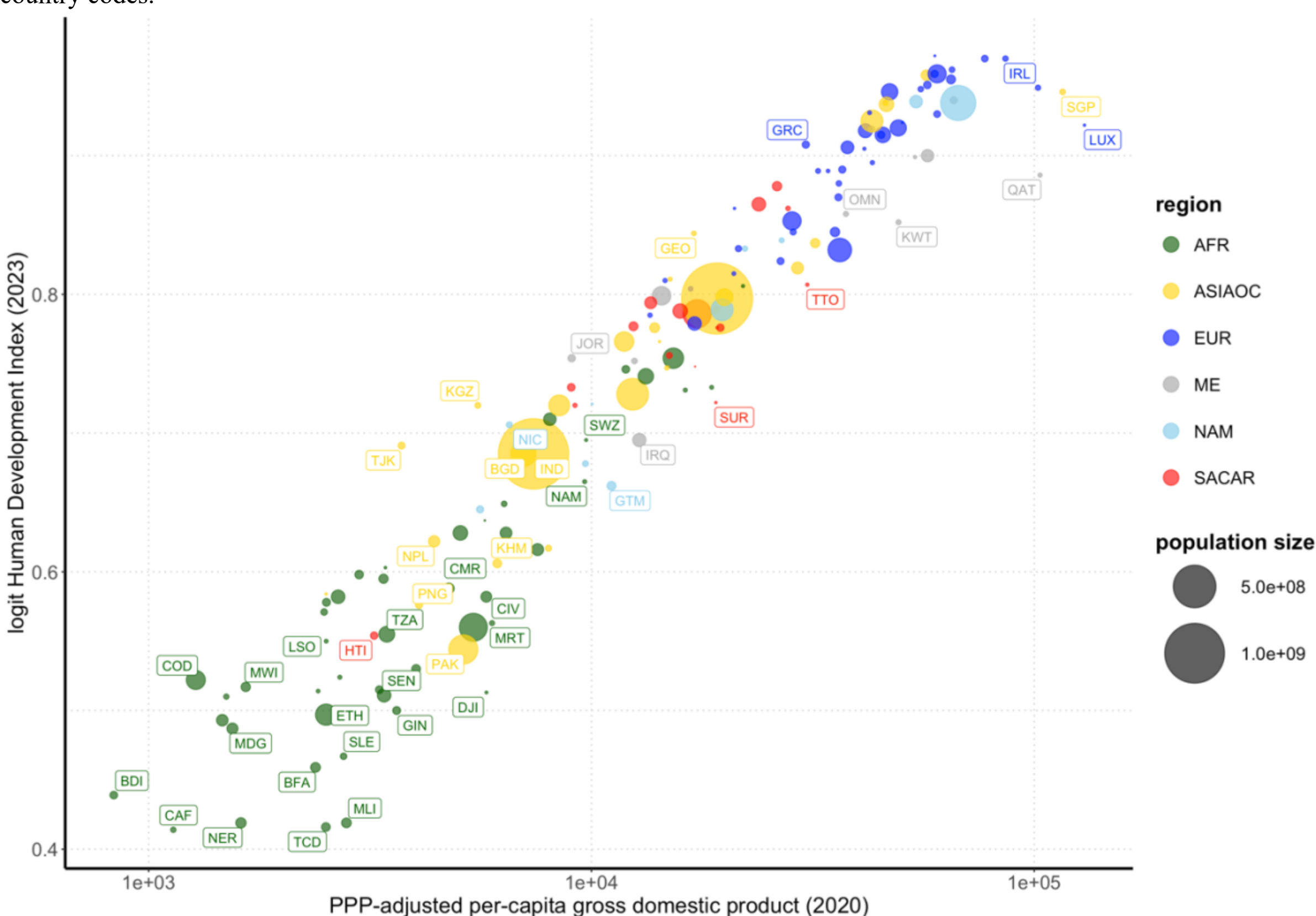

**Fig. S33** | Relationship between the logit of the Human Development Index (2023) (source: United Nations Environment Programme) and logit dependency ratio (2020) (source: United Nations Population Division). Bubble size indicates relative total population size of each country in 2020. Bubbles are colour-coded by major region (**AFR** = Africa; **ASIOC** = Asia + Oceania; **EUR** = Eurasia; **ME** = Middle East; **NAM** = North America; **SACAR** = South America + Caribbean). Labels for a sample of the points are ISO 3-character country codes.

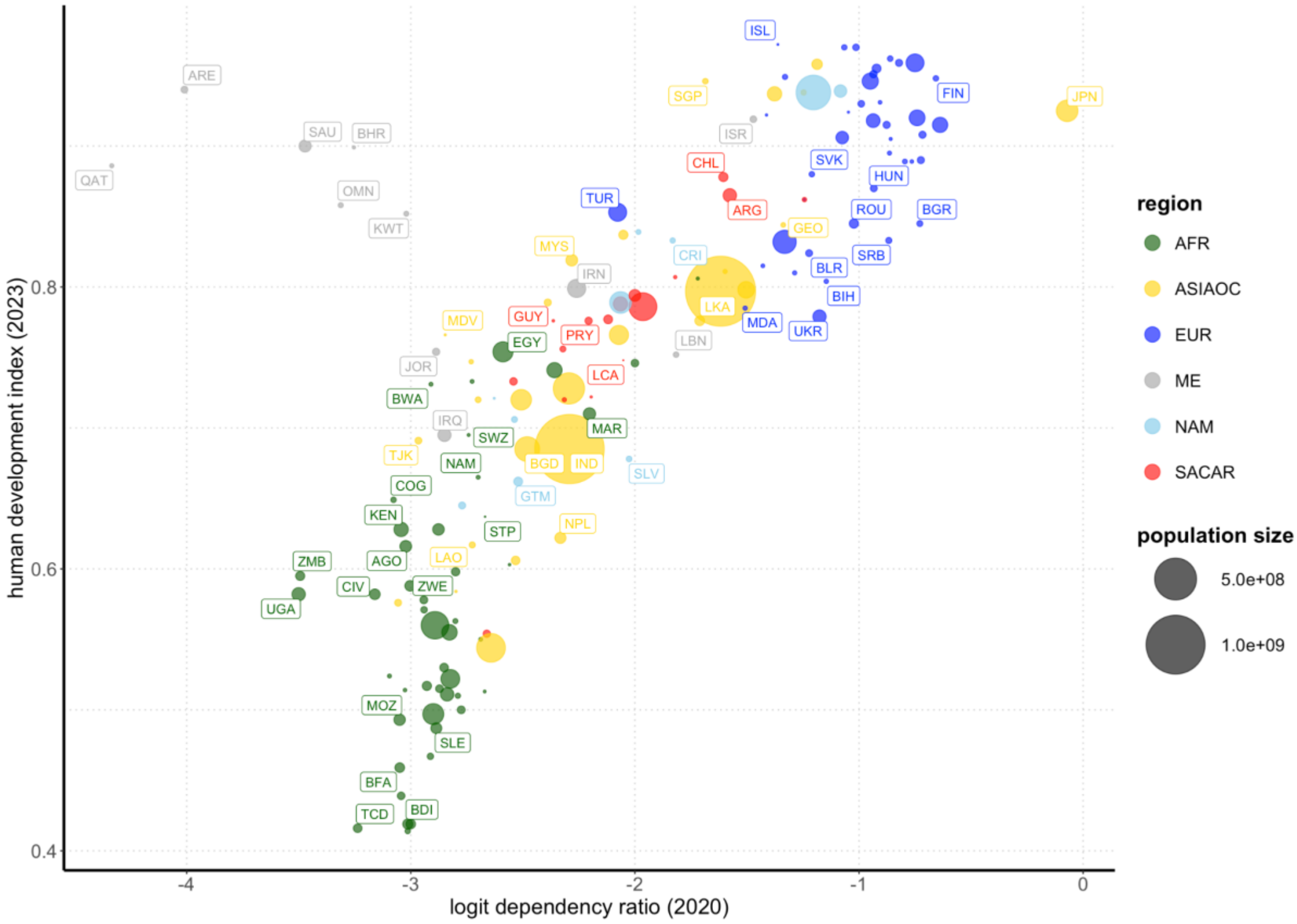

## *Boosted regression trees*

**Fig. S34** | Boosted regression tree results testing the relationships between the logit of the Human Development Index and three hypothesised predictors: logit dependency ratio (2020), $\log_{10}$ total population size in 2020 ($N_{2020}$), and mean rate of population change ($r_{mean}$ for both 1950–2021 in black, and 2012–2021 in grey). Panel **a** shows the mean (± 95% confidence interval) relative contribution (%) of each predictor after 10,000 iterations of the model. The model explained 47.8–74.6% and 47.1–74.3% of the variation ($\beta_{CV}$) in the response for those including $r$ calculated from 1950–2021 and 2012–2021, respectively. Panels **b**–**d**: mean and 95% prediction confidence intervals for each predictor after Gaussian resampling $r_{mean}$ with its associated standard deviation (from 1950–2021 in black and 2012–2021 in grey) and resampling countries to reduce the effects of spatial autocorrelation.

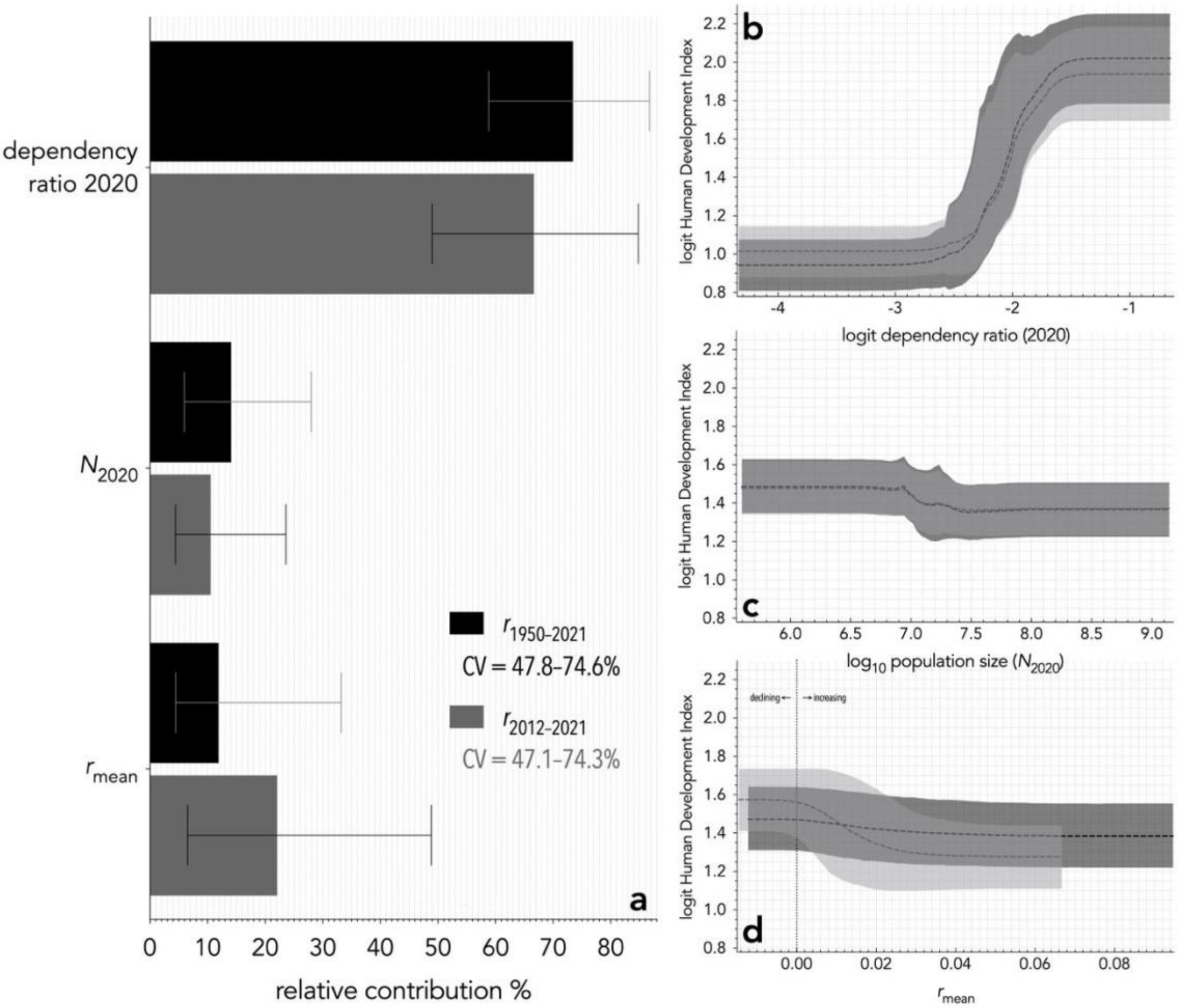

*Planetary pressure-adjusted Human development index*

This adjusts the Human Development Index for planetary pressures in the Anthropocene by discounting the index for pressures on the planet to reflect a concern for intergenerational inequality[15]. The adjusted index is the human development measures adjusted by per-capita carbon-dioxide emissions (production) and material footprint to account for global-scale environmental degradation

**Fig. S35** | Relationship between the logit of the planetary pressure-adjusted Human Development Index (2023) (source: United Nations Environment Programme) and national purchase power parity (PPP)-adjusted per-capita gross domestic product (source: World Bank). Least-squares regression gives $R^2 = 0.432$. Bubble size indicates relative total population size of each country in 2020. Bubbles are colour-coded by major region (**AFR** = Africa; **ASIOC** = Asia + Oceania; **EUR** = Eurasia; **ME** = Middle East; **NAM** = North America; **SACAR** = South America + Caribbean). Labels for a sample of the points are ISO 3-character country codes.

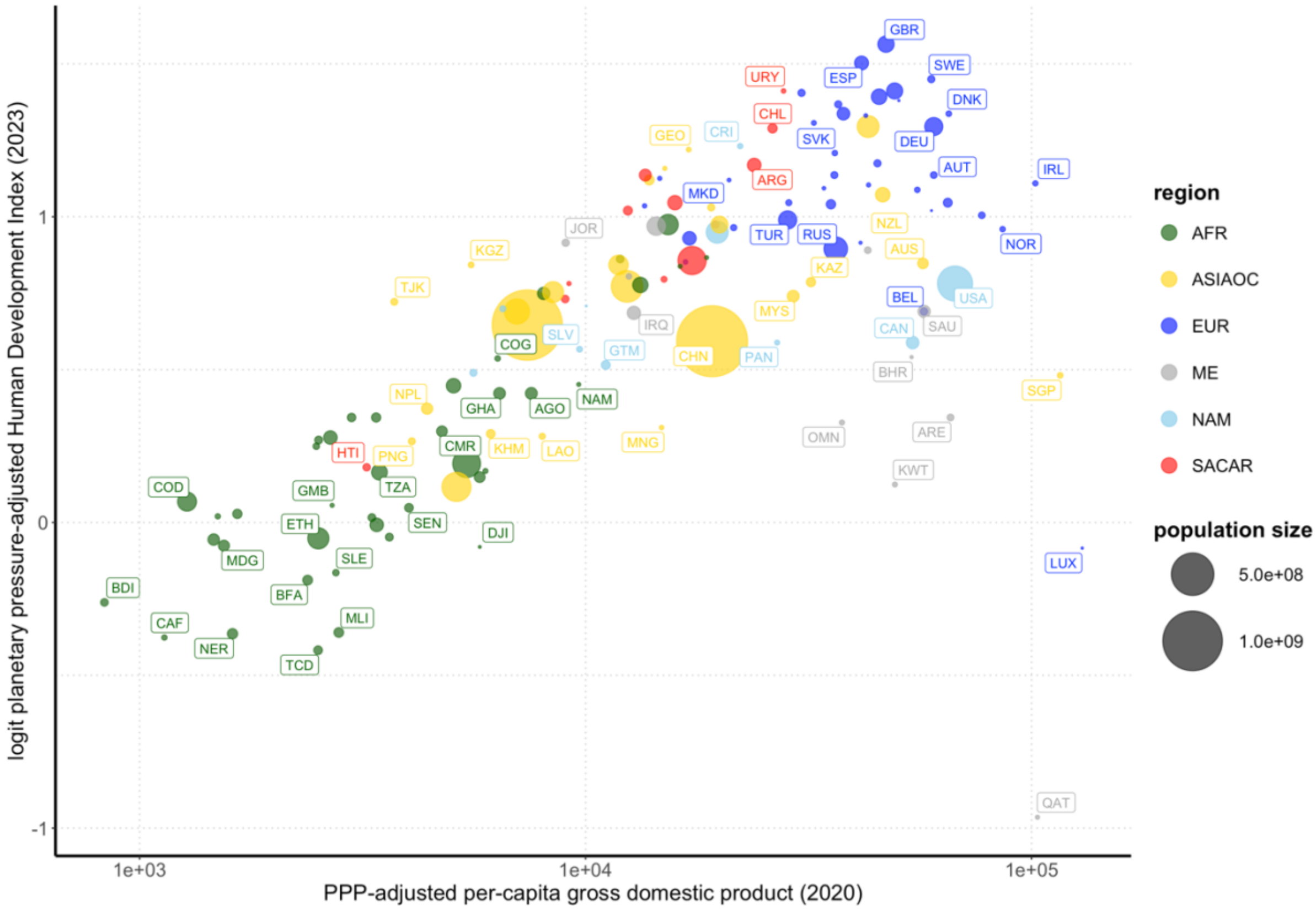

**Fig. S36** | Relationship between the logit of the planetary pressure-adjusted Human Development Index (2023) (source: United Nations Environment Programme) and logit dependency ratio (2020) (source: United Nations Population Division). Bubble size indicates relative total population size of each country in 2020. Bubbles are colour-coded by major region (**AFR** = Africa; **ASIOC** = Asia + Oceania; **EUR** = Eurasia; **ME** = Middle East; **NAM** = North America; **SACAR** = South America + Caribbean). Labels for a sample of the points are ISO 3-character country codes.

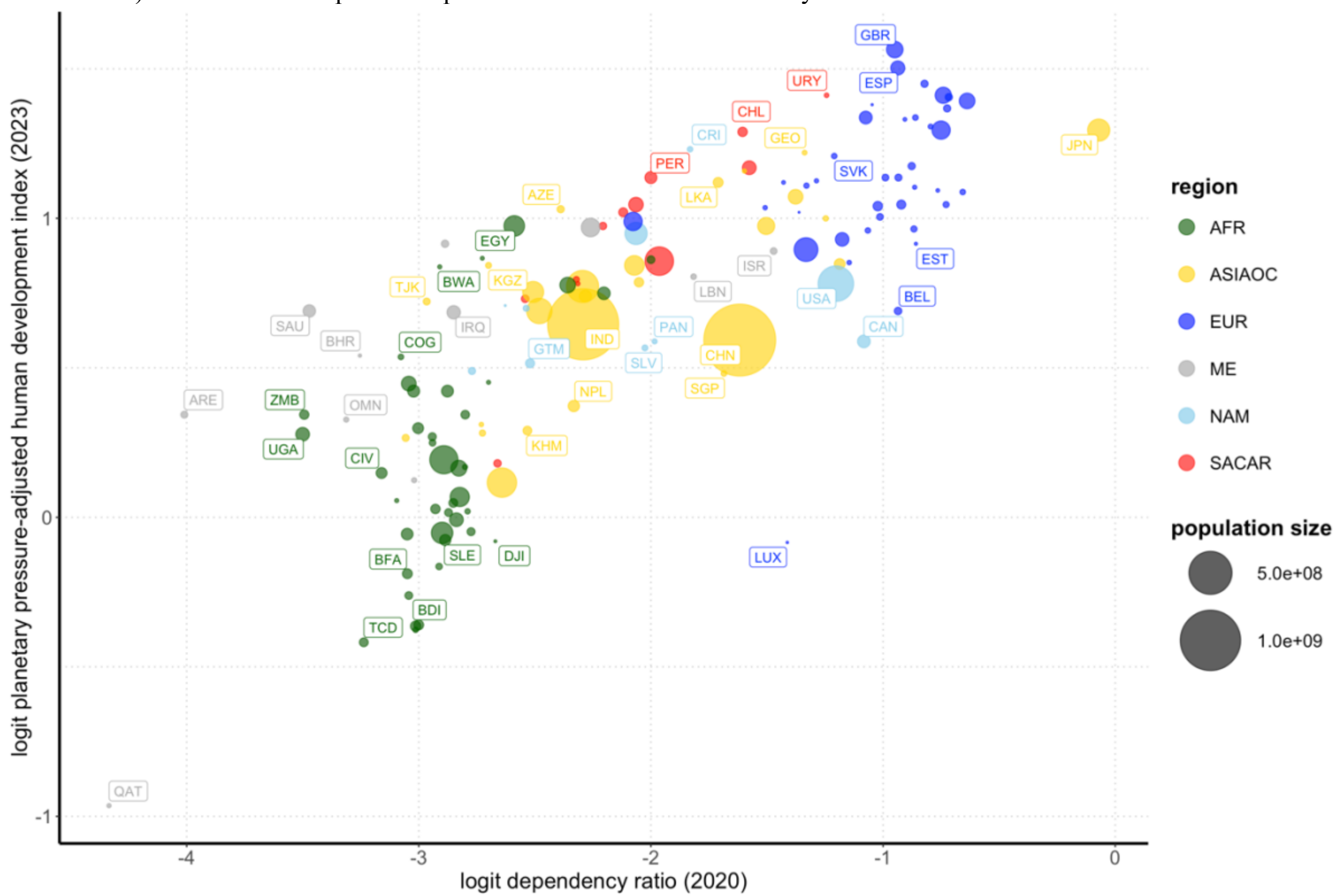

*Boosted regression tree*

The resampled boosted regression tree analysis showed that the dependency ratio explained the most variance in the logit of the planetary pressure-adjusted Human Development Index (Fig. 37a), and there was a threshold effect where a country's well-being declined quickly at a dependency ratio of ~ 0.06–0.10 (-2.2 to -2.7 on the logit scale; Fig. S37b). The population rate of change also predicted a similar change in well-being as for wealth (Fig. S37c), but the effect of total population size on well-being was not evident (Fig. S37d).

**Fig. S37** | Boosted regression tree results testing the relationships between the logit of the planetary pressure-adjusted Human Development Index and three hypothesised predictors: logit dependency ratio (2020), $\log_{10}$ total population size in 2020 ($N_{2020}$), and mean rate of population change ($r_{mean}$ for both 1950–2021 in black, and 2012–2021 in grey). Panel **a** shows the mean (± 95% confidence interval) relative contribution (%) of each predictor after 10,000 iterations of the model. The model explained 57.1–82.4% and 57.0–82.5% of the variation ($\beta_{CV}$) in the response for those including $r$ calculated from 1950–2021 and 2012–2021, respectively. Panels **b**–**d**: mean and 95% prediction confidence intervals for each predictor after Gaussian resampling $r_{mean}$ with its associated standard deviation (from 1950–2021 in black and 2012–2021 in grey) and resampling countries to reduce the effects of spatial autocorrelation.

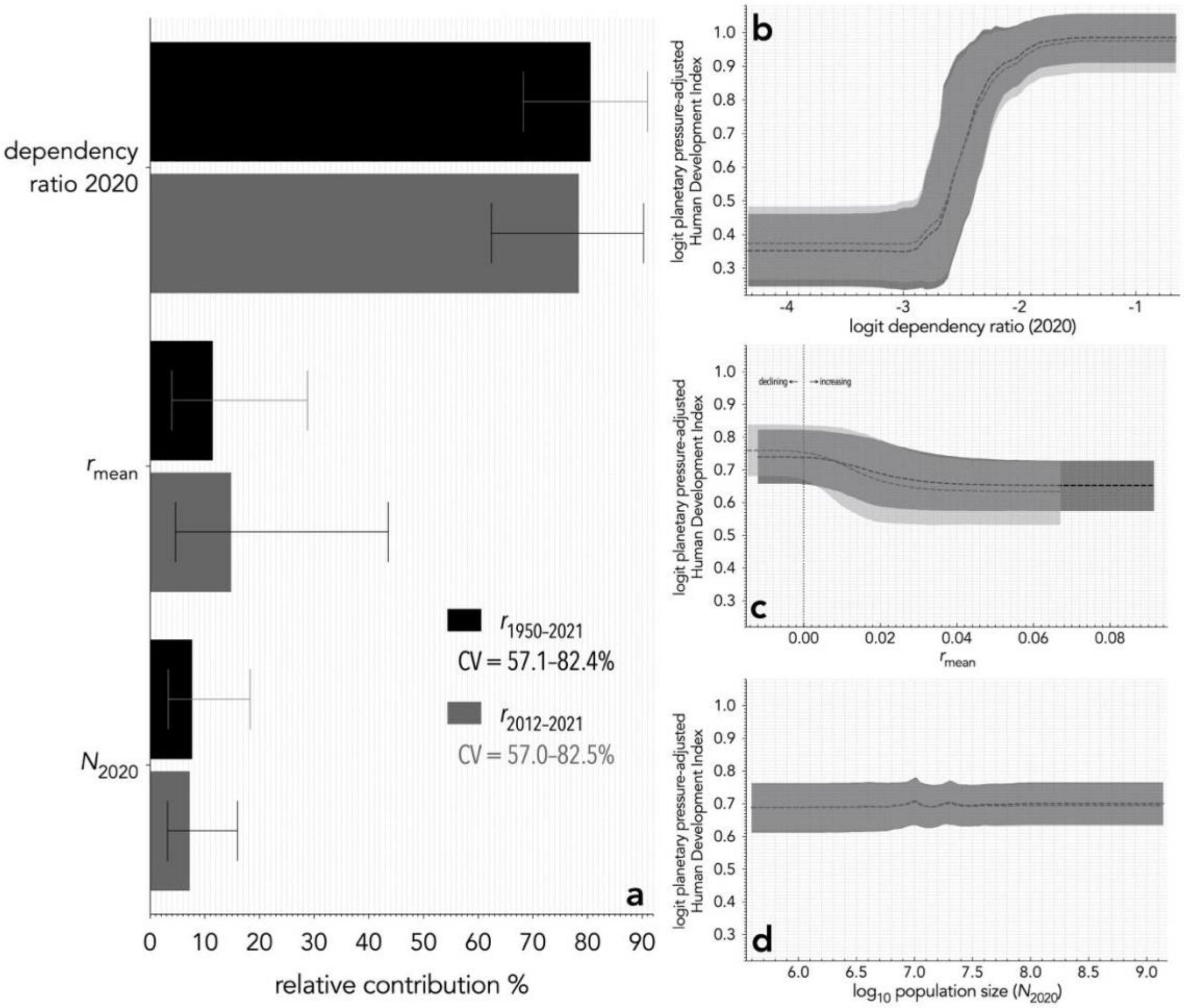

*Total energy supply*

Energy supply is central to economic development[16], even though the legacy of fossil-fuel exploitation is the primary driver[17] of the well-documented ecological and extinction crises gripping the planet[18-20]. We therefore obtained the 2024 national per-capita total energy supply (GJ person$^{-1}$) data for 78 countries from the Energy Institute's *Statistical Review of World Energy* (energyinst.org) and applied the same approach to analyse the relationships to dependency ratio, rate of population change, and total population size.

Among the three predictors we considered, total population size showed the strongest association with variation in per-capita total energy supply, even though the response variable was already expressed on a per-capita basis. (Fig. S38). In other words, countries with large populations have higher per-capita energy supply than countries with smaller populations. Regardless of this primary effect, there was still a positive relationship with the dependency ratio — older-cohort countries had higher per-capita total energy supply on average (Fig. S38). There was little to no effect of rate of population change on total energy supply.

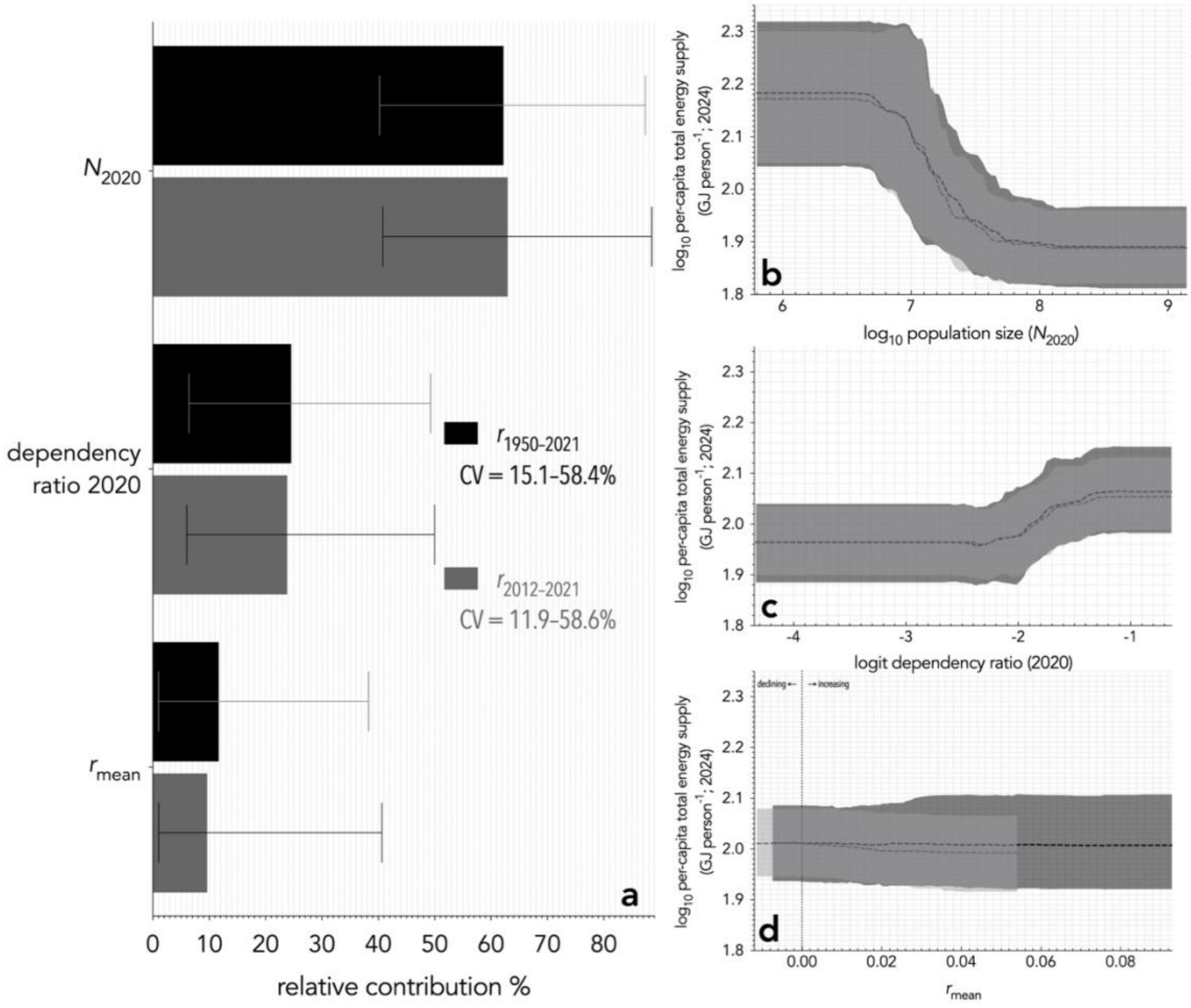

**Fig. S38** | Boosted regression tree results testing the relationships between the $\log_{10}$ of total energy supply per capita (GJ person$^{-1}$) and three hypothesised predictors: logit dependency ratio (2020), $\log_{10}$ total population size in 2020 ($N_{2020}$), and mean rate of population change ($r_{mean}$ for both 1950–2021 in black, and 2012–2021 in grey). Panel **a** shows the mean (± 95% confidence interval) relative contribution (%) of each predictor after 10,000 iterations of the model. The model explained 15.1–58.4% and 11.9–58.6% of the variation ($\beta_{CV}$) in the response for those including $r$ calculated from 1950–2021 and 2012–2021, respectively. Panels **b**–**d**: mean and 95% prediction confidence intervals for each predictor after Gaussian resampling $r_{mean}$ with its associated standard deviation (from 1950–2021 in black and 2012–2021 in grey) and resampling countries to reduce the effects of spatial autocorrelation.

## Appendix VII. Well-being index

To examine an index of well-being that is not at least partially based on economic indicators, we sourced the composite well-being rank derived from the Gallup World Poll[21]. Based on data from approximately four million respondents in 164 countries, the composite ranking is based on four positive well-being measures: life satisfaction, enjoyment, smiling, and being well-rested, and four negative measures: pain, sadness, anger, and worry[21]. We used the overall rank per country (Table 9 in ref[21]), where countries are ranked from highest (lowest rank) to highest (highest rank) well-being. Because the rank includes individual US states, we re-ranked the index for countries after excluding US states. The well-being rank is negatively correlated (lower ranks = higher well-being) with per-capita gross domestic product ($R^2 = 0.228$; Fig. S39).

**Fig. S39** | Relationship between the well-being rank[21] and national purchase power parity (PPP)-adjusted per-capita gross domestic product (source: World Bank). Least-squares regression gives $R^2 = 0.228$. Bubble size indicates relative total population size of each country in 2020. Bubbles are colour-coded by major region (**AFR** = Africa; **ASIOC** = Asia + Oceania; **EUR** = Eurasia; **ME** = Middle East; **NAM** = North America; **SACAR** = South America + Caribbean). Labels for a sample of the points are ISO 3-character country codes.

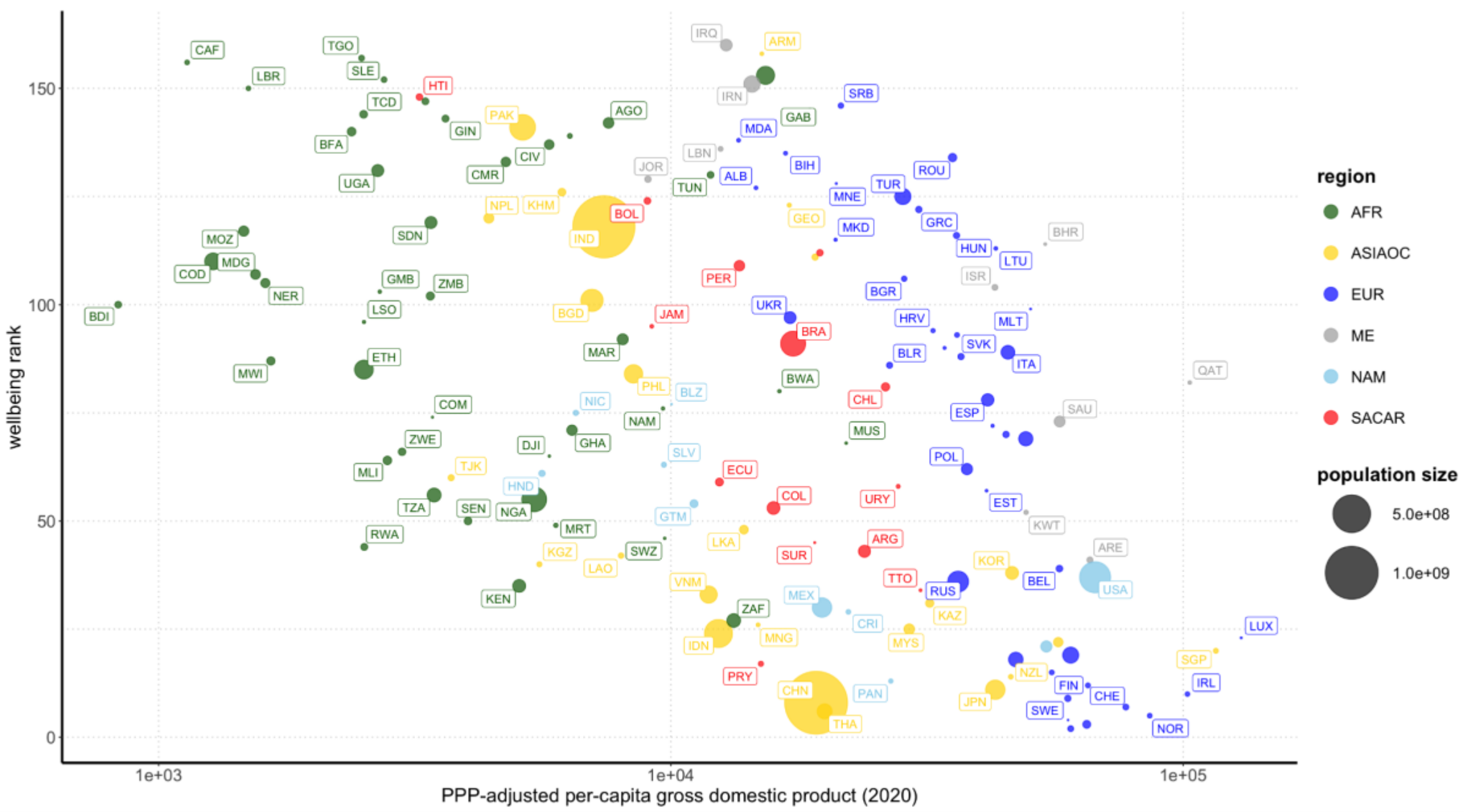

Although there was a general negative trend between the well-being rank and the logit of the dependency ratio, there was substantial clustering in the upper-right sector (Fig. S38).

**Fig. S40** | Relationship between the national composite well-being rank (1 = highest well-being; 160 = lowest well-being; 4 countries missing age-structure data)[21] and the logit dependency ratio ($\Sigma N_{\geq 65} \div \Sigma N_{16–65}$; see Methods). Bubble size indicates relative total population size of each country in 2020. Bubbles are colour-coded by major region (**AFR** = Africa; **ASIOC** = Asia + Oceania; **EUR** = Eurasia; **ME** = Middle East; **NAM** = North America; **SACAR** = South America + Caribbean). Labels for a sample of the points are ISO 3-character country codes.

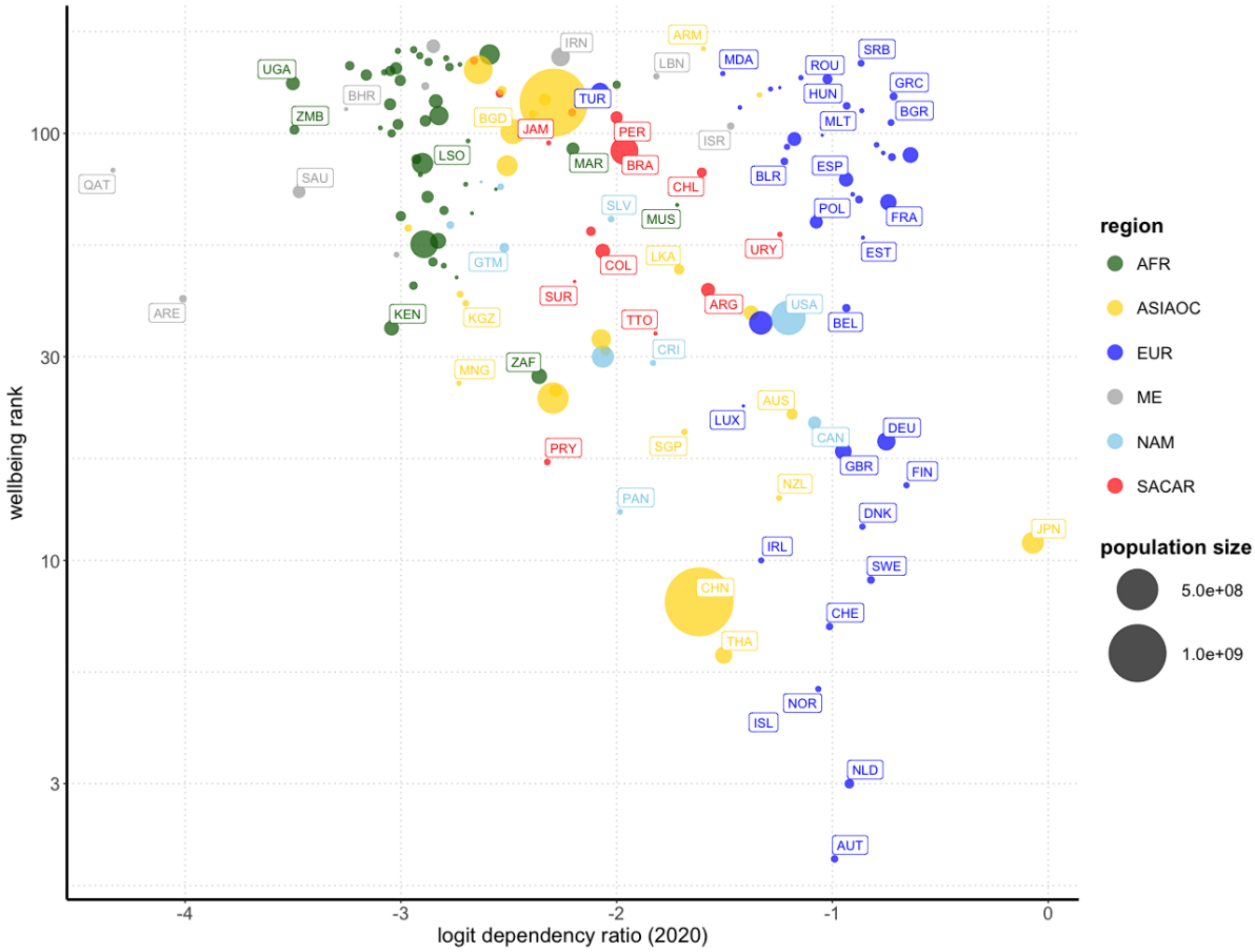

*Boosted regression trees*

When we ran the resampled boosted regression analysis, there was substantial variation in the predicted responses, leading to a low coefficient of variation explained (Fig. S41). While the dependency ratio still had the highest relative contribution (Fig. S41a) and expected relationship to well-being rank (Fig. S41b; i.e., 'older' nations had a lower rank and therefore, higher well-being), there was too much uncertainty to establish a clear relationship.

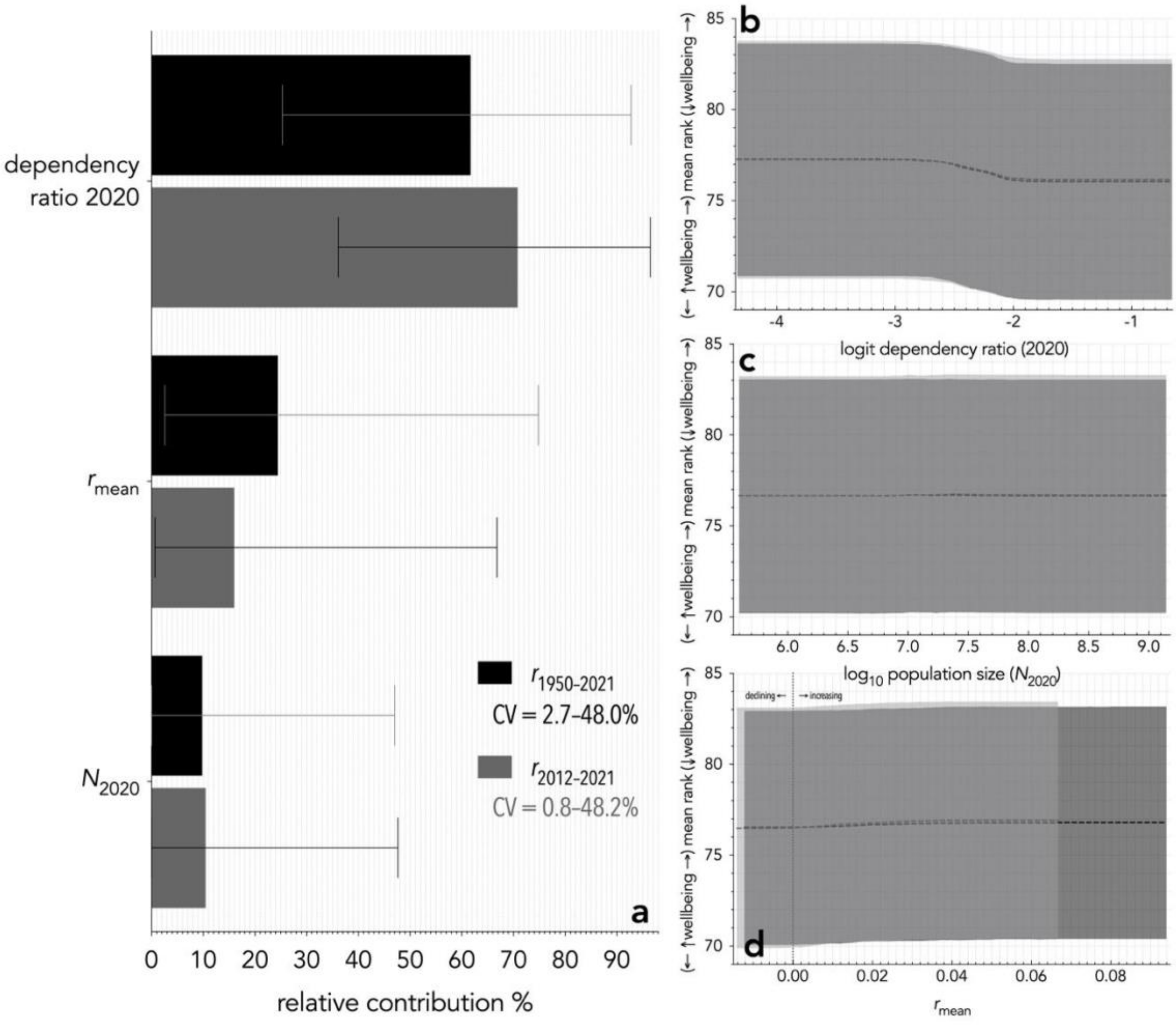

**Fig. S41** | Boosted regression tree results testing the relationships between the composite well-being rank[21] and three hypothesised predictors: logit dependency ratio (2020), log$_{10}$ total population size in 2020 ($N_{2020}$), and mean rate of population change ($r_{mean}$ for both 1950–2021 in black, and 2012–2021 in grey). Panel **a** shows the mean (± 95% confidence interval) relative contribution (%) of each predictor after 10,000 iterations of the model. The model explained 2.7–48.0% and 0.8–48.2% of the variation ($\beta_{CV}$) in the response for those including *r* calculated from 1950–2021 and 2012–2021, respectively. Panels **b**–**d**: mean and 95% prediction confidence intervals for each predictor after Gaussian resampling $r_{mean}$ with its associated standard deviation (from 1950–2021 in black and 2012–2021 in grey) and resampling countries to reduce the effects of spatial autocorrelation.

## Appendix VIII. Healthy life expectancy at birth

This is the average number of years that a person can expect to live in "full health" by taking into account years lived in less than full health due to disease and/or injury (source: World Health Organization).

**Fig. S42** | Relationship between healthy life expectancy at birth (years) (source: World Health Organization) and the logit dependency ratio ($\Sigma N_{\geq 65} \div \Sigma N_{16\text{–}65}$; see Methods). Bubble size indicates relative total population size of each country in 2020. Bubbles are colour-coded by major region (**AFR** = Africa; **ASIOC** = Asia + Oceania; **EUR** = Eurasia; **ME** = Middle East; **NAM** = North America; **SACAR** = South America + Caribbean). Labels for a sample of the points are ISO 3-character country codes.

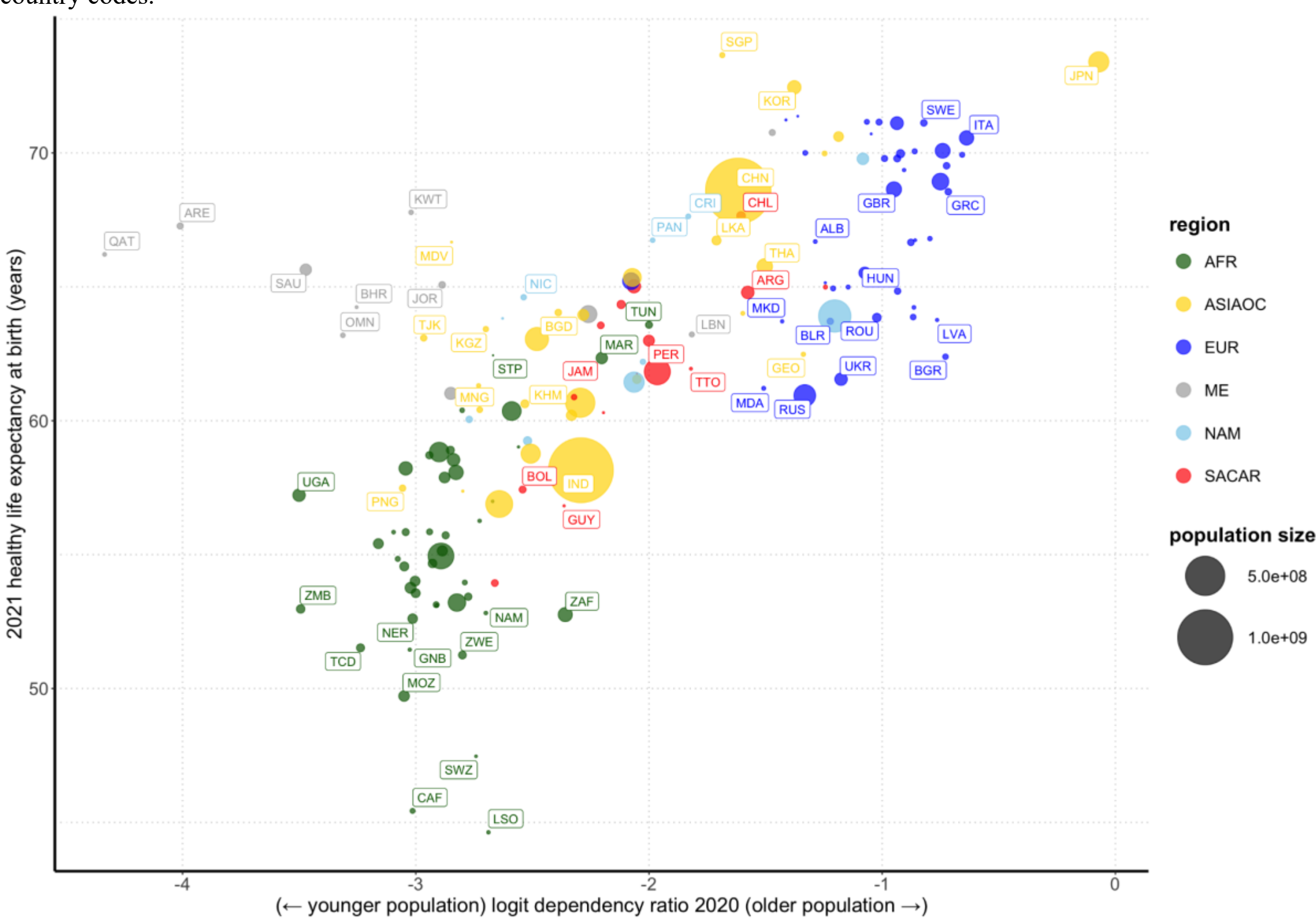

*Boosted regression trees*

The resampled boosted regression tree analysis revealed that of the three predictors we considered, the dependency ratio accounted for the highest proportion of the total variance in the response for the models including either $r_{\text{mean}}$ from 1950–2021 or 2012–2021 (Fig. S43a). The predicted relationship revealed that healthy life expectancy declined rapidly below a dependency ratio of ~ 0.18 (-1.5 on the logit scale) to stabilise at ~ 0.05 (-3.0 on the logit scale) (Fig. S43b). Both rate of population change and population size had weak to no discernible additional effects on life expectancy (Fig. S43c,d).

**Fig. S43** | Boosted regression tree results testing the relationships between the healthy life expectancy at birth (years) and three hypothesised predictors: logit dependency ratio (2020), $\log_{10}$ total population size in 2020 ($N_{2020}$), and mean rate of population change ($r_{\mathrm{mean}}$ for both 1950–2021 in black, and 2012–2021 in grey). Panel **a** shows the mean (± 95% confidence interval) relative contribution (%) of each predictor after 10,000 iterations of the model. The model explained 45.8–72.4% and 44.8–72.0% of the variation ($\beta_{\mathrm{CV}}$) in the response for those including $r$ calculated from 1950–2021 and 2012–2021, respectively. Panels **b**–**d**: mean and 95% prediction confidence intervals for each predictor after Gaussian resampling $r_{\mathrm{mean}}$ with its associated standard deviation (from 1950–2021 in black and 2012–2021 in grey) and resampling countries to reduce the effects of spatial autocorrelation.

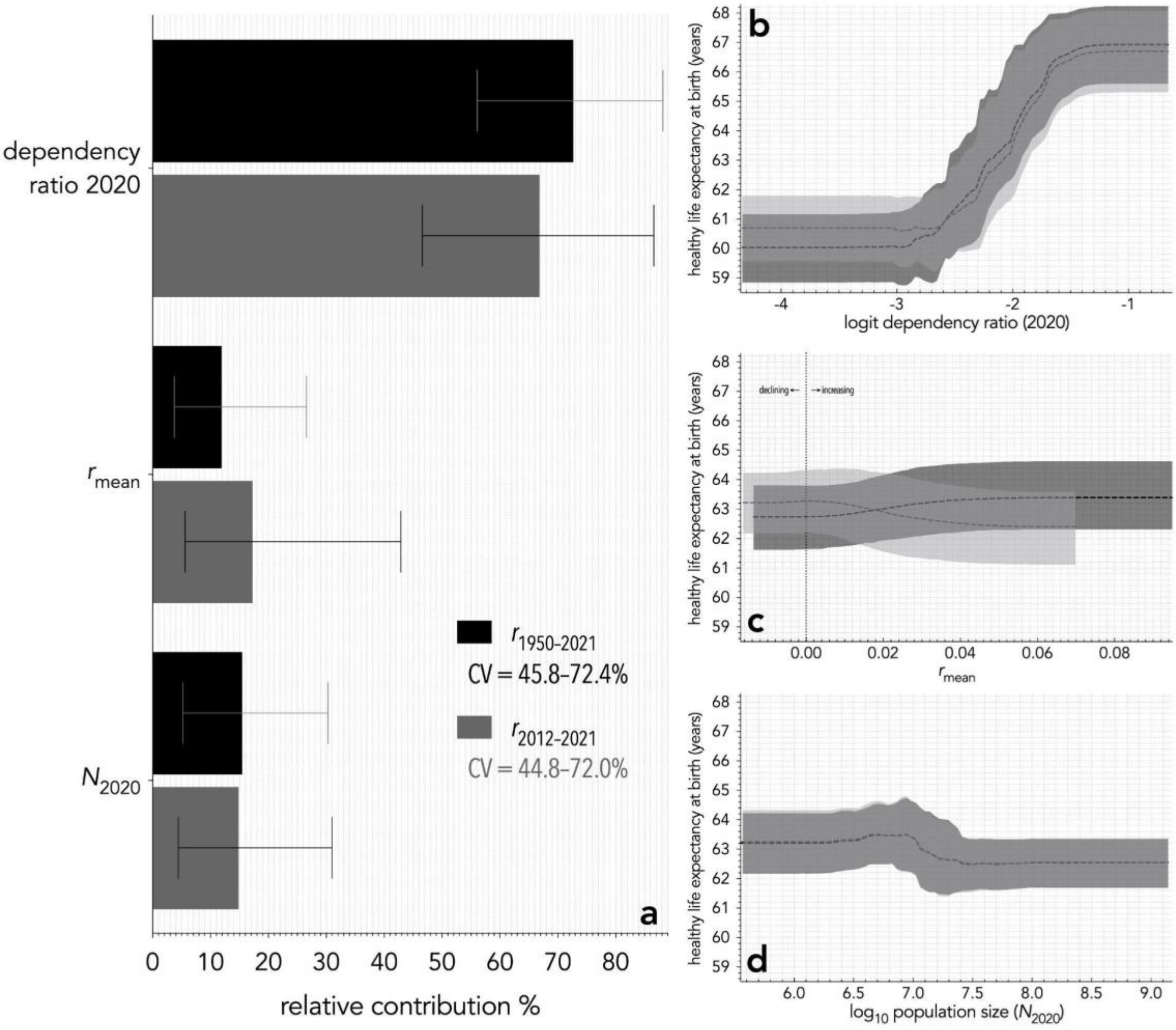

## Appendix IX: Time series

We investigated the relationship between each main response and the dependency ratio per country across years. Within-country time series of each response are temporally autocorrelated, so we calculated the Newey-West heteroscedasticity- and autocorrelation-consistent confidence interval of the slope ($\beta$) between each response and the dependency ratio. If the autocorrelation-consistent confidence interval did not include zero, we deemed there was evidence for a temporal relationship between the response and dependency ratio (with either a positive or negative $\beta$).

Of the countries with worse-outcome responses as their populations aged (higher dependency ratio), 19 had ≥ 3 responses in this category (marked in blue shading in Table S1): Belize (BLZ), Benin (BEN), Bolivia, Burundi (BDI), Costa Rica (CRI), Côte d'Ivoire (CIV), Gabon (GAB), The Gambia (GMB), Indonesia (IDN), Iraq (IRQ), Kyrgystan (KGZ), Mongolia (MNG), Mozambique (MOZ), Oman (OMN), Saudi Arabia (SAU), Sierra Leone (SLE), Slovakia (SVK), Tajikistan (TJK), and Venezuela (VEN). The specific response-dependency ratio relationships for those 19 countries are shown in Figures S44–S62.

**Table S1** | List of countries where an ageing population (higher dependency ratio) was correlated with a worse outcome for at least one of the nine response variables considered (indicated by a ●). Total number of responses with such an outcome indicated in the TOT column. Country codes are the ISO 3-character country codes. DCWI = per-capita domestic comprehensive wealth index; GINI = Gini coefficient (income inequality); R&D = gross expenditure in research and development as a percentage of gross domestic product; PATENTS = per-capita patent applications; HC = human capital index; PP-HDI = planetary pressure-adjusted Human Development index; HALE = healthy life expectancy at birth; FREEDOM = freedom score; CPI = corruption perception index; TOT = total number of responses for which an increasing dependency ratio (older population) was correlated with worse outcome.

| | **response** | | | | | | | | | |
|---|---|---|---|---|---|---|---|---|---|---|
| **country** | **DCWI** | **GINI** | **R&D** | **PATENTS** | **HC** | **PP-HDI** | **HALE** | **FREEDOM** | **CPI** | **TOT** |
| AND | | | | | | | | ● | | 1 |
| ARE | | | | | | | | ● | ● | 2 |
| ARG | | | | ● | | | | | ● | 2 |
| ARM | | | | | | | | | ● | 1 |
| AUT | | ● | | | | | | ● | | 2 |
| AZE | | | | | | | ● | ● | | 2 |
| BDI | | | | | ● | ● | | ● | | 3 |
| BEN | | | | | ● | ● | ● | | | 3 |
| BFA | | | | | | | ● | | | 1 |
| BGD | | | | | | | | ● | | 1 |
| BGR | | ● | | | | | | | ● | 2 |
| BHR | | | | | ● | | | ● | | 2 |
| BLR | | | ● | | | | | | ● | 2 |
| BLZ | | | | | ● | ● | | ● | | 3 |
| BOL | ● | ● | | | | | | | ● | 3 |
| BRA | | | | | | | | ● | | 1 |
| BRB | | | | | | | | ● | | 1 |
| BRN | | | | | | | | ● | | 1 |
| BTN | | | | | | | | | ● | 1 |
| CHE | | | | ● | | | | | | 1 |
| CHL | | | | | | | | ● | | 1 |
| CHN | | | | | | | | ● | ● | 2 |
| CIV | | | | | | ● | ● | ● | | 3 |
| COD | ● | | | | | | | | ● | 2 |
| COG | | | | | | | | ● | | 1 |
| COL | | | | | | | | | ● | 1 |
| COM | | | | | | | | ● | | 1 |
| CMR | | | | | | | ● | | | 1 |
| CPV | | | | | | | | | ● | 1 |
| CRI | | ● | | ● | | | | | ● | 3 |
| CZE | | | | | | | | • | | 1 |
| DEU | | • | | | | | | | | 1 |
| DJI | | | | | | | | • | | 1 |
| DNK | | • | | | | | | • | | 2 |
| DZA | | | | | | | | • | | 1 |
| EGY | • | | | | | | | • | | 2 |
| EST | | | | | | | | | • | 1 |
| ETH | | | | | | | | | • | 1 |
| FIN | | • | | • | | | | | | 2 |
| FSM | | | | | | | | • | | 1 |
| GAB | | | | | • | | • | | • | 3 |
| GBR | | | | • | | | | • | | 2 |
| GEO | | | | | | | | | • | 1 |
| GHA | | | | | | | | • | | 1 |
| GIN | | | | | | • | • | | | 2 |
| GMB | | | | | | • | • | • | | 3 |
| GNQ | | | | | | • | | | | 1 |
| GRC | | | | | | | | | • | 1 |
| GTM | | | | • | | | | • | | 2 |
| GUY | | | | | | | | | • | 1 |
| HND | | | | | | | | • | | 1 |
| HRV | | | | • | | | | • | | 2 |
| HTI | | | | | | | | • | | 1 |
| HUN | | | | • | | | | | | 1 |
| IDN | | • | | | | | | • | • | 1 |
| IRL | | | | | | | | | • | 1 |
| IRQ | | | | | • | • | | | • | 3 |
| ISL | | | | | | | | • | | 1 |
| ISR | | | | | | | | • | | 1 |
| ITA | | • | | | | | | | • | 2 |
| JAM | | | | • | | | | | | 1 |
| KAZ | | | | | | | | | • | 1 |
| KEN | | | | | • | | | • | | 2 |
| KGZ | | • | | | | | | • | • | 3 |
| KHM | | | | | | | | • | | 1 |
| KOR | | | | | | | | • | | 1 |
| LAO | | | | | | | • | | | 1 |
| LBN | | | | | | | | • | | 1 |
| LBR | • | | | | | | | | • | 2 |
| LBY | | | | | | | • | • | | 2 |
| LCA | | | | | | | • | • | | 2 |
| LIE | | | | | | | | • | | 1 |
| LSO | | | | | | | | | • | 1 |
| LTU | | | | | | | | | • | 1 |
| LUX | | • | | | | | | • | | 2 |
| LVA | | | | | | | | | • | 1 |
| MCO | | | | • | | | | • | | 2 |
| MDA | | | | | | | | • | | 1 |
| MDG | | | | | • | • | | | | 2 |
| MDV | | | | | | | • | | | 1 |
| MEX | | | | | | | | • | | 1 |
| MLI | | | | | | | • | | • | 2 |
| MKD | | | | • | | | | | | 1 |
| MLI | | | | | | • | | | | 1 |
| MNE | | | | | | | | • | • | 2 |
| MNG | • | | • | • | • | | | • | | 5 |
| MOZ | | | | • | • | | • | | • | 4 |
| MRT | | | | | | | | | • | 1 |
| MUS | | | | | | | | • | | 1 |
| MWI | | | | • | | | | • | | 2 |
| NGA | | | | | | • | | | | 1 |
| NIC | | | | | | | | • | | 1 |
| NLD | | | | | | | | • | | 1 |
| NPL | | | | | | | | | • | 1 |
| OMN | | | | | | | • | • | • | 3 |
| PHL | | | | | | | | • | | 1 |
| PNG | • | | | | | | | | • | 2 |
| PLW | | | | | | | | • | | 1 |
| POL | | | | | | | | • | | 1 |
| PRK | | | | | | | | | • | 1 |
| PRT | | | | | | | | • | | 1 |
| PRY | | | | • | | | | | • | 2 |
| QAT | | | | | • | | | | | 1 |
| RUS | | | | | | | | | • | 1 |
| SAU | • | | | | • | | • | | | 3 |
| SDN | • | | | | | | | | • | 2 |
| SEN | | | | | | • | • | | | 2 |
| SGP | | | | | | | | • | | 1 |
| SLB | | | | | | | | • | | 1 |
| SLE | | | | | • | • | • | | | 3 |
| SLV | | | | • | | | | • | | 2 |
| SMR | | | | | | | | • | | 1 |
| SOM | | | | | | | | | • | 1 |
| SRB | | | | • | | | | • | | 2 |
| SSD | | | | | | | | • | | 1 |
| STP | | | | | | | • | | • | 2 |
| SVK | | | | • | | | | • | • | 3 |
| SWE | | • | | • | | | | | | 2 |
| SYR | | | | | | | | • | | 1 |
| TCD | | | | | | • | • | | | 2 |
| TGO | | | | | • | | | | | 1 |
| TJK | | | • | | | | • | • | | 3 |
| TKM | | | | | | | | • | | 1 |
| TLS | | | | | | | | • | | 1 |
| TTO | | | | • | | | | | • | 2 |
| TUN | | | | | | | | | • | 1 |
| TUV | | | | | | | | • | | 1 |
| TWN | | | | | | | | | • | 1 |
| TZA | • | | | | | | | • | | 2 |
| UGA | | | | | | • | | | • | 2 |
| UKR | | | • | | | | | | • | 2 |
| URY | | | | • | | | | | | 1 |
| USA | | • | | | | | | • | | 2 |
| UZB | | | | | | | | | • | 1 |
| VEN | | | | • | | | • | • | | 3 |
| VNM | | | | | | | | | • | 1 |
| VUT | | | | | | | • | • | | 2 |
| YEM | | | | • | • | | | | | 2 |
| ZAF | | | | | | | | • | | 1 |
| ZMB | | | | | | • | • | | | 2 |
| ZWE | | | | | • | | | | • | 2 |

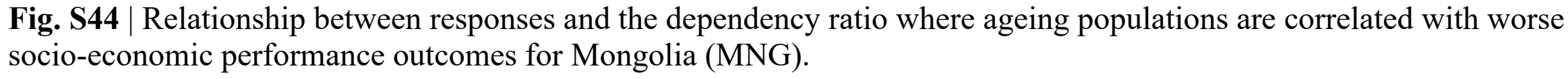

**Fig. S44** | Relationship between responses and the dependency ratio where ageing populations are correlated with worse socio-economic performance outcomes for Mongolia (MNG).

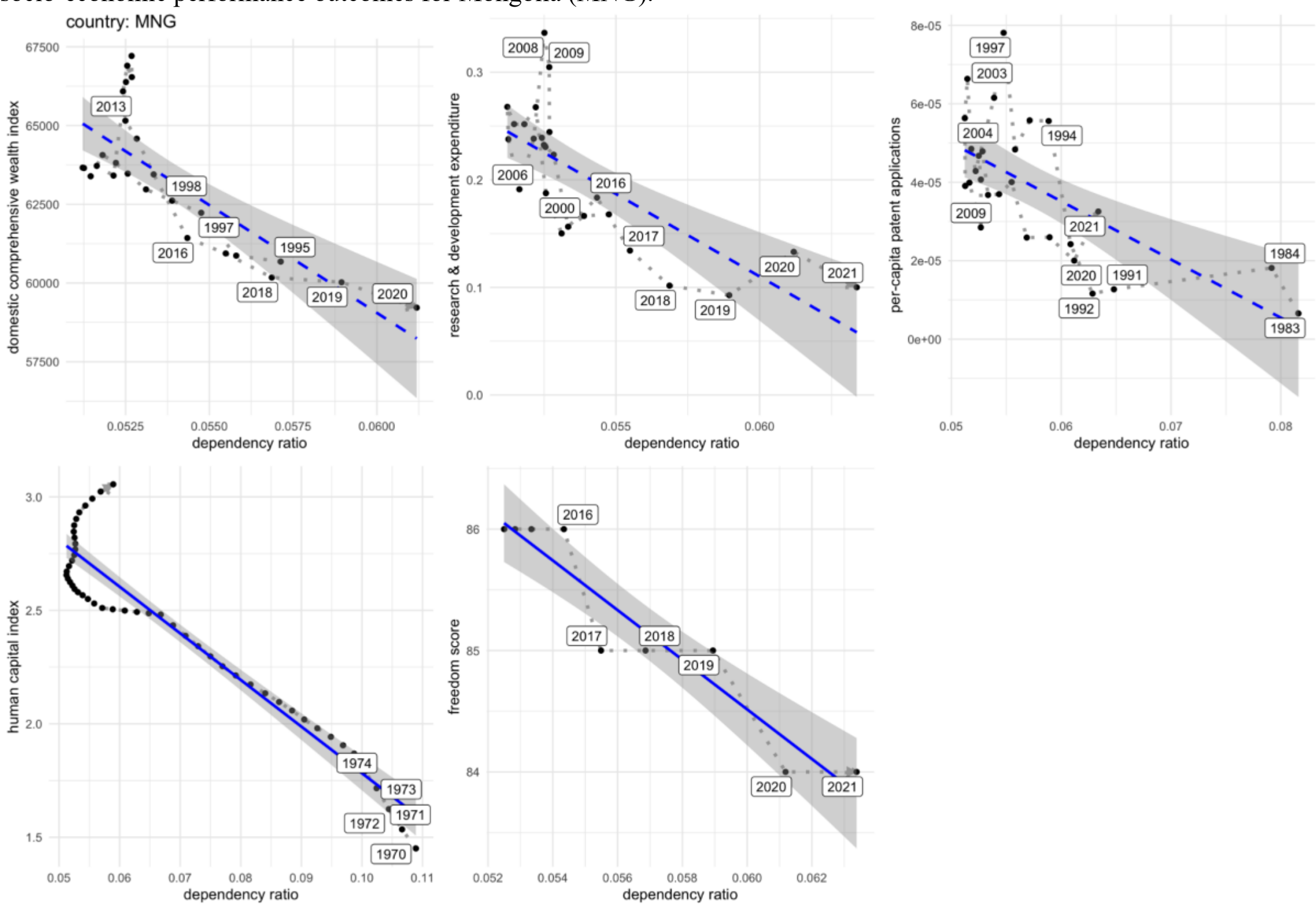

**Fig. S45** | Relationship between responses and the dependency ratio where ageing populations are correlated with worse socio-economic performance outcomes for Mozambique (MOZ).

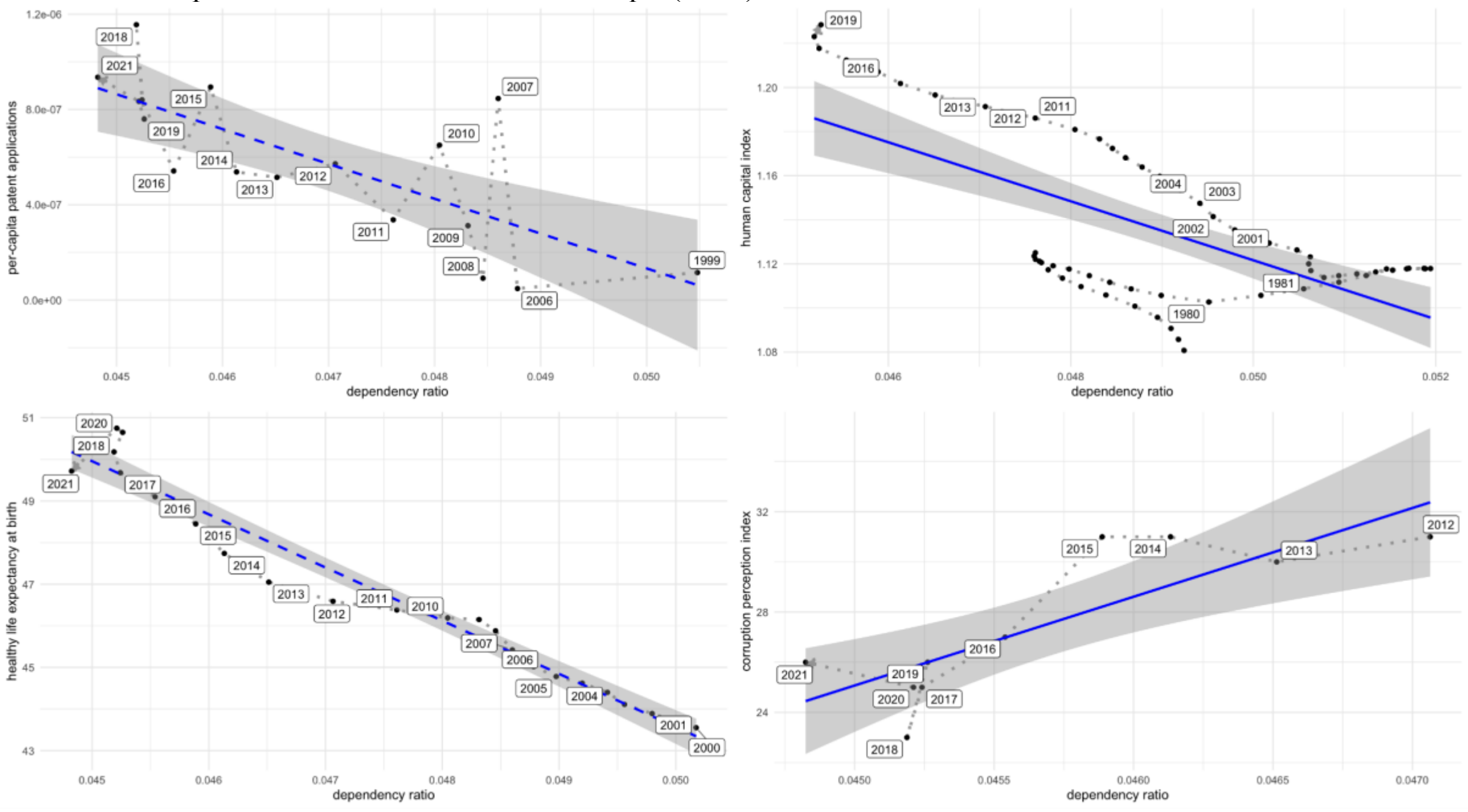

**Fig. S46** | Relationship between responses and the dependency ratio where ageing populations are correlated with worse socio-economic performance outcomes for Burundi (BDI).

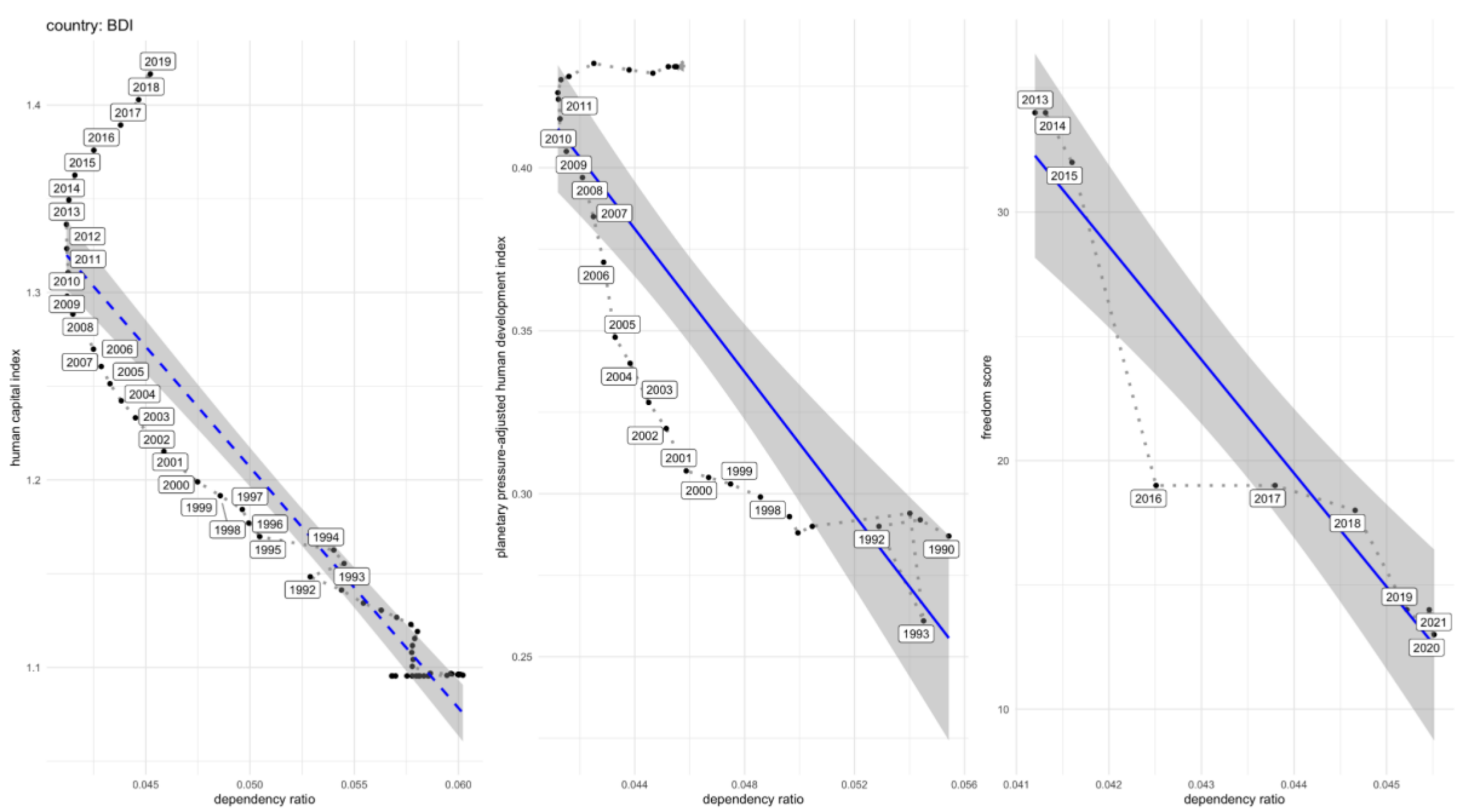

**Fig. S47** | Relationship between responses and the dependency ratio where ageing populations are correlated with worse socio-economic performance outcomes for Benin (BEN).

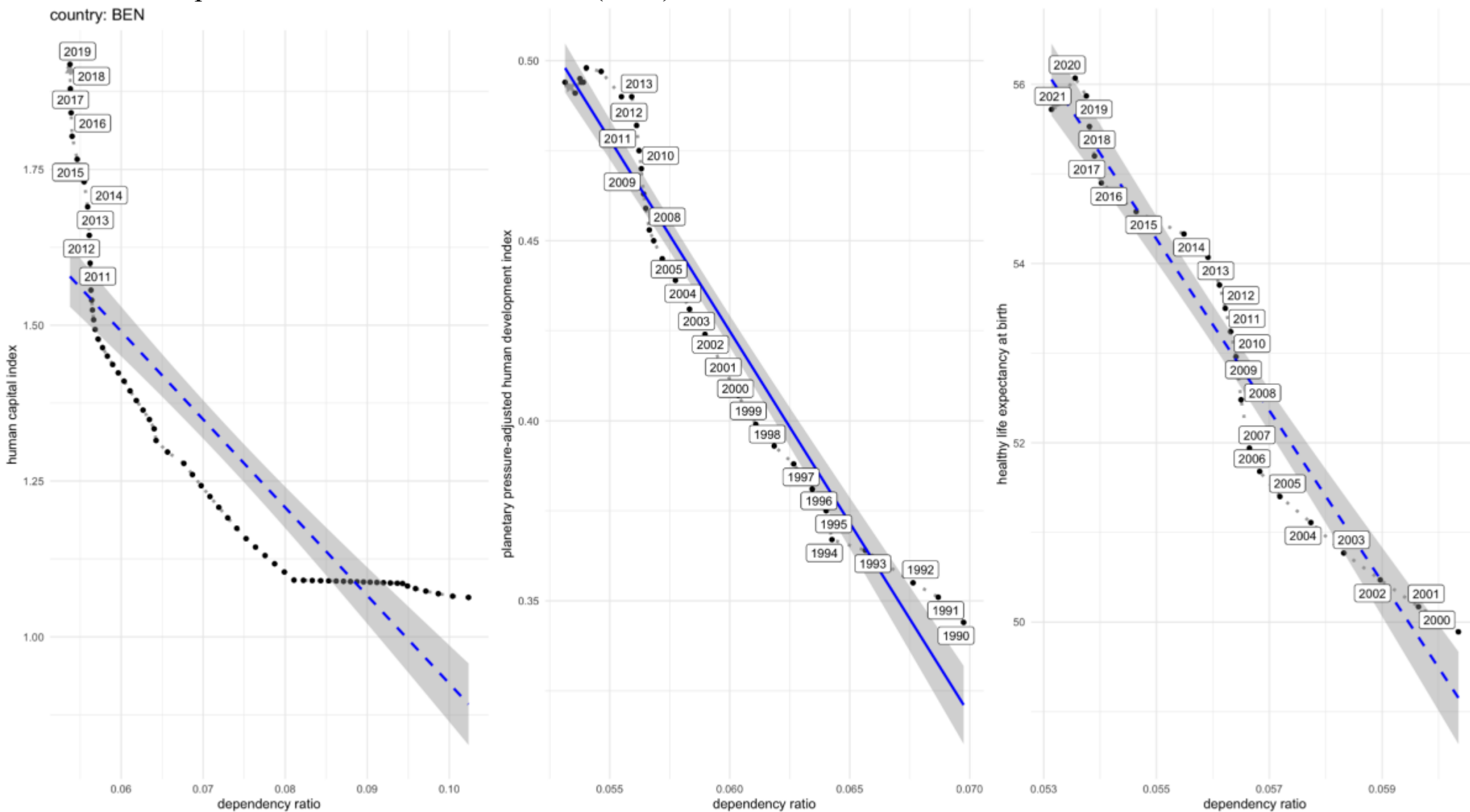

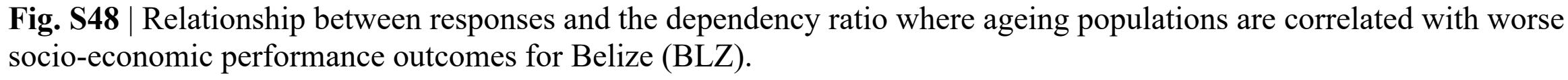

**Fig. S48** | Relationship between responses and the dependency ratio where ageing populations are correlated with worse socio-economic performance outcomes for Belize (BLZ).

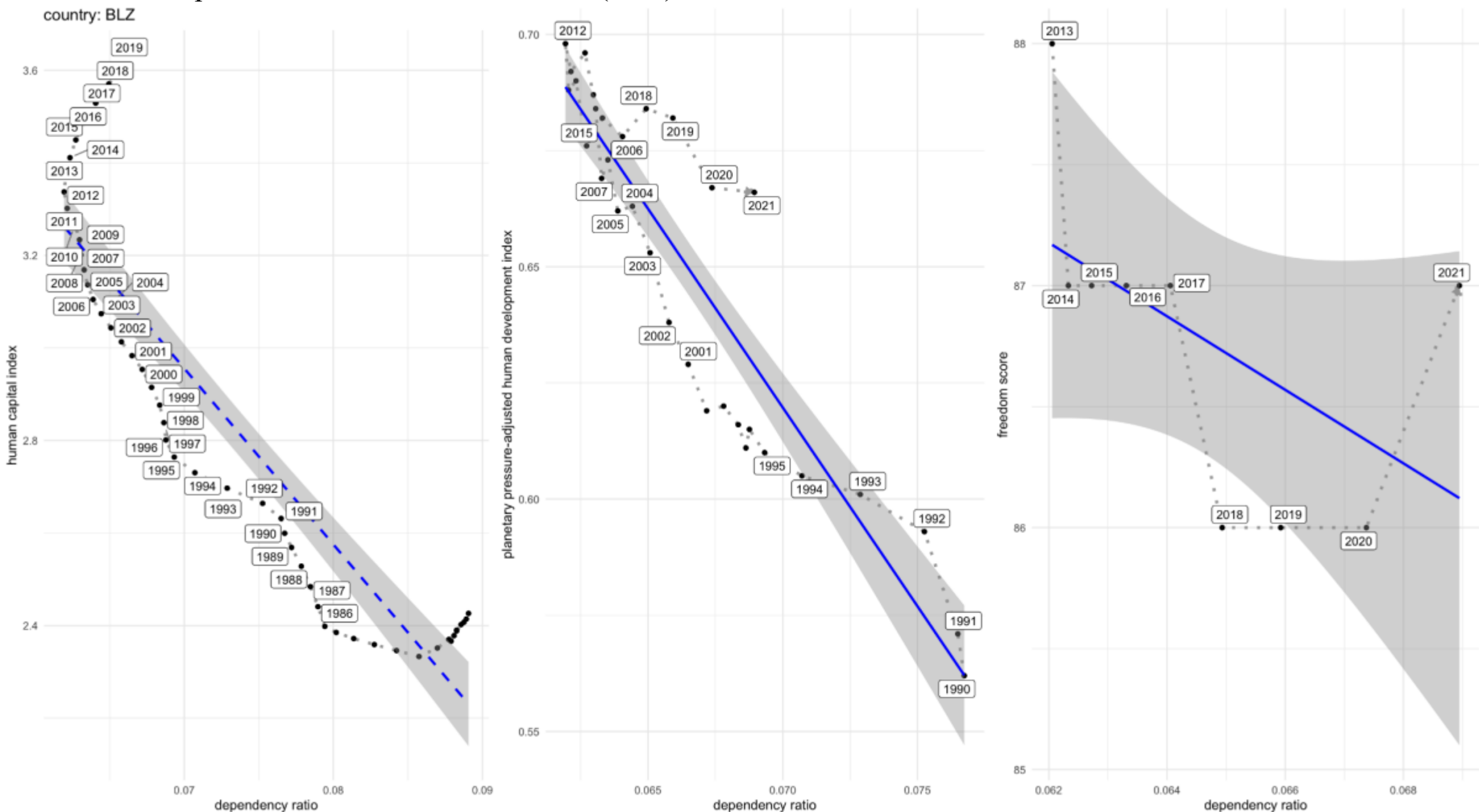

**Fig. S49** | Relationship between responses and the dependency ratio where ageing populations are correlated with worse socio-economic performance outcomes for Bolivia (BOL).

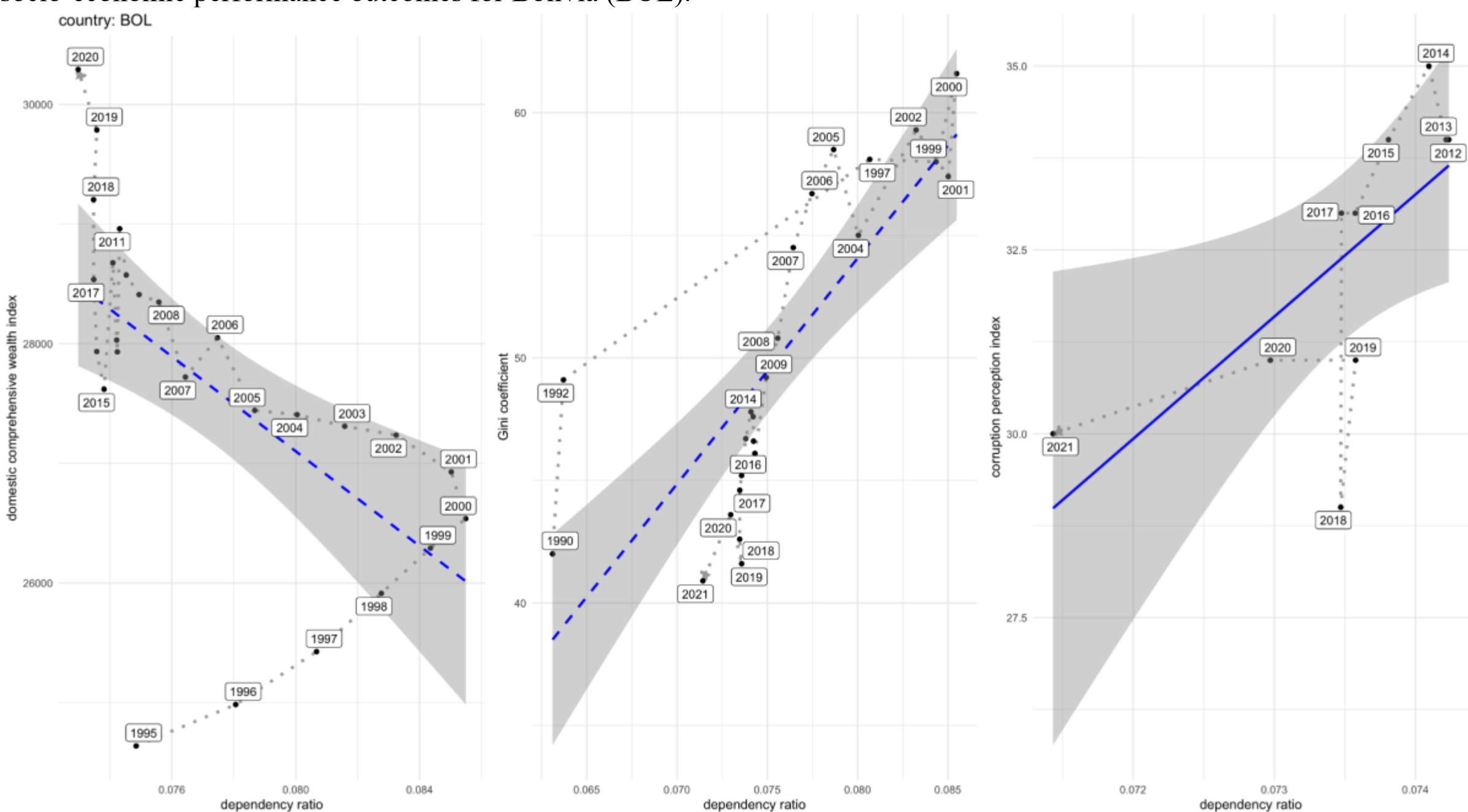

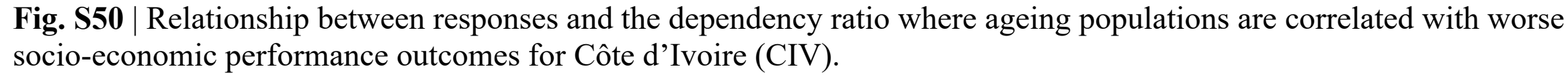

**Fig. S50** | Relationship between responses and the dependency ratio where ageing populations are correlated with worse socio-economic performance outcomes for Côte d'Ivoire (CIV).

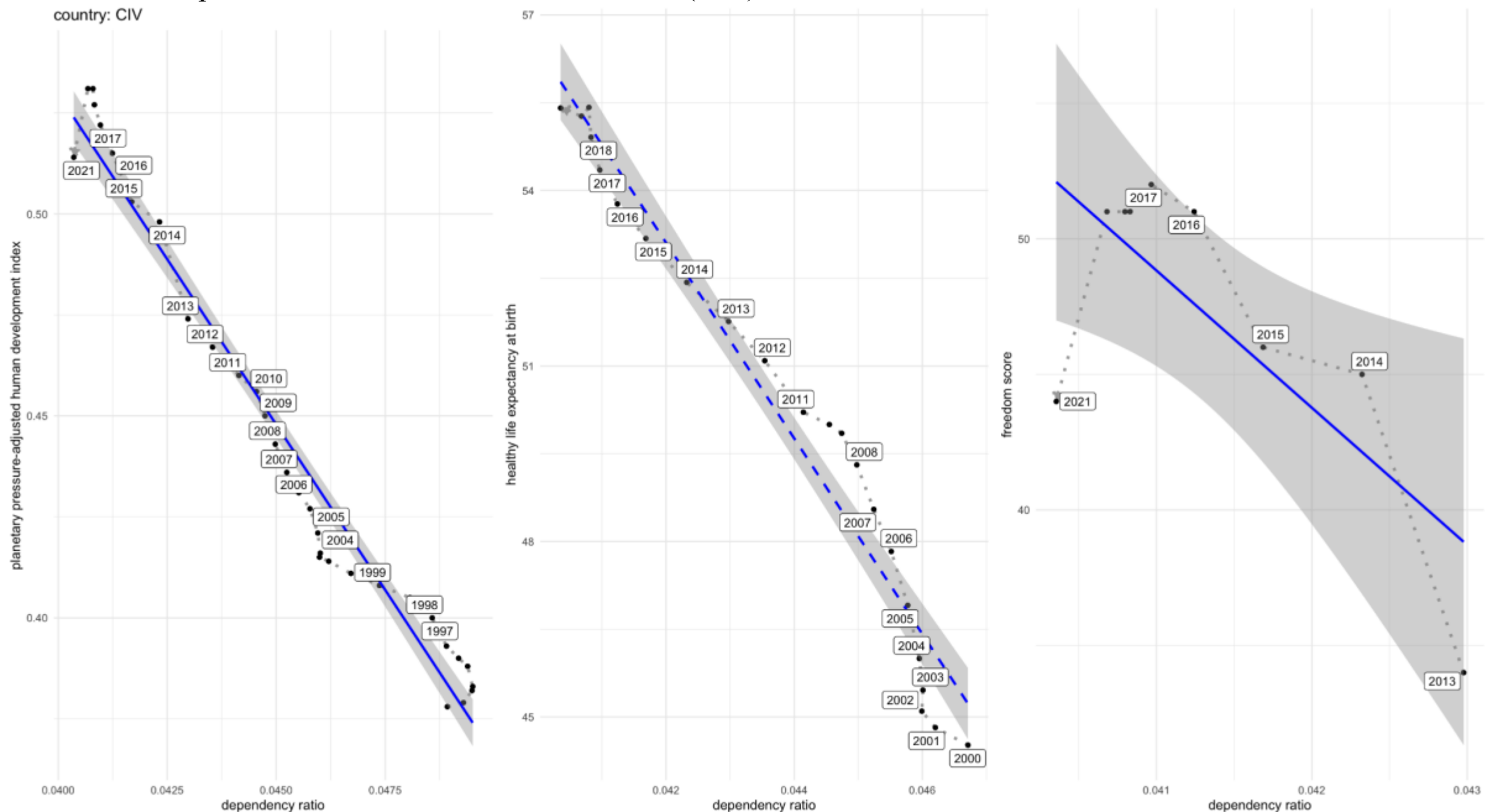

**Fig. S51** | Relationship between responses and the dependency ratio where ageing populations are correlated with worse socio-economic performance outcomes for Costa Rica (CRI).

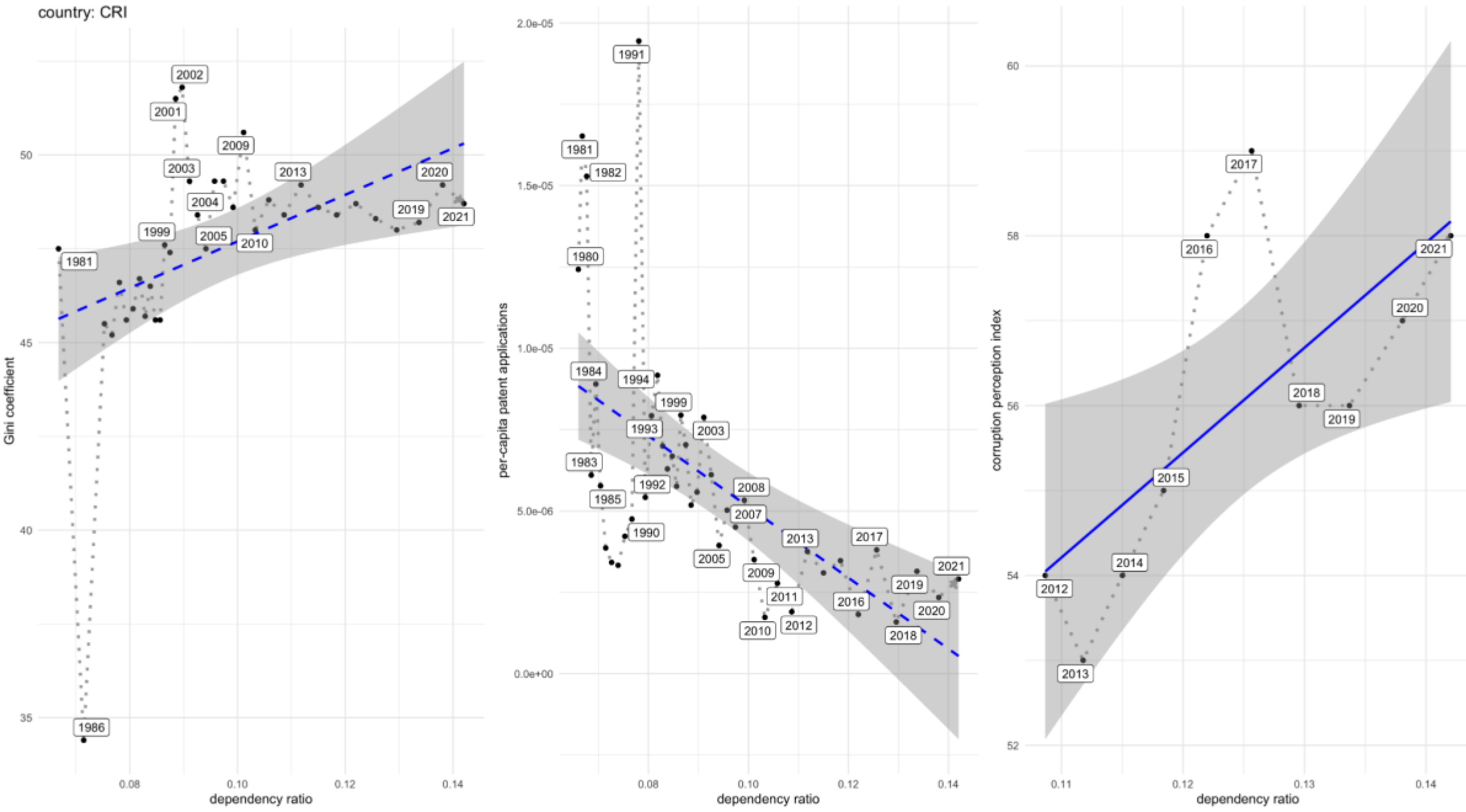

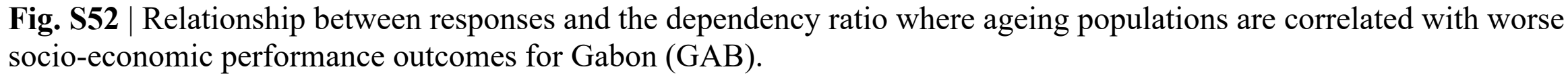

**Fig. S52** | Relationship between responses and the dependency ratio where ageing populations are correlated with worse socio-economic performance outcomes for Gabon (GAB).

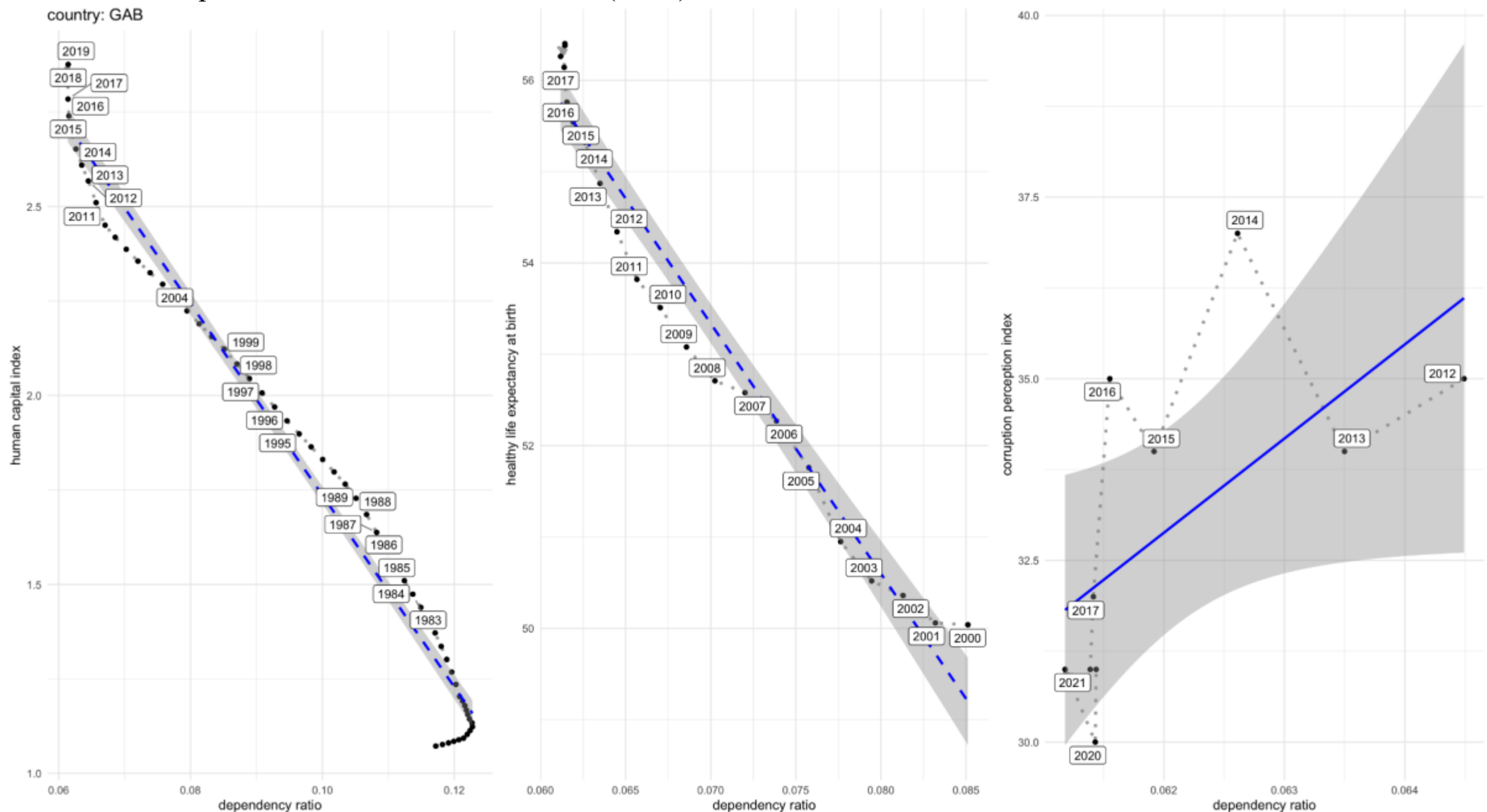

**Fig. S53** | Relationship between responses and the dependency ratio where ageing populations are correlated with worse socio-economic performance outcomes for The Gambia (GMB).

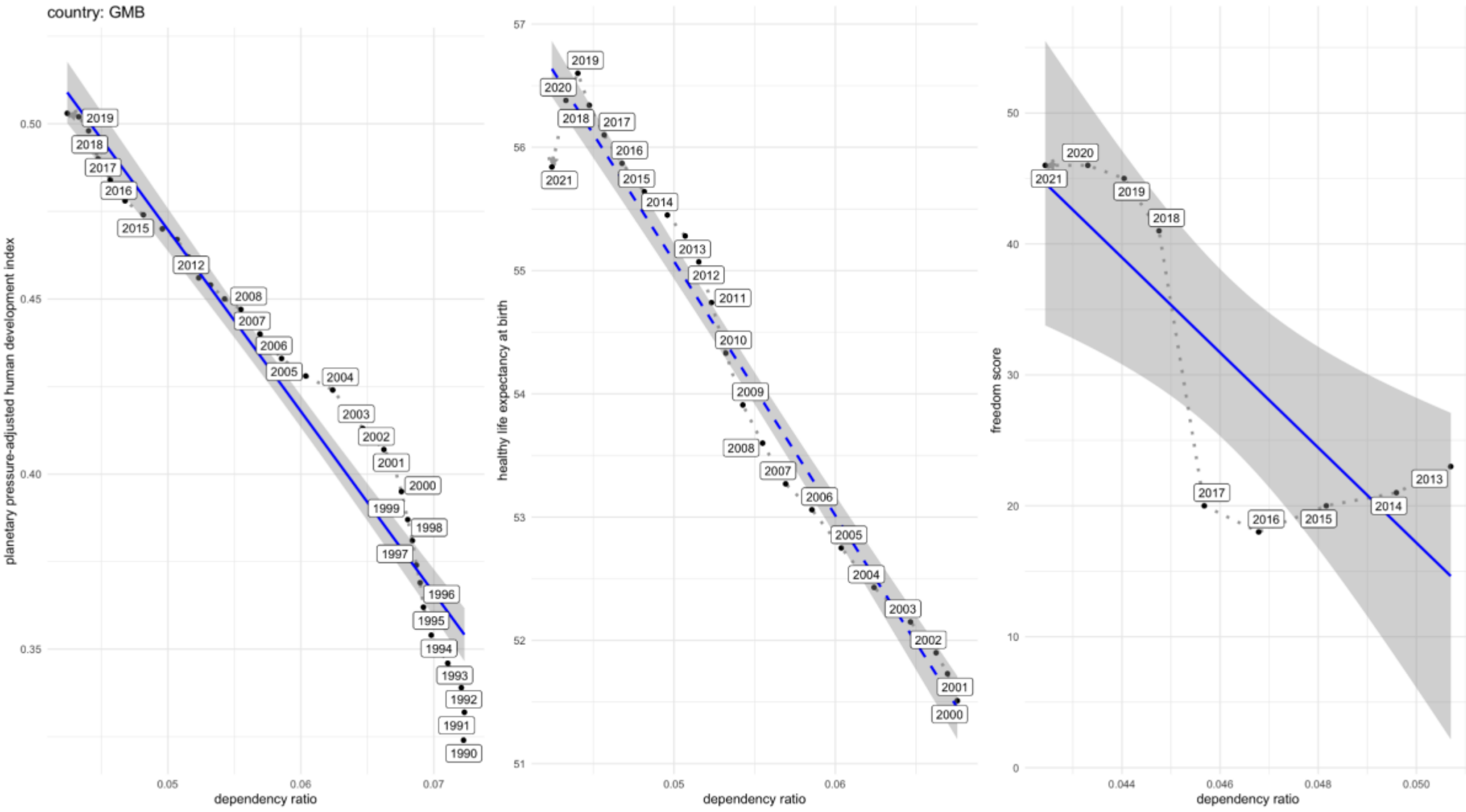

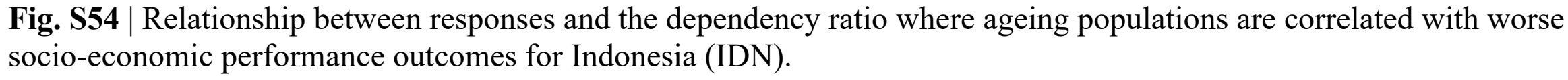

**Fig. S54** | Relationship between responses and the dependency ratio where ageing populations are correlated with worse socio-economic performance outcomes for Indonesia (IDN).

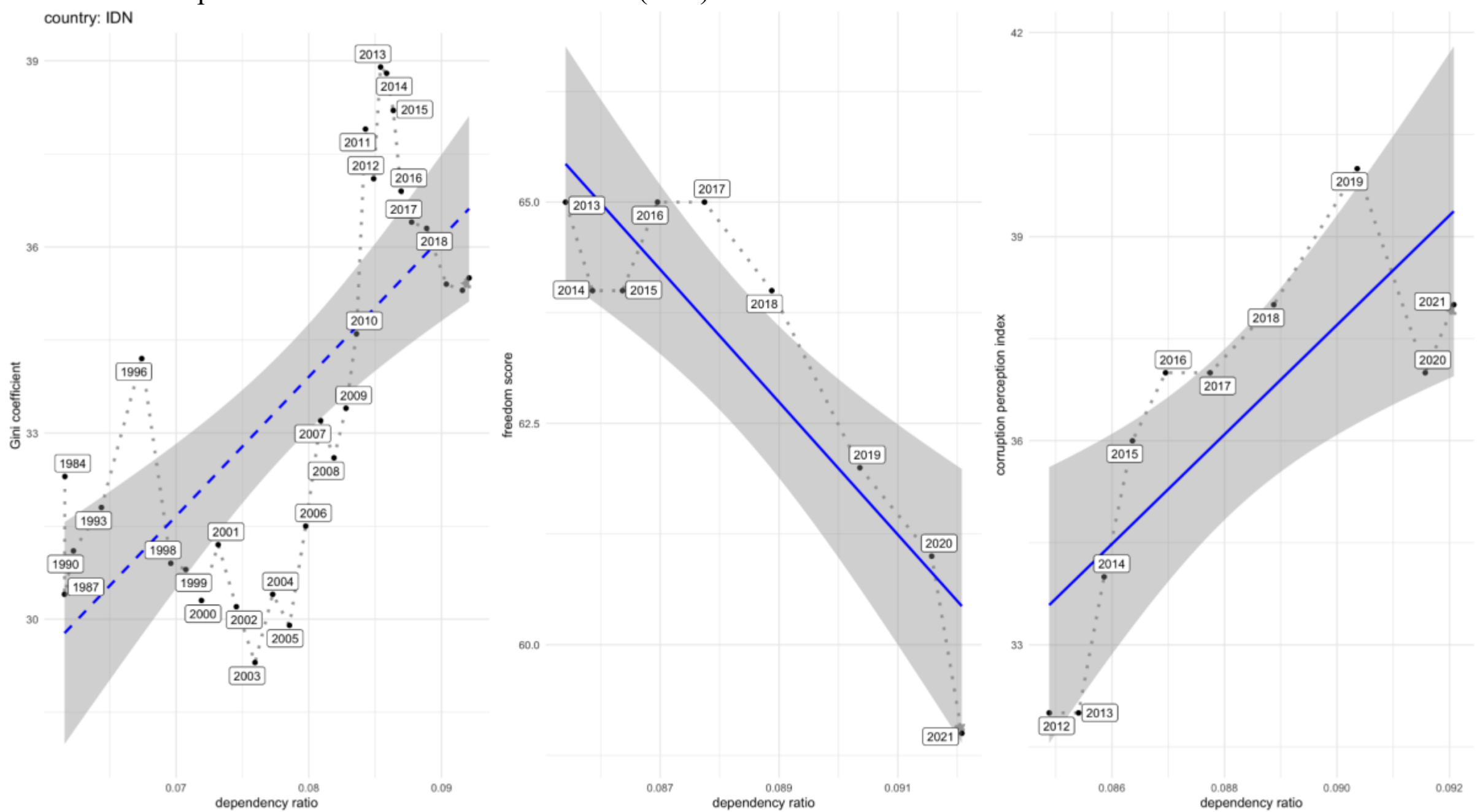

**Fig. S55** | Relationship between responses and the dependency ratio where ageing populations are correlated with worse socio-economic performance outcomes for Iraq (IRQ).

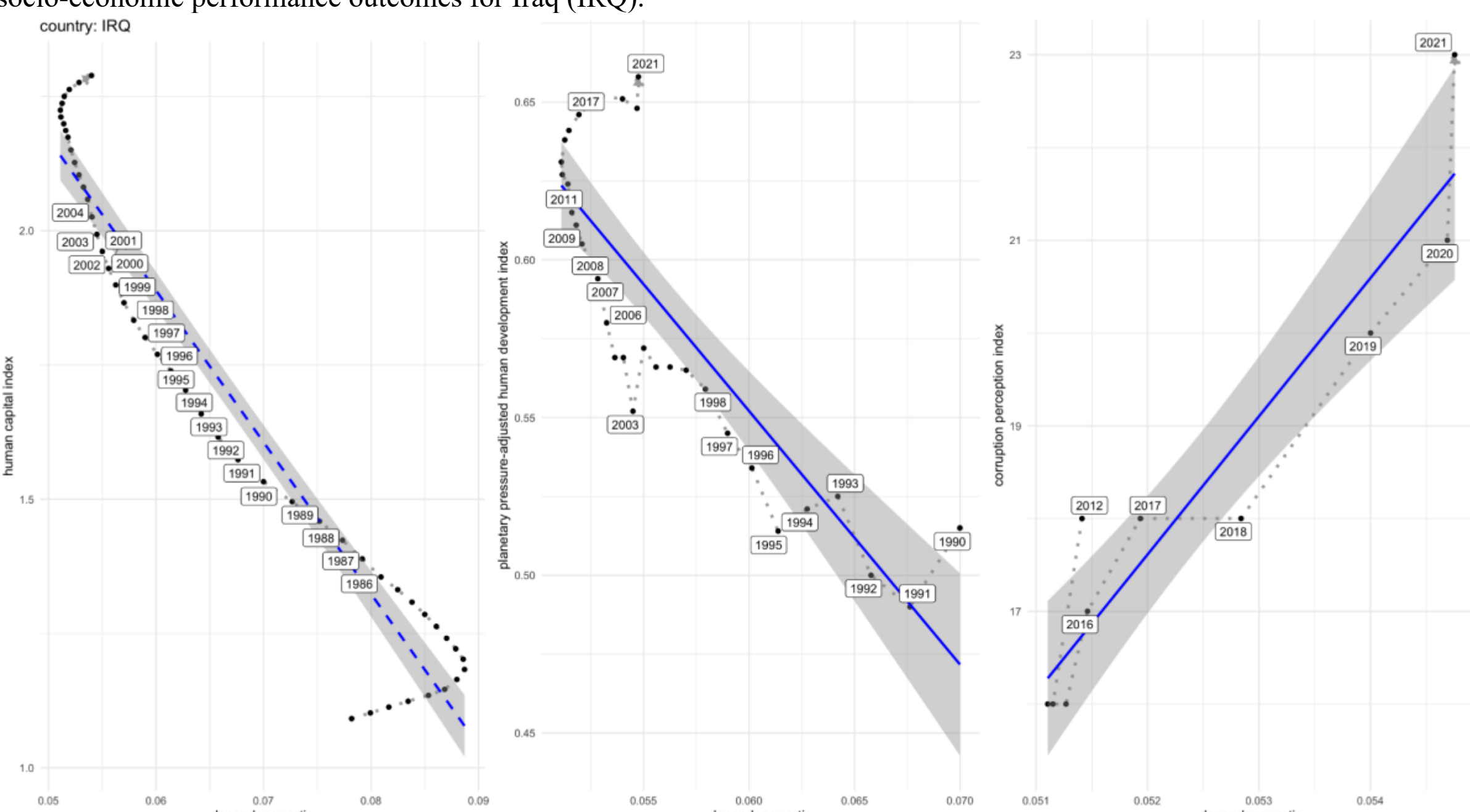

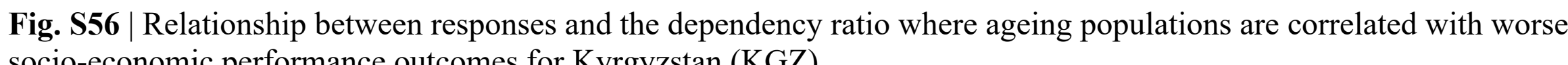

**Fig. S56** | Relationship between responses and the dependency ratio where ageing populations are correlated with worse socio-economic performance outcomes for Kyrgyzstan (KGZ).

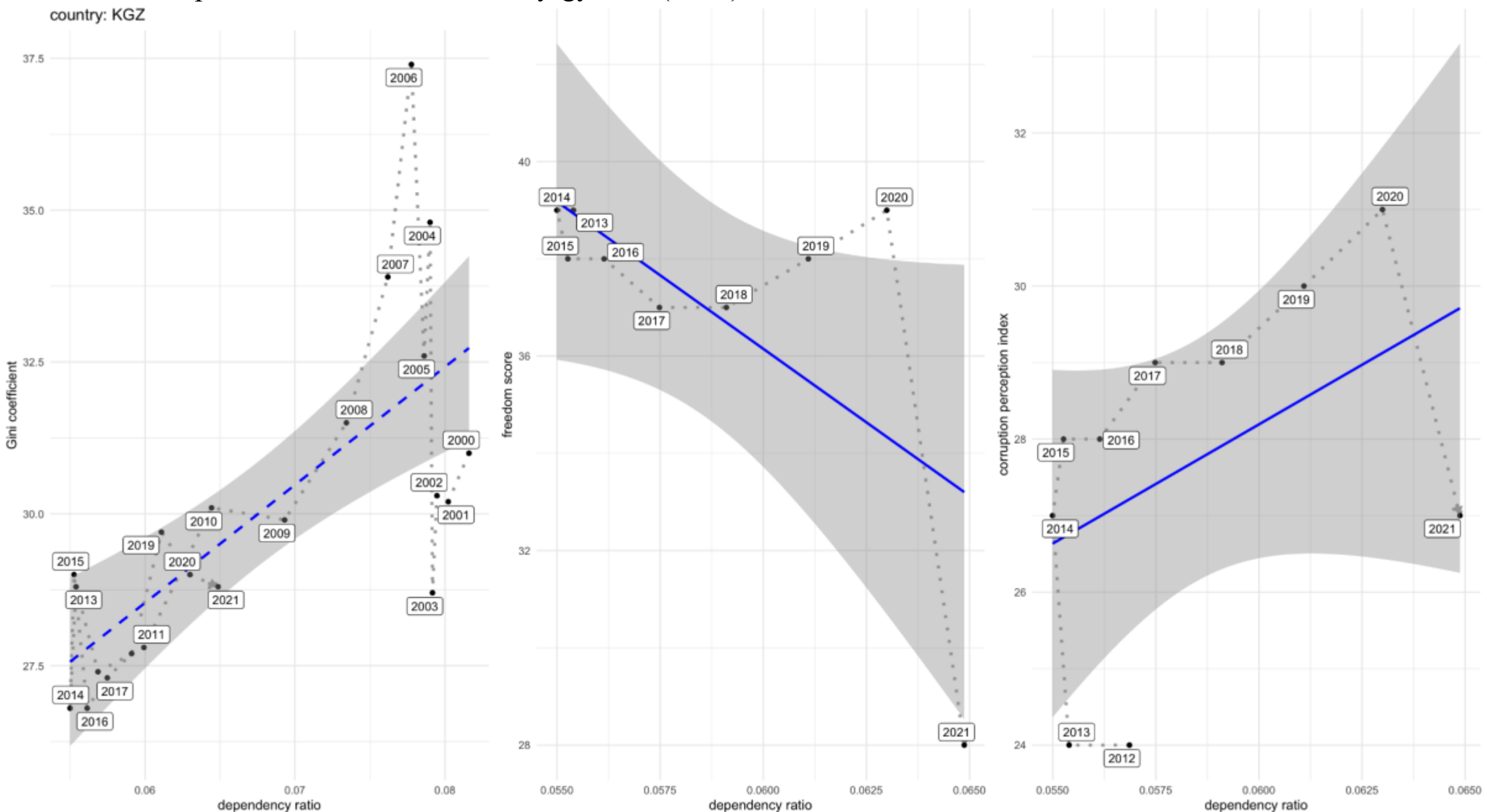

**Fig. S57** | Relationship between responses and the dependency ratio where ageing populations are correlated with worse socio-economic performance outcomes for Oman (OMN).

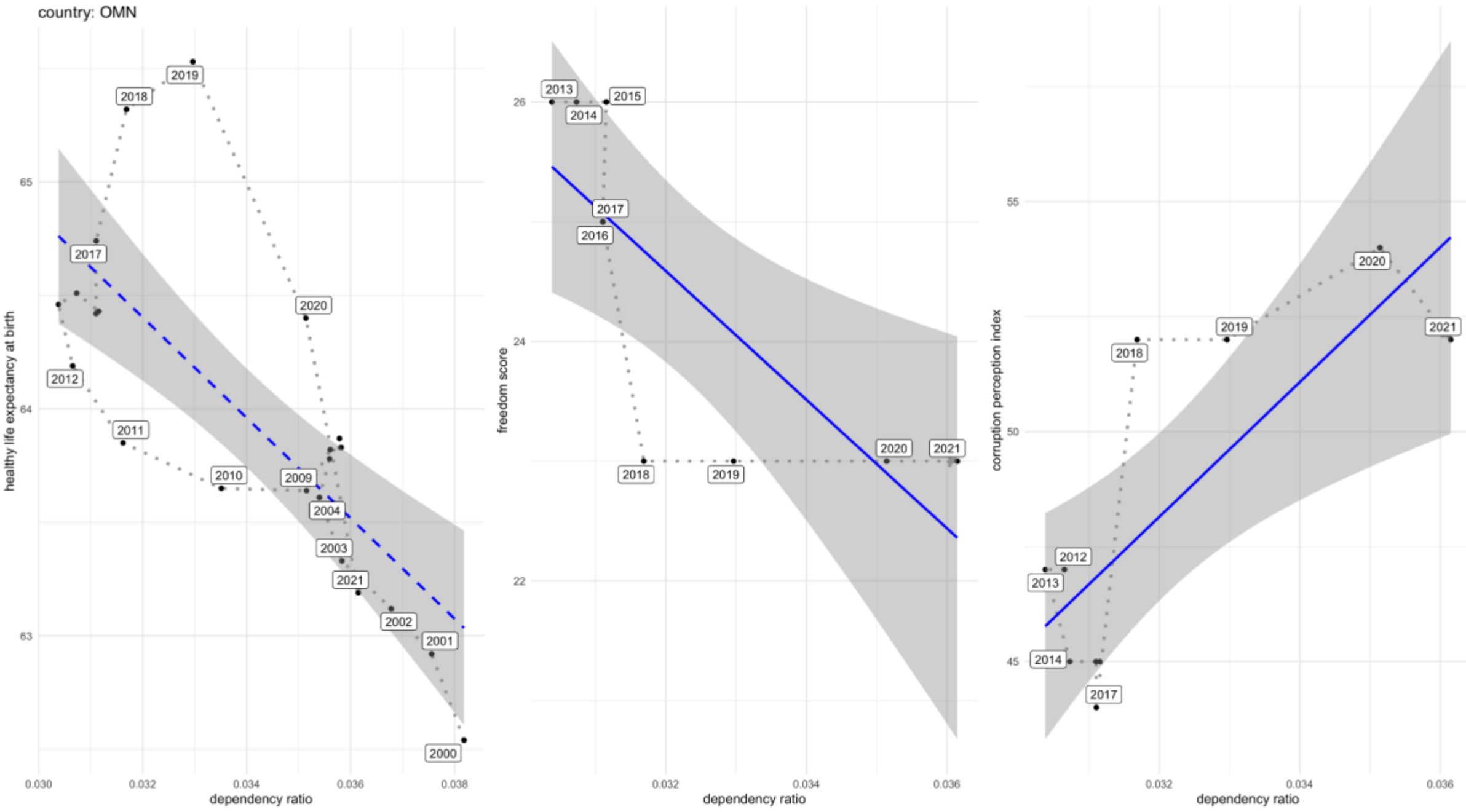

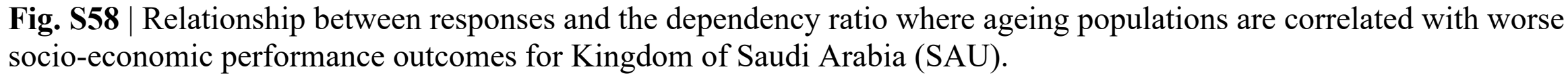

**Fig. S58** | Relationship between responses and the dependency ratio where ageing populations are correlated with worse socio-economic performance outcomes for Kingdom of Saudi Arabia (SAU).

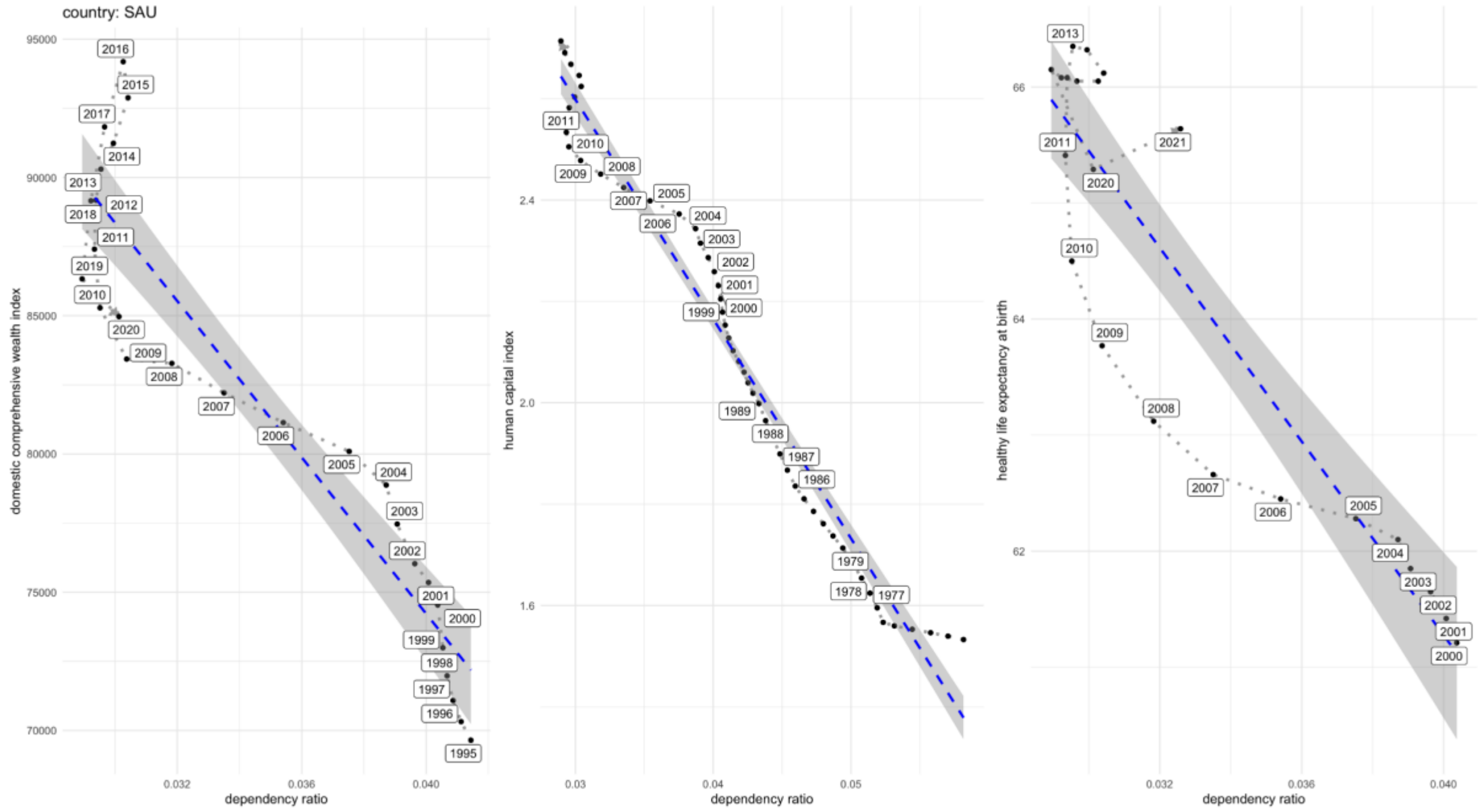

**Fig. S59** | Relationship between responses and the dependency ratio where ageing populations are correlated with worse socio-economic performance outcomes for Sierra Leone (SLE).

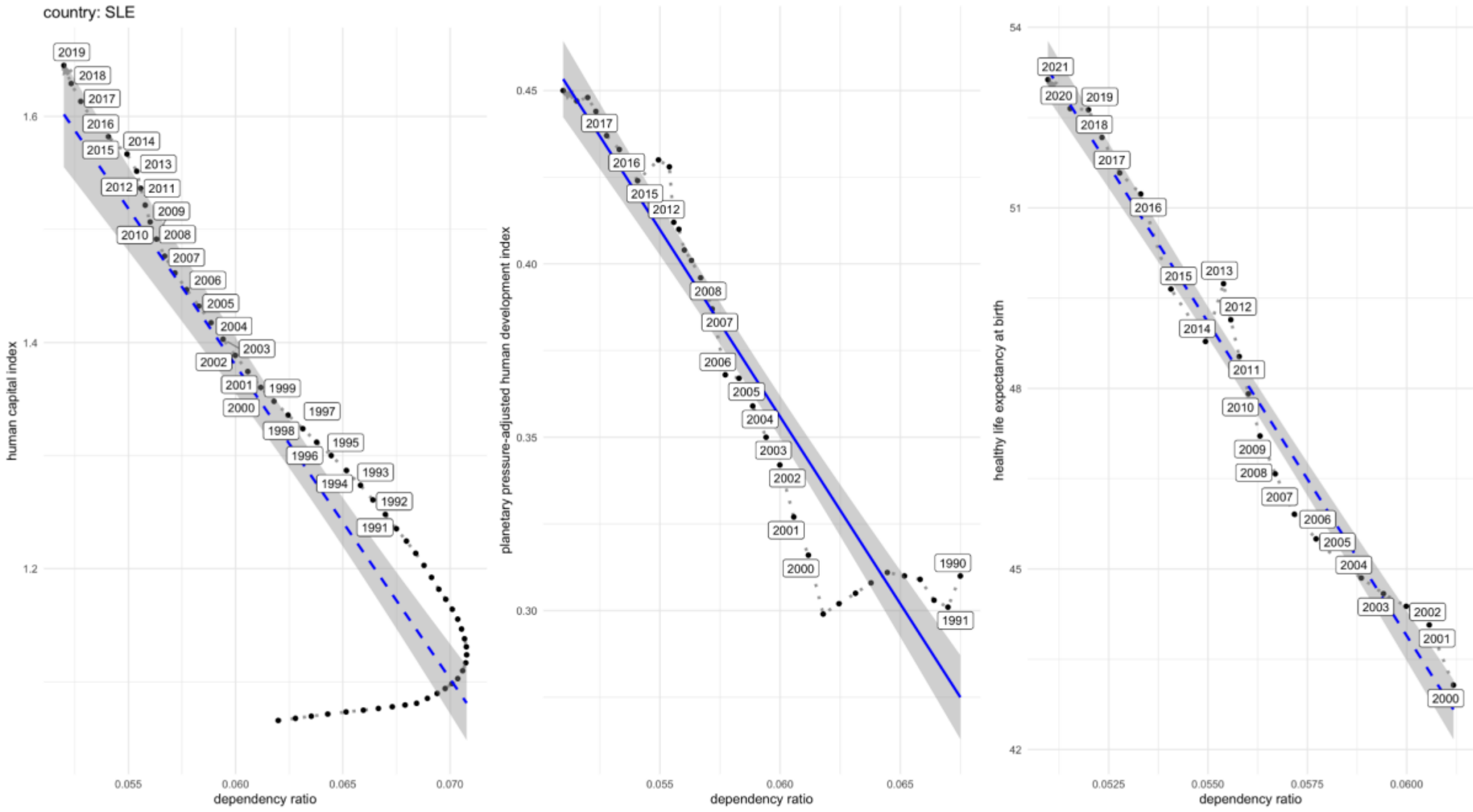

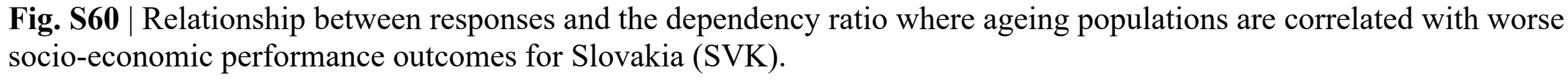

**Fig. S60** | Relationship between responses and the dependency ratio where ageing populations are correlated with worse socio-economic performance outcomes for Slovakia (SVK).

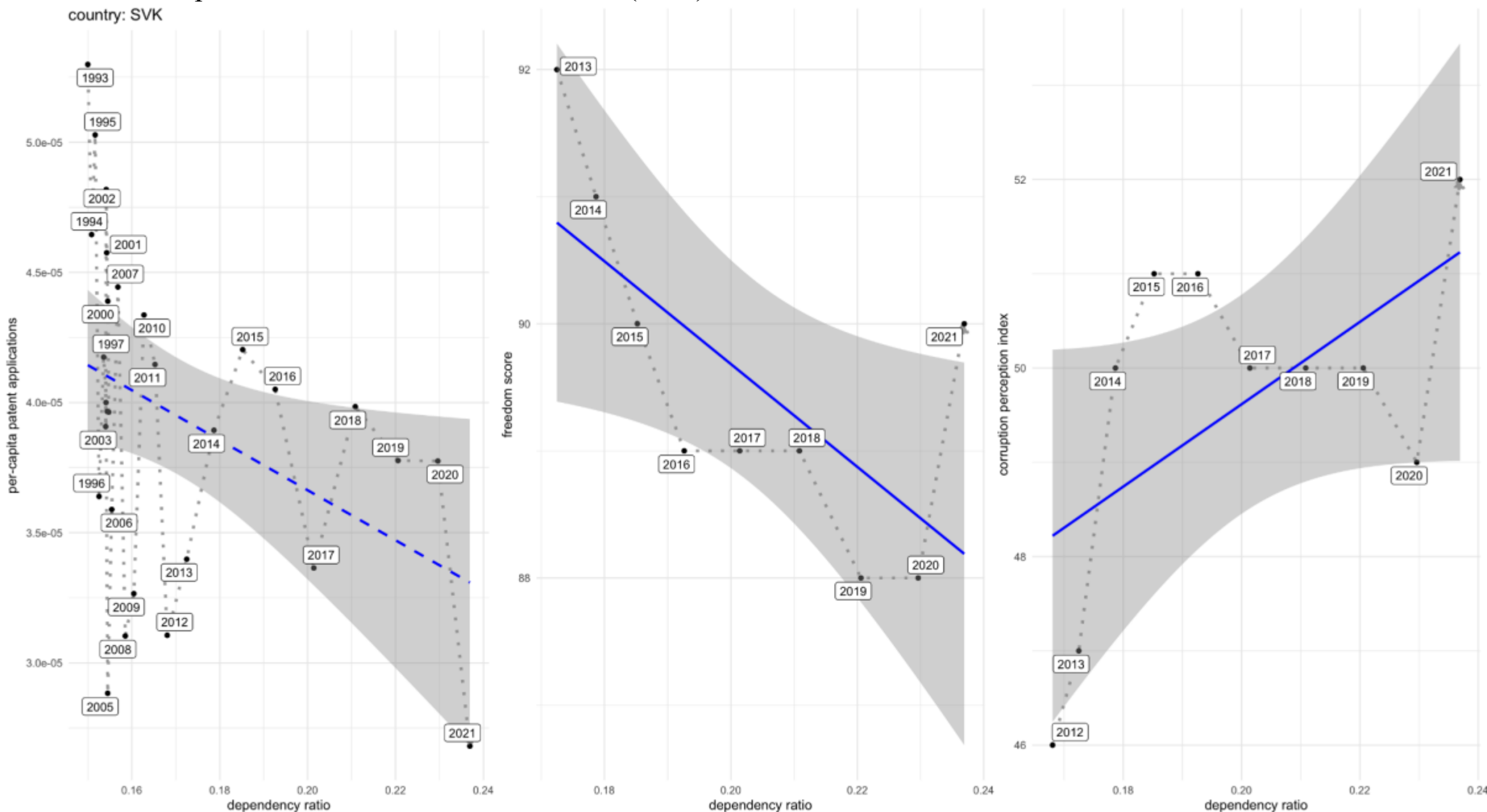

**Fig. S61** | Relationship between responses and the dependency ratio where ageing populations are correlated with worse socio-economic performance outcomes for Tajikistan (TJK).

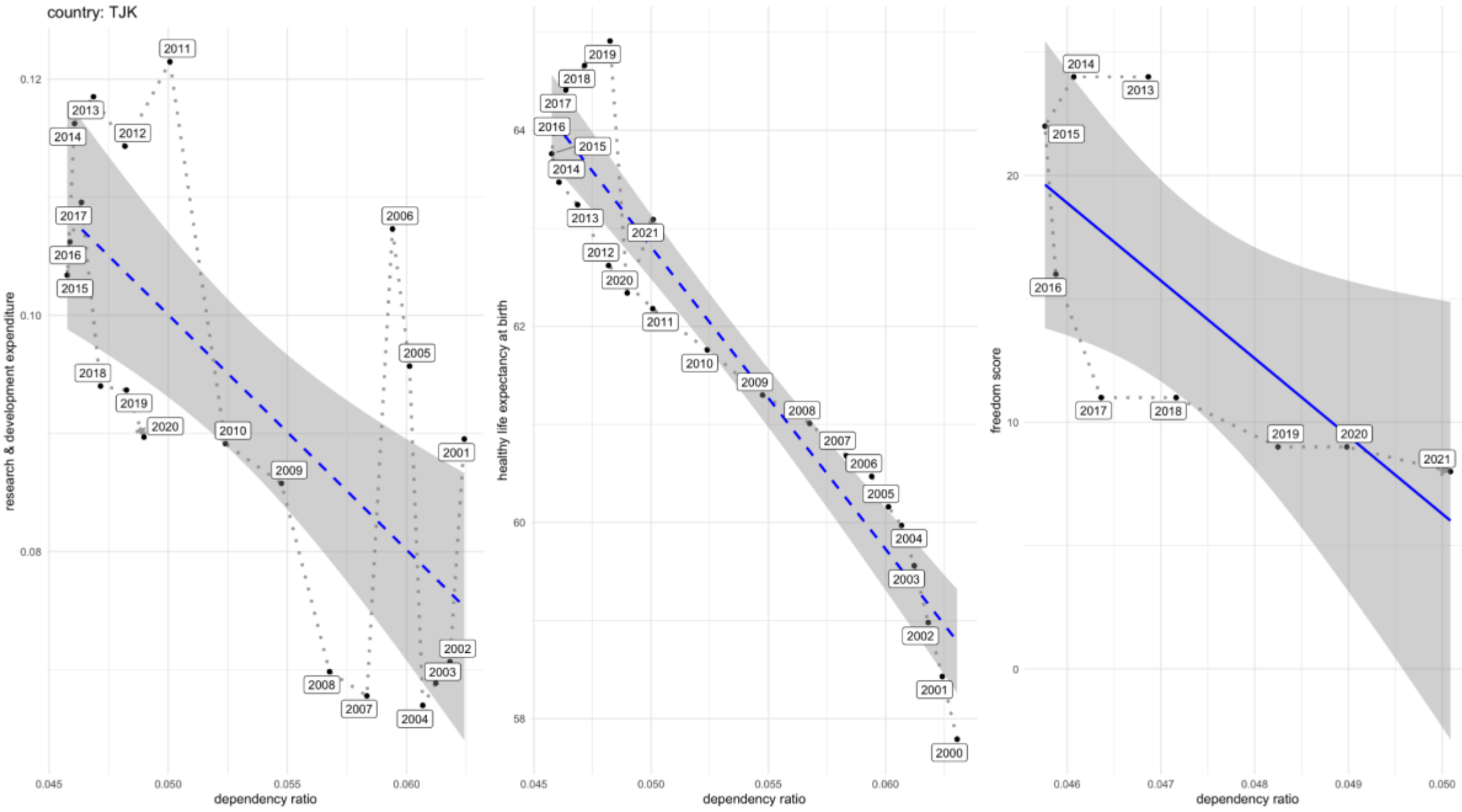

**Fig. S62** | Relationship between responses and the dependency ratio where ageing populations are correlated with worse socio-economic performance outcomes for Venezuela (VEN).

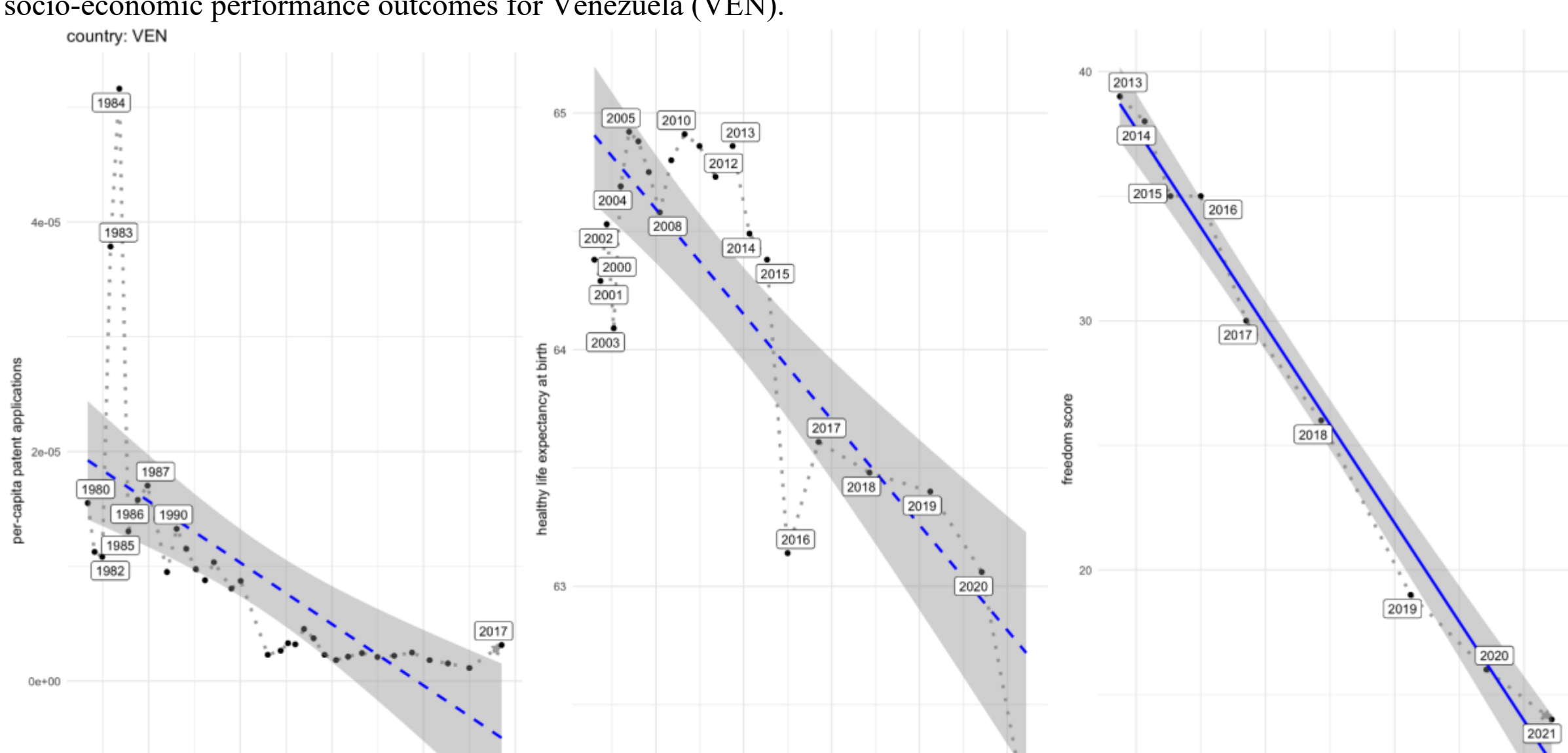

None of the 19 countries is in population decline; in fact, the mean dependency ratio and rate of population change for those countries are clustered around the mean value for African countries (Fig. S63). In other words, these countries are fast-growing and young, so the negative relationships identified above are likely driven by other socio-economic and political conditions.

**Fig. S63** | Relationship between mean dependency ratio (1950–2021) and mean rate of population change ($r$) (1950–2021) for the 19 countries with ≥ 3 responses where an ageing population is correlated with a worse socio-economic outcome. The large, coloured circles represent the means among countries in each major region (Europe in dark blue; South America-Caribbean in red; Asia-Oceania in gold; North America in light blue; Africa in green; Middle East in silver). Three of the four quadrants named: older, slower growth (upper left); younger, slower growth (lower left); younger, faster growth (lower right). Countries: SVK = Slovakia; GAB = Gabon; KGZ = Kyrgyzstan; BOL = Bolivia; MNG = Mongolia; CRI = Costa Rica; BEN = Benin; VEN = Venezuela; BLZ = Belize; IDN = Indonesia; SLE = Sierra Leone; BDI = Burundi; TJK = Tajikistan; IRQ = Iraq; GMB = The Gambia; OMN = Oman; MOZ = Mozambique; CIV = Côte d'Ivoire; SAU = Kingdom of Saudi Arabia.

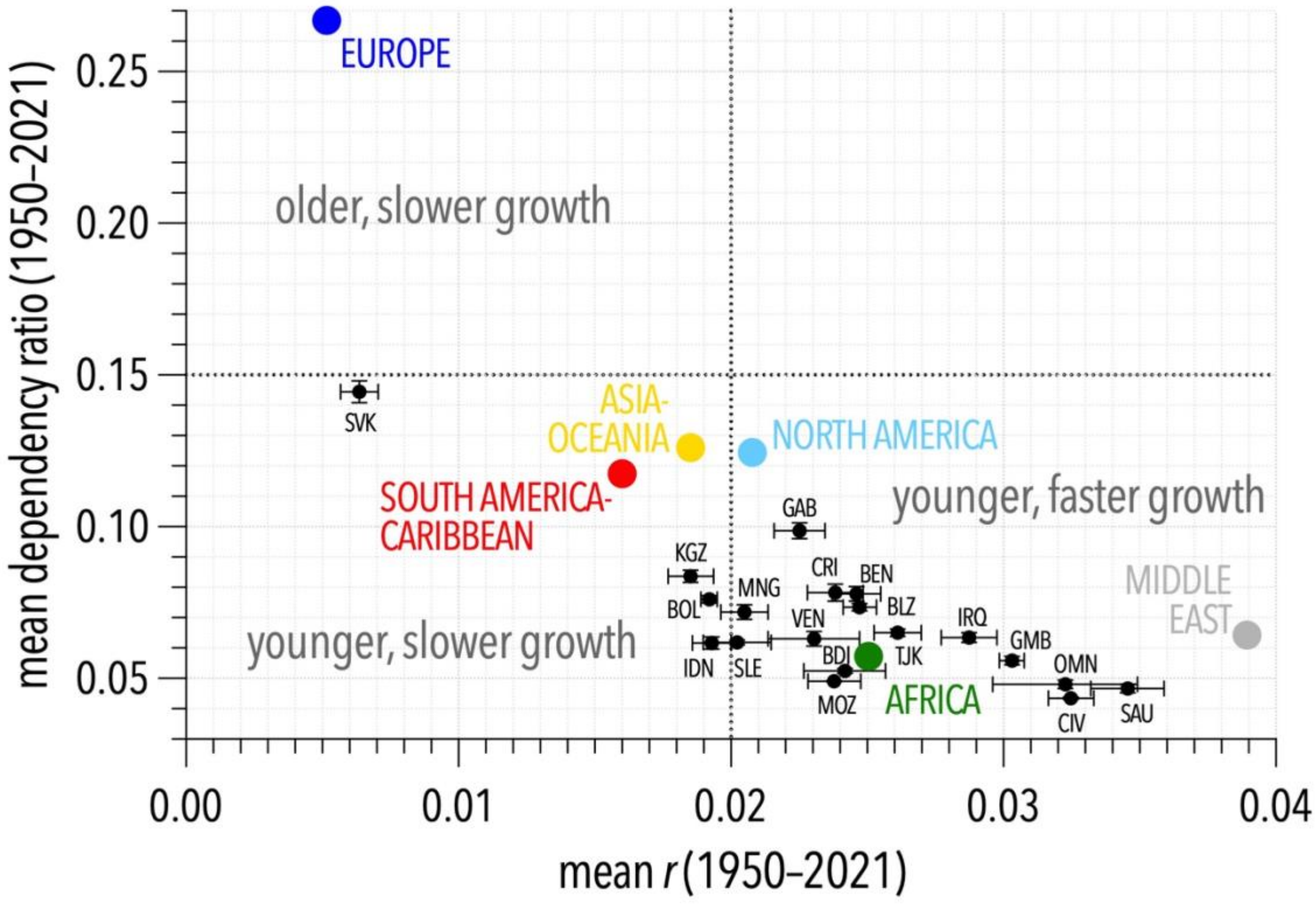

## Appendix X: Within-region comparisons

General relationships between dependency ratio and the principal socio-economic responses within global regions.

**Table S2** | Within-continent linear relationships between logit dependency ratio and transformed socio-economic responses. Results are shown only where the two-sided slope test type I error estimate was < 0.05; all omitted regional relationships had slopes that could not be differentiated statistically from zero. Positive slopes indicate higher values for older population structures, except that negative slopes indicate more favourable outcomes for Gini coefficient (greater equality) and wellbeing rank (better wellbeing). *n* = number of countries; $\beta$ = regression slope; $\rho$ = Spearman's $\rho$; $R^2$ = adjusted coefficient of determination; *P* = estimated type I error. *DCWI* = domestic comprehensive wealth index; *R&D* = research and development expenditure; *PAT* = patent applications; *GINI* = Gini coefficient; *PPHDI* = planetary pressure-adjusted Human Development Index; HLE = healthy life expectancy at birth; *FREE* = freedom score; *CPI* = corruption perception index; *WBR* = wellbeing rank; *GDP* = per-capita GDP. AFR = Africa; ASIOC = Asia–Oceania; EUR = Europe; NAM = North America; SACAR = South America–Caribbean; ME = Middle East.

| response | region | *n* | $\beta$ | $\rho$ | $R^2$ | *P* |
|---|---|---|---|---|---|---|
| *DCWI* | AFR | 46 | 0.39 | 0.40 | 0.18 | 0.0020 |
| | ASIOC | 28 | 0.54 | 0.51 | 0.42 | <0.001 |
| | EUR | 40 | 0.56 | 0.27 | 0.08 | 0.038 |
| | NAM | 10 | 0.90 | 0.84 | 0.85 | <0.001 |
| | SACAR | 16 | 0.59 | 0.71 | 0.54 | <0.001 |
| *R&D* | ASIOC | 25 | 1.61 | 0.78 | 0.57 | <0.001 |
| | EUR | 40 | 1.28 | 0.47 | 0.17 | 0.0050 |
| | NAM | 9 | 2.36 | 0.88 | 0.89 | <0.001 |
| *PAT* | AFR | 30 | 1.08 | 0.46 | 0.23 | 0.0043 |
| | A–O | 26 | 1.42 | 0.75 | 0.56 | <0.001 |
| | NAM | 10 | 1.65 | 0.60 | 0.64 | 0.0035 |
| *GINI* | NAM | 9 | −0.31 | −0.42 | 0.51 | 0.018 |
| *PPHDI* | AFR | 40 | 0.63 | 0.34 | 0.21 | 0.0019 |
| | ASIOC | 26 | 0.33 | 0.72 | 0.47 | <0.001 |
| | EUR | 38 | 0.41 | 0.45 | 0.15 | 0.0092 |
| | ME | 11 | 0.46 | 0.72 | 0.47 | 0.012 |
| | SACAR | 12 | 0.69 | 0.92 | 0.76 | <0.001 |
| *HLE* | AFR | 46 | 5.81 | 0.33 | 0.18 | 0.0022 |
| | ASIOC | 28 | 5.28 | 0.67 | 0.54 | <0.001 |
| | NAM | 10 | 3.77 | 0.67 | 0.35 | 0.043 |
| | SACAR | 16 | 7.46 | 0.79 | 0.59 | <0.001 |
| *FREE* | AFR | 46 | 1.09 | 0.26 | 0.07 | 0.047 |
| | ASIOC | 28 | 1.29 | 0.33 | 0.24 | 0.0045 |
| | EUR | 39 | 2.37 | 0.31 | 0.16 | 0.0072 |
| | NAM | 10 | 1.77 | 0.58 | 0.37 | 0.037 |
| | SACAR | 16 | 2.28 | 0.73 | 0.64 | <0.001 |
| *CPI* | ASIOC | 28 | 0.87 | 0.72 | 0.50 | <0.001 |
| | EUR | 40 | 0.94 | 0.33 | 0.09 | 0.038 |
| | NAM | 9 | 1.56 | 0.98 | 0.83 | <0.001 |
| | SACAR | 16 | 1.39 | 0.57 | 0.55 | <0.001 |
| *WBR* | ASIOC | 25 | −0.30 | −0.44 | 0.20 | 0.015 |
| *GDP* | AFR | 52 | 0.67 | 0.56 | 0.33 | <0.001 |
| | ASIOC | 47 | 0.44 | 0.68 | 0.44 | <0.001 |
| | NAM | 10 | 0.59 | 0.88 | 0.86 | <0.001 |
| | SACAR | 31 | 0.33 | 0.50 | 0.32 | <0.001 |

## Appendix XI: Lag analysis

Within-country lag-analysis (0 to 10 years) results.

**Table S3** | Summary of country-level best-scoring dependency-ratio lags. A non-zero lag was supported when it improved AIC by ≥ 2 or adjusted $R^2$ by ≥ 0.02 relative to the zero-lag model. IQR = interquartile range. $\beta$ = regression slope; $\rho$ = Spearman's $\rho$; $R^2$ = adjusted coefficient of determination. *DCWI* = domestic comprehensive wealth index; *R&D* = research and development expenditure; *PAT* = patent applications; *GINI* = Gini coefficient; *PPHDI* = planetary pressure-adjusted Human Development Index; HLE = healthy life expectancy at birth; *FREE* = freedom score; *CPI* = corruption perception index; *GDP* = per-capita GDP.

| response | *n* countries | supported non-0 lags, *n* (%) | median lag (IQR), years | median supported lag, years | median best $\rho$ | median adjusted $R^2$ |
|---|---|---|---|---|---|---|
| *DCWI* | 151 | 95 (62.9) | 6.0 (8.8) | 8.0 | 0.87 | 0.87 |
| *R&D* | 78 | 56 (71.8) | 4.0 (8.8) | 7.0 | 0.74 | 0.64 |
| *PAT* | 106 | 69 (65.1) | 4.0 (9.0) | 7.0 | −0.10 | 0.56 |
| *GINI* | 57 | 47 (82.5) | 4.0 (8.0) | 6.0 | −0.65 | 0.63 |
| *PPHDI* | 152 | 86 (56.6) | 4.0 (9.0) | 8.0 | 0.81 | 0.84 |
| *HLE* | 185 | 101 (54.6) | 2.0 (7.0) | 7.0 | 0.67 | 0.65 |
| *FREE* | 195 | 64 (32.8) | 0.0 (1.0) | 1.0 | −0.56 | 0.61 |
| *CPI* | 165 | 81 (49.1) | 1.0 (2.0) | 2.0 | 0.30 | 0.50 |
| *GDP* | 197 | 131 (66.5) | 5.0 (10.0) | 9.0 | 0.72 | 0.76 |

**Table S4** | Pooled lag-aligned relationships between dependency ratio and socio-economic responses. Each country aligned using its best-scoring lag. Country-means analysis contains one mean per country after alignment. All two-sided slope tests gave type I error estimate (p) < $1.3\times10^{-8}$. $\rho$ = Spearman's $\rho$; $R^2$ = adjusted coefficient of determination. *DCWI* = domestic comprehensive wealth index; *R&D* = research and development expenditure; *PAT* = patent applications; *GINI* = Gini coefficient; *PPHDI* = planetary pressure-adjusted Human Development Index; HLE = healthy life expectancy at birth; *FREE* = freedom score; *CPI* = corruption perception index; *GDP* = per-capita GDP.

| response | level | observations (countries) | $\beta$ | $\rho$ | $R^2$ |
|---|---|---|---|---|---|
| *DCWI* | aligned years | 3106 (150) | 0.42 | 0.61 | 0.29 |
| | country means | 150 (150) | 0.42 | 0.61 | 0.28 |
| *R&D* | aligned years | 1435 (78) | 1.08 | 0.60 | 0.39 |
| | country means | 78 (78) | 1.11 | 0.63 | 0.40 |
| *PAT* | aligned years | 2900 (106) | 1.00 | 0.72 | 0.51 |
| | country means | 106 (106) | 1.06 | 0.76 | 0.57 |
| *GINI* | aligned years | 1087 (57) | −0.43 | −0.57 | 0.43 |
| | country means | 57 (57) | −0.44 | −0.54 | 0.44 |
| *PPHDI* | aligned years | 3910 (152) | 0.50 | 0.79 | 0.50 |
| | country means | 152 (152) | 0.50 | 0.83 | 0.55 |
| *HLE* | aligned years | 3431 (185) | 5.69 | 0.73 | 0.38 |
| | country means | 185 (185) | 5.53 | 0.72 | 0.37 |
| *FREE* | aligned years | 1633 (190) | 1.10 | 0.63 | 0.35 |
| | country means | 190 (190) | 1.12 | 0.64 | 0.37 |
| *CPI* | aligned years | 1492 (165) | 0.62 | 0.60 | 0.33 |
| | country means | 165 (165) | 0.65 | 0.62 | 0.34 |
| *GDP* | aligned years | 5153 (197) | 0.40 | 0.63 | 0.29 |
| | country means | 197 (197) | 0.40 | 0.63 | 0.29 |

## References cited in the Supplementary Information